\documentclass[aps,twocolumn,amsmath,amssymb,floatfix,reprint,footinbib,superscriptaddress,longbibliography,showkeys]{revtex4-1}
\usepackage{graphicx}
\usepackage{mathrsfs}
\usepackage{mdwlist}
\usepackage{subfigure}
\usepackage{booktabs}
\usepackage{amsmath}
\usepackage{extarrows}
\usepackage{dsfont}
\usepackage{amstext}
\usepackage{amsbsy}
\usepackage{bbm}
\usepackage{bm}
\usepackage{amssymb,mathrsfs,amsmath}
\usepackage{threeparttable}
\usepackage{booktabs}
\usepackage{bbding,pifont}
\usepackage{amsthm}
\usepackage{graphicx}
\usepackage{textcomp}
\usepackage{multirow}%
\usepackage{pifont}
\usepackage{color}
\usepackage{sansmath}
\usepackage{makecell}
\usepackage{tabularx}
\usepackage[colorlinks,citecolor=blue]{hyperref}
\definecolor{Dgreen}{RGB}{0, 100, 0}
\usepackage{url}
\usepackage[colorlinks]{hyperref}
\hypersetup{%
	plainpages=true,
	breaklinks=true,  
	hypertexnames=false,  
	pageanchor=true,
	colorlinks=true,
	linkcolor={blue},
	citecolor={blue},
	urlcolor={blue},
	anchorcolor={black}
}

\newcommand\sst{\scriptscriptstyle}

\begin{document}

\title{Composite pulses for high fidelity population transfer in  three-level systems}

\author{Zhi-Cheng Shi}
\affiliation{Fujian Key Laboratory of Quantum Information and Quantum Optics (Fuzhou University), Fuzhou 350108, China}
\affiliation{Department of Physics, Fuzhou University, Fuzhou 350108, China}

\author{Cheng Zhang}
\affiliation{Fujian Key Laboratory of Quantum Information and Quantum Optics (Fuzhou University), Fuzhou 350108, China}
\affiliation{Department of Physics, Fuzhou University, Fuzhou 350108, China}

\author{Du Ran}
\affiliation{School of Electronic Information Engineering, Yangtze Normal University, Chongqing 408100, China}
\affiliation{Department of Electrical Engineering Physical Electronics, Tel Aviv University, Ramat Aviv 69978, Israel}

\author{Yan Xia}\thanks{xia-208@163.com}
\affiliation{Fujian Key Laboratory of Quantum Information and Quantum Optics (Fuzhou University), Fuzhou 350108, China}
\affiliation{Department of Physics, Fuzhou University, Fuzhou 350108, China}

\author{Reuven Ianconescu}
\affiliation{Department of Electrical Engineering Physical Electronics, Tel Aviv University, Ramat Aviv 69978, Israel}
\affiliation{Shenkar College of Engineering and Design 12, Anna Frank St., Ramat Gan, Israel}

\author{Aharon Friedman}
\affiliation{Schlesinger Family Accelerators Center, Ariel University, Ariel 40700, Israel}

\author{X. X. Yi} \thanks{yixx@nenu.edu.cn}
\affiliation{\mbox{Center for Quantum Sciences and School of Physics, Northeast Normal University, Changchun 130024, China}}

\author{Shi-Biao Zheng}
\affiliation{Fujian Key Laboratory of Quantum Information and Quantum Optics (Fuzhou University), Fuzhou 350108, China}
\affiliation{Department of Physics, Fuzhou University, Fuzhou 350108, China}

\begin{abstract}
In this work,
we propose a composite pulses scheme by modulating phases to achieve high fidelity population transfer in three-level systems.
To circumvent the obstacle that not enough variables are exploited to eliminate the systematic errors in the transition probability,
we put forward a cost function to find the optimal value.
The cost function is independently constructed either in ensuring an accurate population of the target state,
or in suppressing the population of the leakage state, or both of them.
The results demonstrate that population transfer is implemented with high fidelity even when existing the deviations in the coupling coefficients.
Furthermore, our composite pulses scheme can be extensible to arbitrarily long pulse sequences.
As an example,
we employ the composite pulses sequence for achieving the three-atom singlet state in an atom-cavity system with ultrahigh fidelity.
The final singlet state shows robustness against deviations and is not seriously affected by waveform distortions.
Also, the singlet state maintains a high fidelity under the decoherence environment.
\end{abstract}

\maketitle

\section{Introduction}

Implementation of high fidelity quantum coherent control is a top priority in quantum information processing (QIP) \cite{PhysRevLett.74.3776,RevModPhys.70.1003,RevModPhys.73.357,PhysRevLett.96.246601, PhysRevA.61.041801,PhysRevA.65.051803,PhysRevA.102.012221,PhysRevLett.126.153403,PhysRevLett.121.123603,PhysRevLett.111.050404}.
Many works \cite{osti7365050,physchem.52.1.763,RevModPhys.91.045001},
which design different kinds of pulse shapes,
have been devoted to ensuring a remarkable quantum computing performance.
Originally,
the resonant pulse (RP) technique,
where the frequency of radiation exactly matches the transition frequency,
is regarded as a popular tool to accurately achieve coherent control,
due to its fast operation and simple waveform \cite{scully97}.
For example,
one can achieve complete population inversion through a $\pi$-pulse,
or a maximum superposition state through a $\pi/2$-pulse \cite{Gerry2004}.
However,
the resonant pulse is extremely susceptible to external perturbations,
such as fluctuations of control fields.
To overcome this difficulty,
the adiabatic passage (AP) technique has been developed \cite{Shankar1994,RevModPhys.79.53}.
The AP technique selects one eigenstate of the system Hamiltonian as an evolution path and propels the initial state to adiabatically evolve along this path.
In this technique,
the robustness against parameter perturbations is improved at the cost of the slow evolution rate and lower fidelity.
To enjoy both the ultrahigh fidelity of RP and the robustness of AP,
one can adopt the composite pulses (CPs) technique.

The CPs technique, comprised of a well-organized train of constant pulses with determined pulse areas and relative phases,
was conceived in early nuclear magnetic resonance (NMR) \cite{Wimperis1991,Wimperis1994,Levitt1979,Levitt1986}.
To date, this technique has been used extensively in QIP \cite{PhysRevA.70.052318,PhysRevA.101.012321,PhysRevA.83.053420,PhysRevLett.113.043001,PhysRevA.97.043408,PhysRevA.87.052317,PhysRevA.84.065404, PhysRevA.99.013424,PhysRevA.102.013105,PhysRevA.101.013827}.
One unique feature of CPs is that the pulse sequence is available to compensate for the systematic error in any physical parameters (e.g., pulse duration, pulse amplitude, detuning, Stark shift, etc.) \cite{PhysRevLett.113.043001,PhysRevA.70.052318,Levitt1986,PhysRevA.101.012321,PhysRevA.83.053420}.
Previous works \cite{PhysRevA.92.033406,PhysRevA.84.062311,PhysRevA.93.032340,PhysRevA.100.032333,PhysRevA.89.022310, PhysRevA.92.060301,PhysRevA.93.023830} about precise quantum control are mostly based on two-level systems.
For example,
by using a narrow-band composite sequence of laser pulses,
the error-tolerance region of high fidelity local addressing operation in trapped ions and atoms is improved by $20\%$ \cite{Ivanov2011}.
Besides, an arbitrary single-qubit gate, which is robust against pulse area error, could be implemented by a composite $\theta$ pulses sequence \cite{PhysRevA.99.013402}.

At present,
works on CPs in two-level systems are relatively complete and extensive \cite{RevModPhys.76.1037}.
As is well known,
the three-level system is another representative physical model in QIP.
Some interesting phenomena, such as coherent trapping \cite{scully97}, electromagnetically induced transparency \cite{RevModPhys.77.633}, and lasing without inversion \cite{PhysRevLett.62.2813} manifest in three-level systems.
Recently,
works on a general use of CPs scheme in three-level systems are still deficient \cite{Levitt1984,Ramamoorthy1991,PhysRevLett.106.233001,PhysRevA.87.043418, PhysRevA.88.063406,Ishida2018}. The main difficulty is that the analytical expression of the propagator is very complicated for a general three-level system.
One way to deal with this difficulty is to reduce the three-level system into one or more two-level systems.
In Ref.~\cite{PhysRevResearch.2.043235},
the closed-loop three-level system of the chiral molecule is divided into three independent two-level systems \cite{Lehmann2018,Leibscher2019}.
Then, the CPs sequence using RP in these two-level systems achieves the chiral resolution necessary to overcome Rabi frequency errors and detuning errors.
In Refs.~\cite{PhysRevA.84.063413,PhysRevResearch.2.043194},
two powerful tools,
Morris-Shore transformation \cite{PhysRevA.27.906}
and Majorana decomposition \cite{Majorana1932}, are exploited to simplify a three-level system into an effective two-level system, so as to obtain the propagator of the system. Note that some approximations or restrictions are usually adopted in this simplification process.
Specifically, Morris-Shore transformation \cite{PhysRevA.27.906} is applied when the system can be mapped into two sets of states with the same energies (in the rotating-wave approximation) and two coupling coefficients must have the same time dependence, while Majorana decomposition \cite{Majorana1932} always demands that the system have the SU(2) dynamic symmetry.
The common feature of these approaches \cite{PhysRevA.84.063413,PhysRevResearch.2.043194,PhysRevA.27.906,Majorana1932} is to reduce the dynamics of the three-level system to a two-level system dynamics. As a result, one ignores the dynamics of the excited (leakage) state and thus the system does not exist the population leakage. However, the dynamics of the excited state should be considered in the three-level system, and that is what makes a three-level system different from a two-level system.
It is therefore
necessary to conceive a general CPs sequence able to directly achieve precise and robust control in a three-level system.

In three-level systems, there are some issues that must be considered.
For instance,
during the construction process of qubits,
the existence of the leakage state,
which is beyond the computational subspace,
leads to an imperfect quantum operation performance.
This issue is inevitable and poses a great threat to the fidelity of qubit operations.
The works on suppressing leakage in three-level systems are ongoing \cite{PhysRevB.95.241307,PhysRevA.103.052612}. Besides, compared to the two-level system,
the additional modulation parameters in the three-level system may lead to various types of systematic errors.
The existence of these systematic errors has a detrimental effect on the quantum operation accuracies, to various degrees.
On the other hand,
as shown in Refs.~\cite{PhysRevLett.113.043001,PhysRevA.83.053420,PhysRevA.97.043408,PhysRevA.101.012321},
the fidelity improves with the increase in the number of pulses.
However, considering the finite coherence time,
a large number of pulses would cost long evolution time,
which is tremendously detrimental for quantum computation.
Therefore, one generally requires the number of pulses to be as small as possible, hence it is essential to learn how to efficiently eliminate the systematic errors within finite pulses.
This is especially true in the case where the quantum systems have various types of deviations.

In this work,
we achieve arbitrary population transfer by phase modulation CPs in the three-level system. We directly deduce the transition probability of each state, including the excited state. In other words, we concern the dynamics of the excited state as well.
By using the Taylor expansion,
the transition probability of the target (leakage) state can be sorted by different order derivatives.
Then, we construct a cost function to overcome the difficulty that there are not enough variables to eliminate high-order derivatives in the short pulse sequence.
Different cost functions have respective purposes for accurately controlling population transfer or effectively suppressing population leakage, or both of them.
We study minutely two kinds of short pulse sequences as examples to demonstrate the pulse design process, which can easily extend to longer pulse sequences.
For applications,
our CPs scheme further shows the feasibility of achieving the three-atoms singlet state with ultrahigh fidelity in an atoms-cavity system.
The results indicate that the final singlet state is robust against several kinds of defects,
including deviations in coupling coefficients, waveform distortions, and the disturbances of the decoherence environment.

The paper is organized as follows.
In Sec.~\ref{II},
we design the CPs sequence in the three-level system through minimizing the cost function.
In Sec.~\ref{III},
we first employ different cost functions to analytically search for the best group of phases in the two-pulse sequence.
For more than two pulses, we adopt numerical methods to find different phases of the CPs sequence. In principle, the numerical methods are suitable for arbitrary long pulse sequences.
In Sec.~\ref{IV}, we study the feasibility of the CPs sequence through a specific application in the atom-cavity system.
Finally, a conclusion is given in Sec.~\ref{V}.

\section{Toy model and the general theory} \label{II}

Consider a $\Lambda$-type three-level quantum system interacting with two external fields, where the three-level system has
two ground states $|g\rangle$ and $|r\rangle$, and an excited state $|e\rangle$. The ground states $|g\rangle$ and $|r\rangle$, acting as a qubit, cannot be immediately coupled to each other. Hence, we require an extra excited state $|e\rangle$ to construct the indirect coupling between two ground states.
More specifically, the transition $|g\rangle\leftrightarrow|e\rangle$ ($|r\rangle\leftrightarrow|e\rangle$) is driven by the external field with the coupling strength $\Omega_1$ ($\Omega_2$), the phase $\alpha$ ($\beta$), and the detuning $\Delta$. In presence of deviations in the external fields, in the interaction picture, the system Hamiltonian is given by ($\hbar=1$)
\begin{equation}
H\!=\!\Delta|e\rangle \langle e|\!+\!(1+\epsilon_1)\Omega_1e^{i\alpha}|g\rangle\langle e|\!+\!(1+\epsilon_2)\Omega_2e^{i\beta}|r\rangle\langle e|\!+\!\mathrm{H.c.},
\end{equation}
where the deviations $\epsilon_1$ and $\epsilon_2$ are random unknown constants, which would give rise to the systematic errors during quantum operations.

Here, we do not intend to specify a concrete physical system, because the three-level system studied here can be found in very different physics research fields ranging from high precision spectroscopy, to quantum information processing, NMR, and metrology.
For example, this physical model is quite familiar in the quantum system of a three-level atom interacting with two laser fields \cite{scully97}.
In the atomic system, the inhomogeneous distribution of the laser fields leads to different interaction coefficients between the atom and the laser fields. Also, the spatial position of the atom cannot be exactly determined due to its micro-vibration. As a result, the interaction coefficients in the system are susceptible to deviations.
It is worth mentioning that similar energy-level structures can also be found in trapped ions \cite{RevModPhys.75.281}, diamond nitrogen-vacancy centers \cite{Yang2010}, and superconducting circuits \cite{RevModPhys.85.623,PhysRevApplied.7.054022,PhysRevA.94.052311,PhysRevA.96.022304,PhysRevA.97.022332,PhysRevApplied.14.034038}.
In addition, some complicated quantum systems, such as the electrons in semiconductors \cite{PhysRevB.70.235317},  the spin chain \cite{PhysRevA.85.022312,Chen2016}, neutral atoms interactions \cite{PhysRevA.98.062338,PhysRevA.97.042336,Shi:18,PhysRevA.101.042328,PhysRevA.102.053118,PhysRevA.103.012601}, and massive quantum
particles in an optical Lieb lattice \cite{Taie2020}, can be reduced to the three-level physical model as well.

\begin{figure}[t]
	\centering
	\includegraphics[scale=0.065]{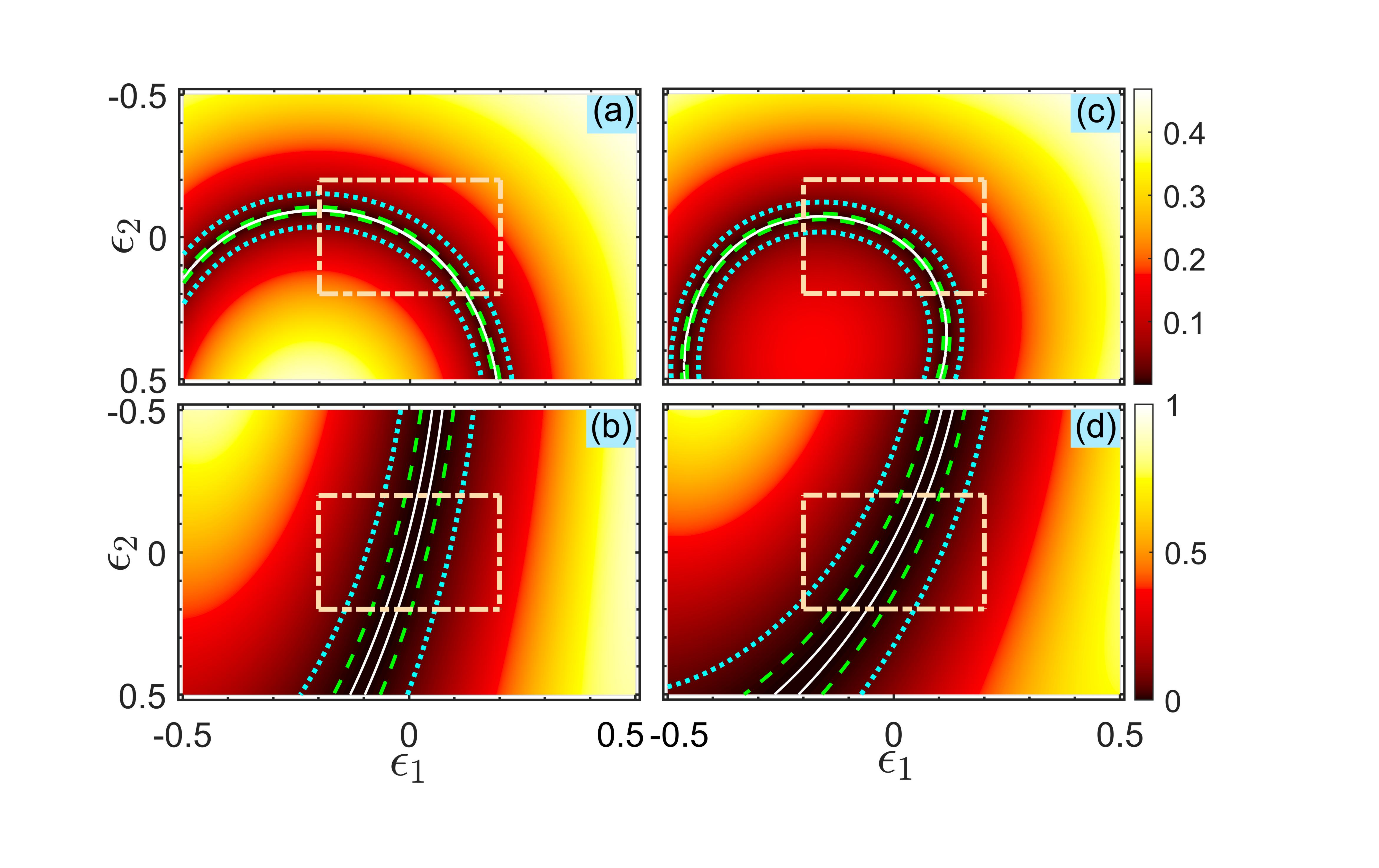}
	\caption{Infidelity $\mathcal{F}_r$ (top panels) and population leakage $P_e^a$ (bottom panels) vs the deviations $\epsilon_1$ and $\epsilon_2$ in the resonant pulses, where the infidelity $\mathcal{F}_r=|P_r^a-\mathcal{P}_r|$ quantifies the population drift in the desired population $\mathcal{P}_r$. Here, $\mathcal{P}_r=1/2$ in the left column, while $\mathcal{P}_r=3/4$ in the right column. Henceforth, the white-solid curves, the green-dashed curves, and the cyan-dotted curves correspond to $\mathcal{F}_r(P_e^a)=0.001$, 0.01, and 0.05, respectively. The rectangular region (labeled by the yellow-dot-dashed lines) are: $|\epsilon_1|\leq0.2$ and $|\epsilon_2|\leq0.2$. The results demonstrate that the population transfer suffers from the deviations in the resonant pulses. Infidelity is particularly noticeable, since the region enclosed by white-solid curves is almost negligible. }  \label{fig:01a}
\end{figure}

Assume that the system is initially in the ground state $|g\rangle$. If there are no deviations in the external fields (i.e., $\epsilon_1=0$ and $\epsilon_2=0$), one can easily achieve perfect population transfer for the qubit. To this end, we choose the evolution time $T=2\pi/\sqrt{\Delta^2+4(\Omega_1^2+\Omega_2^2)}$, then the population of the ground state $|r\rangle$ becomes
\begin{equation}
\mathcal{P}_{r}=\frac{4\Omega_1^2\Omega_2^2\cos^2\frac{\Delta T}{4}}{(\Omega_1^2+\Omega_2^2)^2}.
\end{equation}
Thus, the value of $\mathcal{P}_{r}$ can be modulated by altering either the detuning or the ratio of two coupling coefficients.
However, in presence of deviations in the external fields, the actual population $P_r^a$ would keep away from the desired value $\mathcal{P}_{r}$. This is verified in Figs.~\ref{fig:01a}(a) and \ref{fig:01a}(c), which demonstrate that the infidelity $\mathcal{F}_r=|P_r^a-\mathcal{P}_r|$ is high even for small deviations.
Note that Figs.~\ref{fig:01a}(a) and \ref{fig:01a}(c) also demonstrate that the value of infidelity becomes extremely low in some regions where the deviations are large, e.g., $\mathcal{F}_r\approx8\times10^{-5}$ when $\epsilon_1\approx-0.46$ and $\epsilon_2\approx0.42$ in Fig.~\ref{fig:01a}(c). However, it is meaningless in practice, because the values of the deviations $\epsilon_1$ and $\epsilon_2$ are unknown in principle and they are usually independent of each other.
On the other hand, population leakage from the ground states to the excited state invariably does happen in the presence of deviations, which can be observed in Figs.~\ref{fig:01a}(b) and \ref{fig:01a}(d).
Hence, the goal of this work is to precisely achieve the predefined population $\mathcal{P}_{r}$, and meanwhile, to dramatically suppress the population leakage by designing a composite $N$-pulse sequence.

First, we require to calculate the population of the ground state $|r\rangle$ in absence of the deviations.
Assume that the Hamiltonian of the $n$th pulse has the following form ($n=1,\dots,N$)
\begin{eqnarray}  \label{h3}
H_n&=&\Delta_n|e\rangle \langle e|+(1+\epsilon_1)\Omega_{1n}e^{i\alpha_n}|g\rangle\langle e|  \nonumber\\[1ex]
&&+(1+\epsilon_2)\Omega_{2n}e^{i\beta_n}|r\rangle\langle e|+\mathrm{H.c.}
\end{eqnarray}
For the pulse duration $T_n=2\pi/\sqrt{\Delta_n^2+4(\Omega_{1n}^2+\Omega_{2n}^2)}$, the corresponding propagator, labelled as $U(\theta_n,\gamma_n)$, can be written as (up to a global phase)
\begin{eqnarray}
U(\theta_n,\gamma_n)=\left(
                               \begin{array}{ccc}
                                 \cos\theta_ne^{i\Phi_n} & \sin\theta_ne^{-i\gamma_n} & 0 \\[1ex]
                                 \!-\sin\theta_ne^{i\gamma_n} & \cos\theta_ne^{-i\Phi_n} & 0 \\[1ex]
                                 0 & 0 & \!\!ie^{-i\Theta_n} \\
                               \end{array}
                             \right),    \nonumber
\end{eqnarray}
where
\begin{eqnarray}
\Theta_n&=&\frac{\pi\Delta_n}{2\sqrt{\Delta_n^2+4(\Omega_{1n}^2+\Omega_{2n}^2)}}, \nonumber\\[1ex] \Phi_n&=&\arctan\frac{(\Omega_{1n}^2-\Omega_{2n}^2)\cos\Theta_n}{(\Omega_{1n}^2+\Omega_{2n}^2)\sin\Theta_n}, \nonumber\\[1ex]
\theta_n&=&\arcsin\frac{2\Omega_{1n}\Omega_{2n}\cos\Theta_n}{\Omega_{1n}^2+\Omega_{2n}^2}, \nonumber\\[1ex]
\gamma_n&=&\beta_n-\alpha_n-\pi/2. \nonumber
\end{eqnarray}
Then, the total propagator of the composite $N$-pulse sequence can be written as
\begin{eqnarray}
U(\theta_N,\gamma_N)U(\theta_{N-1},\gamma_{N-1})\cdots U(\theta_2,\gamma_2)U(\theta_1,\gamma_1).  \nonumber
\end{eqnarray}
The detailed derivation of the total propagator is given in Appendix~\ref{a3}.
After some algebraic manipulations, the population of the ground state $|r\rangle$, labelled as $P^{\sst(0)}_{r}$, becomes
\begin{eqnarray}  \label{5}
P^{\sst(0)}_r=\sin^2\vartheta_{N},
\end{eqnarray}
where the expression of $\vartheta_{N}$ is also presented in Appendix~\ref{a3}.
For simplicity, we set $\Delta_n=0$ in the following, which means that the three-level system works in the resonant regime. Apparently, the following analysis can be easily generalized to the two-photon resonant regime.
Furthermore, we choose $\theta_n=\pi/4$ ($n=1,\dots,N$), which means that two coupling coefficients $\Omega_1$ and $\Omega_2$ remain unchanged in the $N$-pulse sequence. Note that it is also suitable for the other values of $\theta_n$. As a result, we only modulate the phases $\alpha_n$ and $\beta_n$, and the form of CPs now becomes
\begin{eqnarray}
U\left(\frac{\pi}{4},\gamma_N\right)U\left(\frac{\pi}{4},\gamma_{N-1}\right)\cdots U\left(\frac{\pi}{4},\gamma_2\right)U\left(\frac{\pi}{4},\gamma_1\right).   \nonumber
\end{eqnarray}

In presence of deviations in the external fields, by the Taylor expansion, the actual population $P_r^a$ of the ground state $|r\rangle$ can be written as
\begin{eqnarray}  \label{9}
P_r^a&=&P^{\sst(0)}_r+P^{\sst(1)}_r+P^{\sst(2)}_r+\cdots \nonumber\\[0.1ex]
     &=&C^{\sst(0)}_{\!N\!,1}+C^{\sst(1)}_{\!N\!,1}\epsilon_1+C_{\!N\!,2}^{\sst(1)}\epsilon_2+ C^{\sst(2)}_{\!N\!,1}\epsilon_1\epsilon_2+C^{\sst(2)}_{\!N\!,2}\epsilon_1^2 +C^{\sst(2)}_{\!N\!,3}\epsilon_2^2 \nonumber\\[0.1ex]
     && +{O}(\epsilon_1\epsilon_2^2,\epsilon_1^2\epsilon_2,\epsilon_1^3,\epsilon_2^3),
\end{eqnarray}
where $C^{\sst(l)}_{\!N\!,k}$ ($k=1,2,\dots$) are the coefficients of the $l$th-order term in the $N$-pulse sequence.
To achieve the predefined population $\mathcal{P}_{r}$, one should design the phases $\alpha_n$ and $\beta_n$ ($n=1,\dots,N$) to guarantee the validity of the following equation
\begin{eqnarray}  \label{10}
P^{\sst(0)}_r=C^{\sst(0)}_{\!N\!,1}=\mathcal{P}_{r}=\sin^2\theta.
\end{eqnarray}
That is, the zeroth-order term of the actual population $P_r^a$ should be equal to the predefined population.
This is the first condition that the phases $\alpha_n$ and $\beta_n$ must satisfy.
Furthermore, the global phase of the propagator is inessential for the system dynamics, and what really matters is the phase difference $\alpha_{mn}=\alpha_{m}-\alpha_{n}$ and $\beta_{mn}=\beta_{m}-\beta_{n}$ ($m,n=1,\dots,N$).
Hence, one can set the phases $\alpha_1$ and $\beta_1$ of the first pulse to be arbitrary.
Nevertheless, they become extremely useful when the composite pulses are designed for single qubit gates, because the values of $\alpha_1$ and $\beta_1$ can be used to modulate the relative phase between the ground states.
As a result, apart from Eq.~(\ref{10}), there still remain $(2N-3)$ free phases (variables) in the $N$-pulse sequence, which can be designed to compensate for the systematic errors caused by the deviations $\epsilon_1$ and $\epsilon_2$.

Unlike the case of two-level systems \cite{PhysRevA.102.013105,PhysRevA.101.013827,PhysRevA.92.033406,PhysRevA.84.062311,PhysRevA.93.032340,PhysRevA.100.032333, PhysRevA.89.022310,PhysRevA.92.060301, PhysRevA.93.023830}, there exists a population leakage from the ground states to the excited state in the three-level system. Thus, we need to suppress population leakage (i.e., the population of the excited state $|e\rangle$) as well. Similar to the treatment of $P_r^a$, the actual population $P_e^a$ of the excited state $|e\rangle$ can be expanded by the Taylor expansion, as follows
\begin{eqnarray}  \label{P_e_a}
P_e^a\!&=&\!P^{\sst(0)}_e+P^{\sst(1)}_e+P^{\sst(2)}_e+\cdots \nonumber\\[0.1ex]
     \!&=&\!D^{\sst(2)}_{\!N\!,1}\epsilon_1\epsilon_2\!+\!D^{\sst(2)}_{\!N\!,2}\epsilon_1^2\!+\!D^{\sst(2)}_{\!N\!,3}\epsilon_2^2\!+\! {O}(\epsilon_1\epsilon_2^2,\epsilon_1^2\epsilon_2,\epsilon_1^3,\epsilon_2^3),  ~~~~~~
\end{eqnarray}
where $D^{\sst(l)}_{\!N\!,k}$ ($k=1,2,\dots$) are the coefficients of the $l$th-order term in the $N$-pulse sequence.
Contrary to Eq.~(\ref{9}), Eq.~(\ref{P_e_a}) does not contain the first-order term.

Therefore, to obtain high fidelity population transfer in the qubit, the general method is to satisfy the following conditions: ($i$) eliminate the higher-order terms in $P_r^a$ as much as possible. Namely, $C^{\sst(1)}_{\!N\!,1}=C^{\sst(1)}_{\!N\!,2}=\cdots=0$; ($ii$) demand that the value of $P_e^a$ be as small as possible, i.e., $D^{\sst(2)}_{\!N\!,1}=D^{\sst(2)}_{\!N\!,2}=\cdots=0$.
Note that the design procedure of the phases is more complicated in three-level systems than that in two-level systems \cite{PhysRevA.102.013105,PhysRevA.101.013827,PhysRevA.92.033406,PhysRevA.84.062311,PhysRevA.93.032340,PhysRevA.100.032333, PhysRevA.89.022310,PhysRevA.92.060301, PhysRevA.93.023830}, since there are two coefficients [$C^{\sst(1)}_{\!N\!,1}$ and $C^{\sst(1)}_{\!N\!,2}$] in the first-order term while there are six coefficients [$C^{\sst(2)}_{\!N\!,k}$ and $D^{\sst(2)}_{\!N\!,k}$, $k=1,2,3$] in the second-order term, and so forth.
Particularly, when considering the small number of pulses, which is often the practical case, it does not have sufficient phases to eliminate the systematic errors up to the desired order.

To solve this issue, we can proceed as follows.
First, it is worth noting that the influence of the higher-order term of systematic errors on dynamical evolution would gradually diminish. Therefore, the first-order terms have the most serious effect on dynamical evolution, and we should give priority to making these terms vanish by designing suitable phases, i.e., $C^{\sst(1)}_{\!N\!,1}=C^{\sst(1)}_{\!N\!,2}=0$.
Then, the remaining phases are designed to minimize the following cost function:
\begin{eqnarray}
F&=&\sum_{l=2}^{\infty}\left[A_lF_c^{\sst(l)}+B_lF_d^{\sst(l)}\right], \label{11}\\[0.5ex]
F_c^{\sst(l)}&=&\sum\nolimits_{k=1}^{l+1}|C^{\sst(l)}_{N,k}|^2, \\[1ex]
F_d^{\sst(l)}&=&\sum\nolimits_{k=1}^{l+1}|D^{\sst(l)}_{N,k}|^2,
\end{eqnarray}
where $A_l$ and $B_l$ are the weighting coefficients, satisfying $0\leq A_{l-1}\leq A_l$ and $0\leq B_{l-1}\leq B_l$.
Physically, the cost function $F$ represents the trade-off between the accuracy of population transfer and the leakage to the excited state $|e\rangle$.
The conditions $0\leq A_{l-1}\leq A_l$ and $0\leq B_{l-1}\leq B_l$ ensure that the coefficients of the low order terms have a high weight, and thus are preferentially eliminated.
Remarkably, different weighting coefficients have different effects.
Setting $B_l=0$, we have the cost function $F=\sum_{l=2}^{\infty}A_lF_c^{\sst(l)}$ aimed to keep the accuracy of the population transfer,
while setting $A_l=0$ we have the cost function $F=\sum_{l=2}^{\infty}B_lF_d^{\sst(l)}$ aimed to reduce the leakage to the excited state $|e\rangle$.
Note that the minimum value of the cost function is zero, corresponding to satisfy the equations: $C^{\sst(l)}_{N,k}=D^{\sst(l)}_{N,k}=0$.

\section{Robust implementation of arbitrary population transfer by composite pulses}   \label{III}

\begin{figure*}[htbp]
	\centering
	\includegraphics[scale=0.086]{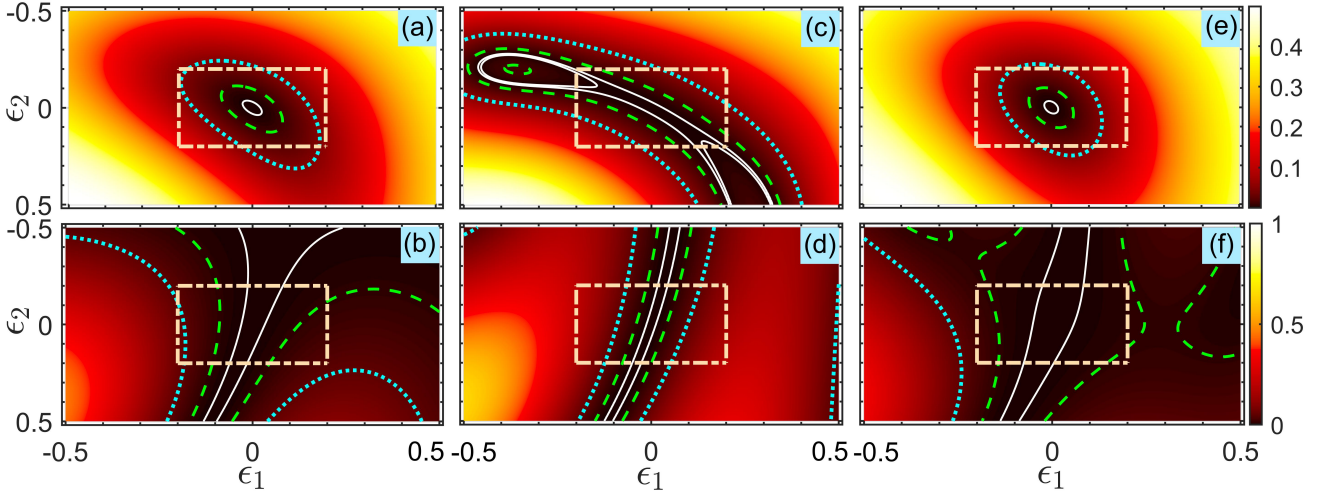}
	\caption{Infidelity $\mathcal{F}_r$ (top panels) and population leakage $P_e^a$ (bottom panels) vs the deviations $\epsilon_1$ and $\epsilon_2$ with different cost functions in the two-pulse sequence, where $\mathcal{P}_{r}=1/2$. (a-b) $F=F_c^{\sst(2)}+F_d^{\sst(2)}$. (c-d) $F=F_c^{\sst(2)}$. (e-f) $F=F_d^{\sst(2)}$. The results demonstrate that the robust behaviors are different by choosing different cost functions. }  \label{fig:01}
\end{figure*}

In this section, we present the design process of the phases $\alpha_n$ and $\beta_n$ ($n=2,...,N$) to achieve high fidelity  population transfer by the $N$-pulse sequence. For the two-pulse sequence, the phases are derived analytically, while for the three-pulse sequence we combine both the analytical method and the numerical method to obtain the phases. For more than three pulses, we carry out the four(five)-pulse sequence by numerical calculations to demonstrate the feasibility of eliminating the higher-order terms of the systematic errors.
It is worth mentioning that the cycle of the phases is $2\pi$, thus we restrict the values of all phases in the interval $[0,2\pi)$ following from here.

\subsection{Two pulses}

The form of the two-pulse sequence is expressed by
\begin{eqnarray}
U\left(\frac{\pi}{4},\gamma_2\right)U\left(\frac{\pi}{4},\gamma_1\right),  \nonumber
\end{eqnarray}
and the zeroth-order coefficient in Eq.~(\ref{9}) becomes
\begin{eqnarray}  \label{11c}
C^{\sst(0)}_{2,1}=\sin^2\frac{\alpha_{12}-\beta_{12}}{2},
\end{eqnarray}
where $\alpha_{12}=\alpha_{1}-\alpha_{2}$ and $\beta_{12}=\beta_{1}-\beta_{2}$.
As a result, the solution of Eq.~(\ref{10}) can be written as
\begin{eqnarray}  \label{12a}
\beta_{12}=\alpha_{12}-2\theta.
\end{eqnarray}
It is worth mentioning that the first-order coefficients $C^{\sst(1)}_{2,1}$ and $C^{\sst(1)}_{2,2}$ in Eq.~(\ref{9}) automatically vanish in the two-pulse sequence, i.e.,
\begin{eqnarray}
C^{\sst(1)}_{2,1}=C^{\sst(1)}_{2,2}=0.  \nonumber
\end{eqnarray}
Therefore, we can further reduce the detrimental effect of the second-order terms, where the corresponding coefficients $C^{\sst(2)}_{2,k}$ and $D^{\sst(2)}_{2,k}$ ($k=1,2,3$) are presented in Appendix~\ref{a1}.

Note that there are only two
controllable variables $\alpha_{12}$ and $\beta_{12}$, and $\beta_{12}$ is required to satisfy Eq.~(\ref{12a}).
Only single variable $\alpha_{12}$ is left to eliminate the second-order terms of the systematic errors, and thus it is enough to consider the second-order coefficients in the cost function given by Eq.~(\ref{11}).
In the following, we give out the optimal value of the variable $\alpha_{12}$ for three types of cost functions (see Appendix~\ref{a1} for details):

($i$) The cost function is chosen as $F=F_c^{\sst(2)}$. That is, we aim merely at eliminating the deviation in the actual population $P_r^a$.
It is not hard to ascertain that $F_c^{\sst(2)}$ is minimal when
\begin{equation} \label{14a}
\alpha_{12}\!=\!\left\{
\begin{aligned}
    \displaystyle\!&\theta\!+\!\arcsin\!\frac{\!\sqrt{2}(16\!+\!19\pi^2)\sin\theta}{19\pi^2},  ~0\leq\theta\leq\theta', \\[1.1ex]
    \displaystyle\!&\theta\!+\!\pi/2,  ~~~~~~~~~~~~~~~~~~~~~~~~~~~~\theta'<\theta\leq\frac{\pi}{2}, \\
\end{aligned}
\right.
\end{equation}
where $\theta'=\arcsin\big[\!{19\pi^2}/{\!\sqrt{2}(16\!+\!19\pi^2)}\big]$.

($ii$) The cost function is chosen as $F=F_d^{\sst(2)}$. That is, we aim merely at suppressing the population leakage $P_e^a$.
Note that $F_d^{\sst(2)}$ is minimal when
\begin{eqnarray}  \label{14b}
\alpha_{12}=\pi-\displaystyle\arctan\frac{(\sqrt{2}-1)\sin 2\theta}{(\sqrt{2}-1)\cos2\theta+1}.
\end{eqnarray}

($iii$) The cost function is chosen as $F=F_c^{\sst(2)}+F_d^{\sst(2)}$. That is, we make an equal weight between the elimination of the deviation in the actual population $P_r^a$ and the suppression of the population leakage $P_e^a$.
Note that $F=F_c^{\sst(2)}+F_d^{\sst(2)}$ is minimal when
\begin{eqnarray}  \label{13}
\alpha_{12}=2\arctan \Theta.
\end{eqnarray}
Here, $\Theta$ satisfies the following quartic equation:
\begin{eqnarray}  \label{18}
a_1\Theta^4+a_2\Theta^3+a_3\Theta^2+a_4\Theta+a_5=0,   \nonumber
\end{eqnarray}
where the expressions for the coefficients $a_k$ ($k=1,\dots,5$) are given in Appendix~\ref{a1}.

Figure~\ref{fig:01} demonstrates the infidelity $\mathcal{F}_r$ and the population leakage $P_e^a$ as a function of the deviations $\epsilon_1$ and $\epsilon_2$ by different cost functions.
Compared Figs.~\ref{fig:01}(a)-\ref{fig:01}(b) with Fig.~\ref{fig:01a}, both the robust behaviors and the population leakage are slightly improved when we choose the cost function $F=F_c^{\sst(2)}+F_d^{\sst(2)}$.
An inspection of Figs.~\ref{fig:01}(c)-\ref{fig:01}(d) demonstrates that population transfer would be more accurate when the phases are determined by Eq.~(\ref{14a}), since the region enclosed by white-solid curves in Fig.~\ref{fig:01}(c) is much larger than that in Fig.~\ref{fig:01}(a). Nevertheless, there are plenty of population leakages in this situation. When adopting the phases determined by Eq.~(\ref{14b}), the population leakage is suppressed at a very low level in a wide region, as shown in Fig.~\ref{fig:01}(f).
These results show that the robust behaviors are quite different when we choose different cost functions.

\subsection{Three pulses}

The form of the three-pulse sequence is expressed by
\begin{eqnarray}
U\left(\frac{\pi}{4},\gamma_3\right)U\left(\frac{\pi}{4},\gamma_2\right)U\left(\frac{\pi}{4},\gamma_1\right),  \nonumber
\end{eqnarray}
and the zeroth-order coefficient in Eq.~(\ref{9}) becomes
\begin{eqnarray}  \label{20}
C^{\sst(0)}_{3,1}=\frac{1}{2}\big[1-\sin(\alpha_{12}-\beta_{12})\sin(\alpha_{23}-\beta_{23})\big],
\end{eqnarray}
where $\alpha_{mn}=\alpha_{m}-\alpha_{n}$ and $\beta_{mn}=\beta_{m}-\beta_{n}$, $m,n=1,2,3$.
The first-order coefficients $C^{\sst(1)}_{3,1}$ and $C^{\sst(1)}_{3,2}$ in Eq.~(\ref{9}) are
\begin{eqnarray}   \label{14}
C^{\sst(1)}_{3,1}=C^{\sst(1)}_{3,2}&=&\frac{1}{\sqrt{2}}\big[\cos(\alpha_{12}-\beta_{12})+\cos(\alpha_{23}-\beta_{23})- \cr
&&\cos(\alpha_{12}-\beta_{12})\cos(\alpha_{23}-\beta_{23})\big].
\end{eqnarray}
For the second-order coefficients, since the expressions are too complicated to present here, we give them in Appendix~\ref{a}.

It is easily found that one solution of equations $\left \{
\begin{array}{ll}
    \!\!C^{\sst(0)}_{3,1}&\!\!=\sin^2\theta \\[.8ex]
    \!\!C^{\sst(1)}_{3,1}&\!\!=0 \\
\end{array}
\right.$ reads
\begin{eqnarray}   \label{23a}
\beta_{12}\!&=&\!\alpha_{12}\!-\!(\!-\!1\!)^s\!\arccos\frac{\!\sqrt{70\!-\!72\cos4\theta\!+\!2\cos8\theta}\!-\!4\!\sin^22\theta}{16}, \nonumber\\[1.1ex]
\beta_{23}\!&=&\!\alpha_{23}\!-\!\arccos\frac{2\cos4\theta\!-\!2-\!\sqrt{70\!-\!72\cos4\theta\!+\!2\cos8\theta}}{16}, \nonumber\\
\end{eqnarray}
where $s=0$ when $0<\theta<\pi/4$, while $s=1$ when $\pi/4\leq\theta\leq\pi/2$.
Note that there are four variables ($\alpha_{12}$, $\alpha_{23}$, $\beta_{12}$, and $\beta_{23}$) in the three-pulse sequence, and the values of the variables $\beta_{12}$ and $\beta_{23}$ are given by Eq.~(\ref{23a}). Therefore, only two variables $\alpha_{12}$ and $\alpha_{23}$ can be designed to eliminate the influence of the deviations.
In the following, the cost function is chosen as $F=F_c^{\sst(2)}+F_d^{\sst(2)}$, that is, the impact of the accuracy of population transfer and the leakage to the excited state are equally weighted in the three-pulse sequence.

\begin{figure}[t]
	\centering
	\includegraphics[scale=0.065]{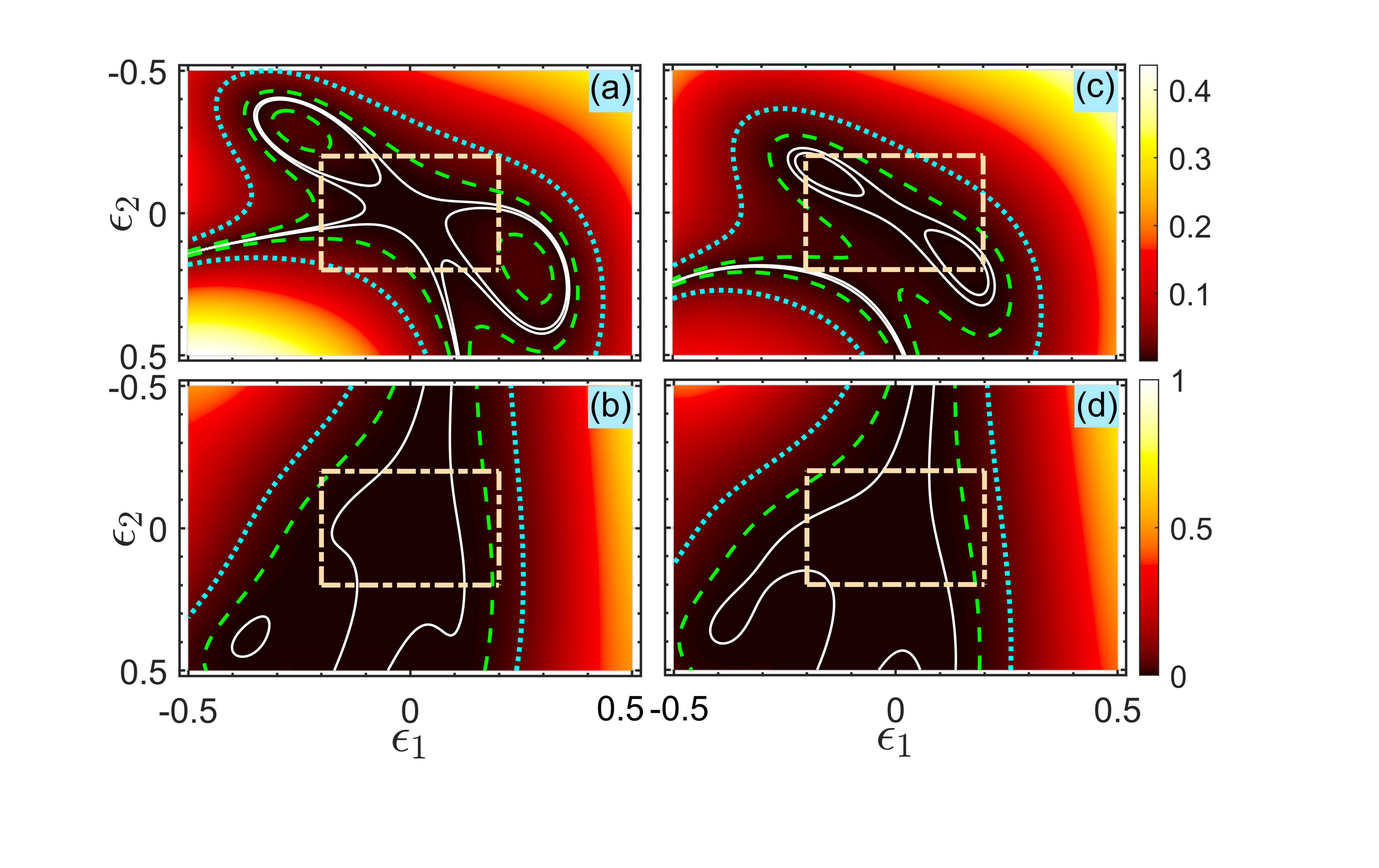}
	\caption{Infidelity $\mathcal{F}_r$ (top panels) and population leakage $P_e^a$ (bottom panels) vs the deviations $\epsilon_1$ and $\epsilon_2$ in the three-pulse sequence, where $\mathcal{P}_{r}=1/2$ in the left column while $\mathcal{P}_{r}=3/4$ in the right column. The phases can be found in Table~\ref{t1}. Compared with Fig.~\ref{fig:01a}, both the infidelity and the population leakage maintain a small value ($\leq0.001$, cf. the white-curves) in a very wide region. }  \label{fig:03}
\end{figure}

Figure \ref{fig:03} shows the performance of the infidelity $\mathcal{F}_r$ and the population leakage $P_e^a$ in the three-pulse sequence. It can be seen that both the infidelity $\mathcal{F}_r$ and the population leakage $P_e^a$ keep a very small value in the presence of strong deviations.
Compared with the case of the resonant pulses (cf. Fig.~\ref{fig:01a}), the region $\mathcal{F}_r(P_e^a)\leq0.001$ becomes much wider in the three-pulse sequence, and thus this sequence is more robust against the deviations $\epsilon_1$ and $\epsilon_2$.
Noting that it is hard to obtain the analytical expressions for the phases $\alpha_n$ and $\beta_n$ in the three-pulse sequence, we present some numerical solutions for different populations (i.e., different $\theta$) in Table~\ref{t1}.

\renewcommand\arraystretch{1.2}
\begin{table*}[htbp]
	\centering
	\caption{The phases of the $N$-pulse sequence to achieve the predefined population $\mathcal{P}_{r}=\sin^2\theta$, where we set $\alpha_1=\beta_1=0$.}
	\label{t1}
	\begin{tabular}{cccc}
		\hline
		\hline
           & \multirow{2}{*}{3-pulse sequence} & \multirow{2}{*}{4-pulse sequence} & \multirow{2}{*}{5-pulse sequence}  \\&&&\\
		\cline{2-2} \cline{3-3} \cline{4-4}
       \cmidrule(lr){2-2} \cmidrule(lr){3-4}
	 $\theta$ &$~~\alpha_2$~~~~~$\alpha_3$~~~~~~$\beta_2$~~~~~~$\beta_3~~~$ & $\alpha_2$~~~~~$\alpha_3$~~~~~ $\alpha_4$~~~~~$\beta_2$~~~~~ $\beta_3$~~~~~ $\beta_4$~ &  $\alpha_2$~~~~~$\alpha_3$~~~~~ $\alpha_4$~~~~~ $\alpha_5$~~~~~ $\beta_1$~~~~~ $\beta_2$~~~~~ $\beta_3$~~~~~ $\beta_4$\\
		\hline
         $\pi$/2 & 3.911~ 3.142~ 2.340~ 3.142~~  & 3.195~ 4.817~ 1.831~ 1.630~ 0.110~ 4.983~~   & 5.291~ 2.472~ 1.925~ 6.089~ 1.325~ 5.943~ 4.238~ 6.095  \\
         $\pi$/3 & 2.663~ 5.027~ 1.562~ 0.189~~  & 2.769~ 5.550~ 1.526~ 5.762~ 5.401~ 2.275~~   & 2.707~ 5.876~ 2.868~ 5.175~ 6.032~ 2.260~ 1.730~ 5.963  \\
         $\pi$/4 & 2.497~ 4.859~ 1.450~ 0.670~~  & 3.554~ 2.374~ 6.030~ 2.768~ 4.729~ 6.027~~   & 3.575~ 1.214~ 2.962~ 6.015~ 4.985~ 4.848~ 2.674~ 4.759\\
         $\pi$/5 & 3.828~ 1.466~ 4.894~ 5.313~~  & 3.514~ 1.885~ 5.420~ 2.786~ 4.299~ 5.850~~   & 3.285~ 0.744~ 4.382~ 1.603~ 4.232~ 4.050~ 5.108~ 3.656  \\
         $\pi$/6 & 3.870~ 1.508~ 4.971~ 5.155~~  & 3.514~ 1.047~ 4.429~ 2.852~ 3.527~ 5.200~~   & 3.317~ 0.802~ 4.516~ 1.688~ 4.266~ 4.106~ 5.013~ 3.469  \\
         $\pi$/7 & 3.911~ 1.550~ 5.048~ 5.070~~  & 5.431~ 3.456~ 3.162~ 4.786~ 5.953~ 4.117~~   & 2.677~ 4.582~ 0.254~ 3.172~ 3.787~ 3.449~ 3.913~ 5.049  \\
        $\pi$/10 & 3.953~ 1.550~ 5.179~ 4.883~~  & 3.408~ 0.733~ 4.039~ 2.840~ 3.307~ 5.417~~   & 5.948~ 2.591~ 4.891~ 3.048~ 3.871~ 5.556~ 5.753~ 5.134  \\
        $\pi$/20 & 1.664~ 3.267~ 3.038~ 0.174~~  & 6.283~ 3.351~ 3.665~ 5.853~ 6.063~ 5.633~~   & 1.837~ 0.585~ 4.787~ 4.277~ 3.287~ 0.220~ 3.388~ 0.247   \\
		\hline
		\hline
	\end{tabular}
\end{table*}

\subsection{More than three pulses}

The form of the four-pulse sequence reads
\begin{eqnarray}
U\left(\frac{\pi}{4},\gamma_4\right)U\left(\frac{\pi}{4},\gamma_3\right)U\left(\frac{\pi}{4}, \gamma_2\right)U\left(\frac{\pi}{4},\gamma_1\right),   \nonumber
\end{eqnarray}
and the zeroth-order coefficient in Eq.~(\ref{9}) becomes
\begin{eqnarray}
\noindent C^{\sst(0)}_{4,1}&=&\frac{1}{2}\big[1\!+\!\cos(\alpha_{23}\!-\!\beta_{23})\sin(\alpha_{12}\!-\!\beta_{12}) \sin(\alpha_{34}\!-\!\beta_{34}) \nonumber\\[.1ex]
&&-\cos(\alpha_{12}-\beta_{12})\cos(\alpha_{34}-\beta_{34})\big],
\end{eqnarray}
where $\alpha_{mn}=\alpha_{m}-\alpha_{n}$ and $\beta_{mn}=\beta_{m}-\beta_{n}$, $m,n=1,2,3,4$.
The first-order coefficients $C^{\sst(1)}_{4,1}$ and $C^{\sst(1)}_{4,2}$ in Eq.~(\ref{9}) are expressed by
\begin{eqnarray}
C^{\sst(1)}_{4,1}&=&-C^{\sst(1)}_{4,2}=-2\sqrt{2}\sin\frac{\alpha_{12}-\beta_{12}}{2}\sin\frac{\alpha_{34}-\beta_{34}}{2} \nonumber\\[0.1ex]
&&\times\sin(\alpha_{23}\!-\!\beta_{23})\sin\frac{\alpha_{12}\!+\!\alpha_{34}\!-\!\beta_{12}\!-\!\beta_{34}}{2}.
\end{eqnarray}

Thus, one solution of equations $\left \{
\begin{array}{ll}
    \!\!C^{\sst(0)}_{4,1}&\!\!=\sin^2\theta \\[.8ex]
    \!\!C^{\sst(1)}_{4,1}&\!\!=0 \\
\end{array}
\right.$ can be written as
\begin{eqnarray}
\beta_{23}&=&\alpha_{23}-\pi, \nonumber\\[0.1ex]
\alpha_{34}&=&\alpha_{12}-\beta_{12}+\beta_{34}-2\theta. \nonumber
\end{eqnarray}
As a result, there are still four variables available to eliminate the influence of the deviations in the four-pulse sequence, and some numerical solutions for different $\theta$ are presented in Table ~\ref{t1}.

Similarly, the zeroth-order in the five-pulse sequence can be expressed by
\begin{eqnarray}
C^{\sst(0)}_{5,1}&=&\frac{1}{2} \Big\{1\!-\!\cos ({\alpha _{45}}\!-\!{\beta _{45}})\sin ({\alpha _{12}}\!-\!{\beta _{12}}) \sin ({\alpha _{23}}\!-\!{\beta _{23}}) \nonumber\\[.1ex]
&&\!-\!\sin ({\alpha _{45}}\!-\!{\beta _{45}}) \big[\cos ({\alpha _{12}}\!-\!{\beta _{12}}) \sin ({\alpha _{34}}\!-\!{\beta _{34}}) \nonumber\\[.1ex]
&&\!+\!\sin ({\alpha _{12}}\!-\!{\beta _{12}})\!\cos ({\alpha _{23}}\!-\!{\beta _{23}})\! \cos ({\alpha _{34}}\!-\!{\beta _{34}})\big]\!\!\Big\}.
\end{eqnarray}
Due to the complex expressions of the first-order coefficients, we present it in Appendix~\ref{c}. After satisfying $\left \{
\begin{array}{ll}
    \!\!C^{\sst(0)}_{5,1}&\!\!=\sin^2\theta \\[.8ex]
    \!\!C^{\sst(1)}_{5,1}&\!\!=0 \\
\end{array}
\right.$, there are still six variables to eliminate the influence of the deviations in the five-pulse sequence, and some numerical solutions for different $\theta$ are presented in Table ~\ref{t1}.

\begin{figure}[htbp]
	\centering
	\includegraphics[scale=0.065]{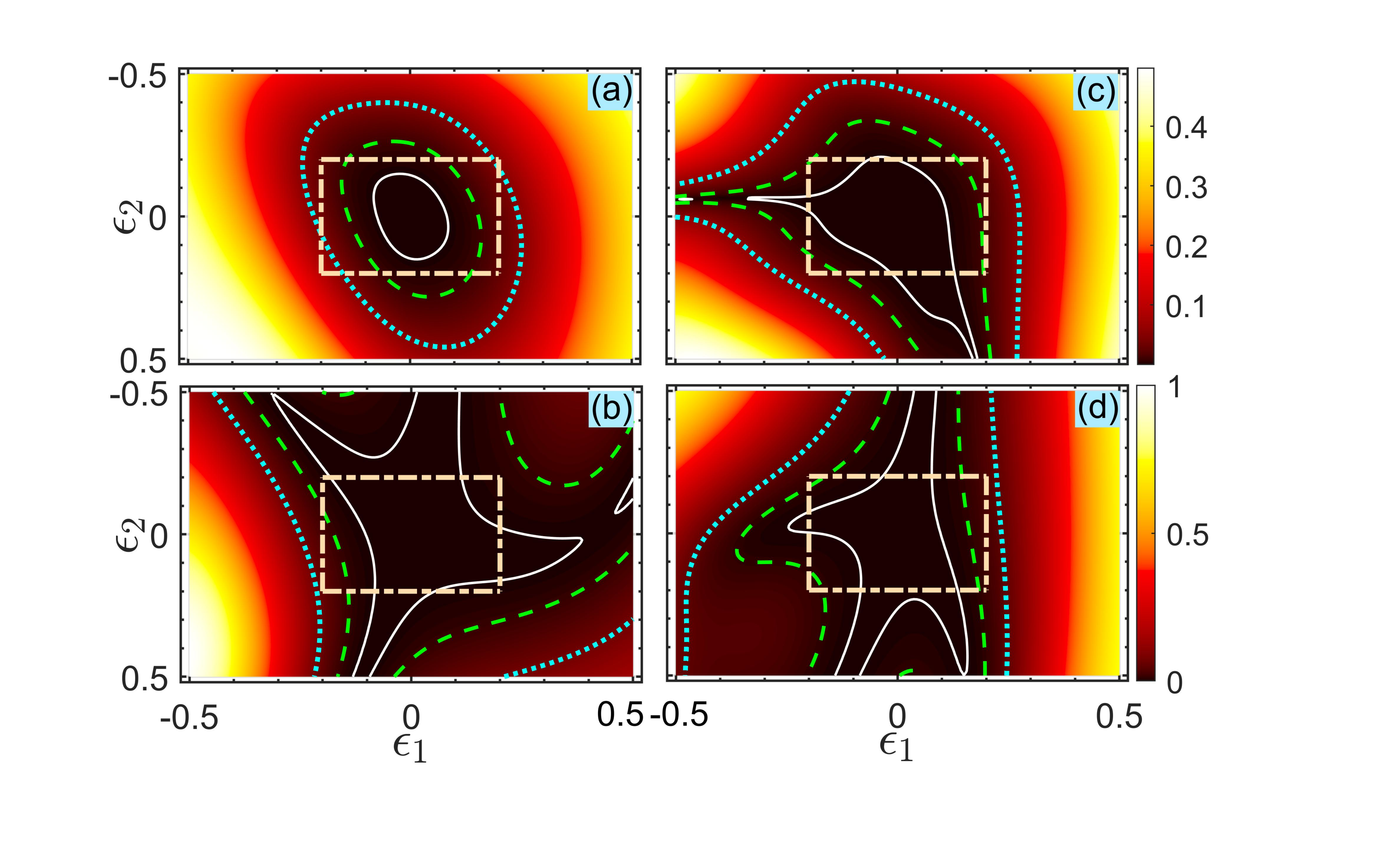}
	\caption{Infidelity $\mathcal{F}_r$ (top panels) and population leakage $P_e^a$ (bottom panels) vs the deviations $\epsilon_1$ and $\epsilon_2$ with different pulse numbers, where $\mathcal{P}_{r}=1/2$.  Left column: the four-pulse sequence. Right column: the five-pulse sequence. The phases can be found in Table~\ref{t1}. The results demonstrate that the infidelity and the population leakage maintain a small value ($\leq0.001$, cf. the white-curves) in a wider region with the increase of pulse numbers. }  \label{fig:04}
\end{figure}

Figure~\ref{fig:04} shows the infidelity and the population leakage as a function of the deviations in the four(five)-pulse sequence.
One can find from Fig.~\ref{fig:04} that the infidelity and the population leakage are small in a wider region as the number of pulses increase. Apparently, the population transfer would be more accurate when more pulses are taken into account.
Note that the cost function is chosen as $F=F_c^{\sst(2)}+F_d^{\sst(2)}$ in Fig.~\ref{fig:04}, namely, the impact of the accuracy of the population transfer and the leakage to the excited state are equally weighted.
We present in Appendix~\ref{b} the detailed discussion on the performance of the accuracy and the leakage by choosing different forms of cost functions.

Next, we demonstrate the evolutionary trajectory of the qubit (i.e., the ground states $|g\rangle$ and $|r\rangle$) on the Bloch sphere under the $N$-pulse sequence. Taking the four-pulse sequence as an example, suppose
the initial state of this system is $|g\rangle$. Then, the final state of this system after the four-pulse sequence can be expressed in the basis $\{|g\rangle, |r\rangle, |e\rangle\}$ by employing four angle variables:
\begin{eqnarray}
U\!\left(\frac{\pi}{4},\!\gamma_4\right)\cdots U\!\left(\frac{\pi}{4},\!\gamma_1\right)|g\rangle\!=\!\!\left(\!\!
     \begin{array}{c}
       \cos\vartheta'\cos\theta \\
       \cos\vartheta'\sin\theta\exp(i\phi') \\
       \sin\vartheta'\exp(i\varphi') \\
     \end{array}
   \!\!\right)\!.~~~~~
\end{eqnarray}
Since we focus on the evolutionary trajectory of the qubit, the phase $\varphi'$ is irrelevant and we can simply write the time evolution of qubit as: $|\Psi\rangle=\cos\vartheta'\big[\cos\theta|g\rangle+\sin\theta\exp(i\phi')|r\rangle\big]$.
When $|\cos\vartheta'|=1$, the evolutionary trajectory is on the Bloch sphere. While it is inside the Bloch sphere when $|\cos\vartheta'|\neq1$.
Here, the state inside the Bloch sphere does not represent the mixed state, and means the population leakage from ground states to the excited state.
Note that the smaller the value of $|\cos\vartheta'|$ is, the larger the population leakage will be. The center of the Bloch sphere, i.e., $\cos\vartheta'=0$, represents the excited state $|e\rangle$.

\begin{figure*}[htbp]
	\centering
	\includegraphics[scale=0.1]{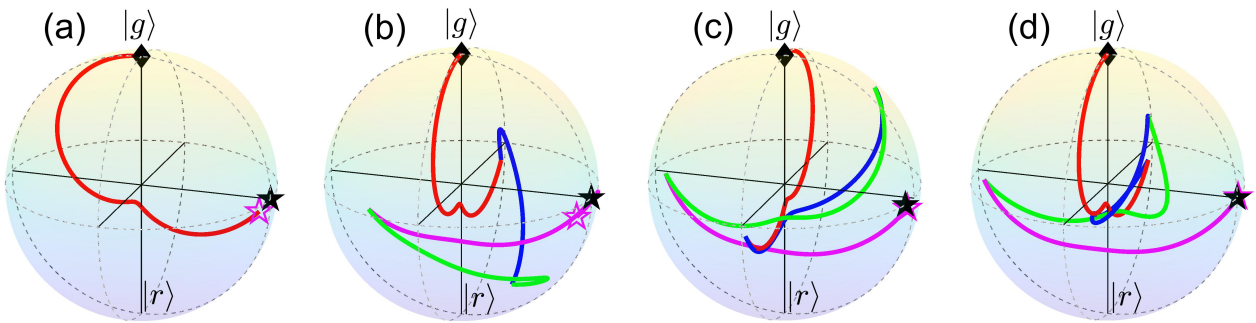}
	\caption{Evolutionary trajectories of the qubit on Bloch spheres with different cost functions in the four-pulse sequence. $\epsilon_1=\epsilon_2=-0.1$, and the other parameters can be found in Table \ref{bt1}. (a) The resonant pulses. (b) $F=F_c^{\sst(2)}+F_c^{\sst(3)}$. (c) $F=F_d^{\sst(2)}+F_d^{\sst(3)}+F_d^{\sst(4)}$. (d) $F=F_c^{\sst(2)}+F_d^{\sst(2)}$. The initial state $|g\rangle$ is labeled by $\blacklozenge$, the target state $1/\sqrt{2}[|g\rangle+\exp(i\phi')|r\rangle]$ is labeled by $\text{\ding{72}}$, and the final state of the qubit after the four-pulse sequence (or the resonant pulses) is labeled by $\text{\ding{73}}$. The trajectory inside the Bloch sphere means the population leakage from the qubit to the excited state. }  \label{fig:04a}
\end{figure*}

Figures \ref{fig:04a}(b)-\ref{fig:04a}(d) show the evolutionary trajectories of the qubit for different cost functions in the four-pulse sequence. By contrast, we plot in Fig.~\ref{fig:04a}(a) the evolutionary trajectory for the resonant pulses.
We can observe from Fig.~\ref{fig:04a} that the population leakage always exists during the evolution process, since the trajectories are inside the Bloch sphere in some evolution stages. Nevertheless, the population leakage is strongly suppressed and the system state approaches the target state at the final time, as shown in Fig.~\ref{fig:04a}(d).
It is shown in Fig.~\ref{fig:04a}(b) that the cost function $F=F_c^{\sst(2)}+F_c^{\sst(3)}$ only guarantees the accuracy of $P_r^a$, while there exists leakage to the excited state. As a result, the final state deviates from the target state.
A similar situation is also found in Fig.~\ref{fig:04a}(c), because the cost function $F=F_d^{\sst(2)}+F_d^{\sst(3)}+F_d^{\sst(4)}$ only prevents the population leakage to the excited state. However, Fig.~\ref{fig:04a}(d) demonstrates that the system is almost driven into the target state, because the cost function $F=F_c^{\sst(2)}+F_d^{\sst(2)}$ is a tradeoff between the accuracy and the population leakage. Note that the corresponding robust region designed by $F=F_c^{\sst(2)}+F_d^{\sst(2)}$ (e.g., the region enclosed by the white-curves in Fig.~\ref{fig:04}) is generally smaller than those designed by the other two cost functions. For details, one can see Appendix~\ref{b}.

\section{Applications: Robust preparation of three-atom singlet state by composite pulses}  \label{IV}

In above section, we have showed the design process of composite pulses in a general physical system, where the three-level structure can be found in atomic systems \cite{scully97}, trapped ions \cite{RevModPhys.75.281}, diamond nitrogen-vacancy centers \cite{Yang2010}, and superconducting circuits \cite{RevModPhys.85.623,PhysRevApplied.7.054022,PhysRevA.94.052311,PhysRevA.96.022304,PhysRevA.97.022332,PhysRevApplied.14.034038}, etc.
In this section, we demonstrate that the CPs scheme can be also generalized to complicated systems for performing different quantum tasks, e.g., coherent conversion between two qubits \cite{yang2021realizing} or preparation of entangled states \cite{Shao2010}, etc.
The fundamental is to reduce complicated systems to the familiar three-level physical model.
As an example, in presence of deviations, we next prepare the three-atom singlet state with ultrahigh fidelity in an atom-cavity system, and this approach is easily extended to other complicated systems.

\begin{figure}[htbp]
	\centering
	\includegraphics[scale=0.35]{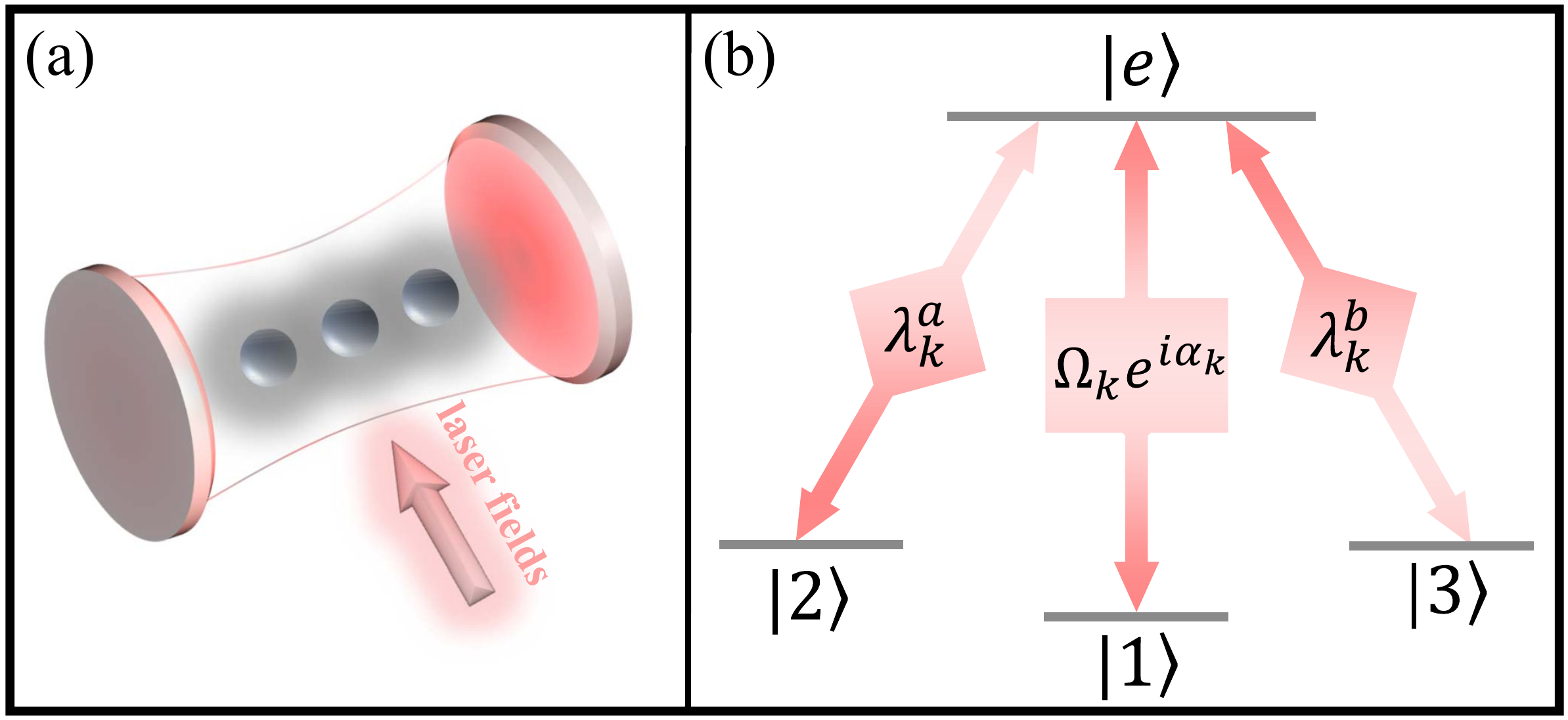}
	\caption{(a) The schematic diagram of the cavity-atom system. (b) The level configuration of the atoms. }  \label{fig:05}
\end{figure}

As shown in Fig.~\ref{fig:05}(a), consider the atom-cavity system, where three identical four-level atoms are trapped in a bimodal cavity. The four-level atom has three ground states $|1\rangle$, $|2\rangle$, and $|3\rangle$, and an excited state $|e\rangle$, as shown in Fig.~\ref{fig:05}(b). The transition $|2\rangle_{k}\leftrightarrow|e\rangle_{k}$ ($|3\rangle_{k}\leftrightarrow|e\rangle_{k}$) is resonantly coupled by the cavity mode $a$ ($b$) with the coupling constant $\lambda_{k}^{a}$ ($\lambda_{k}^{b}$), where the subscript $k$ represents the $k$th atom.
The transition $|1\rangle_{k}\leftrightarrow|e\rangle_{k}$ is resonantly coupled by the laser field with the coupling strength $\Omega_k$ and the phase $\alpha_k$.
In the interaction picture, the Hamiltonian of the atom-cavity system reads ($\hbar=1$)
\begin{eqnarray}
H&\!=\!&\sum_{k=1}^3\Big[(1\!+\!\epsilon_{k})\Omega_k\exp(i\alpha_k)|e\rangle_{kk}\langle 1|\!+\!\lambda_{k}^{a}(1\!+\!\zeta_k^a)|e\rangle_{kk}\langle 2|\hat{a} \nonumber\\[.1ex] &&+\lambda_{k}^{b}(1+\zeta_k^b)|e\rangle_{kk}\langle 3|\hat{b}+\mathrm{H.c.}\Big],
\end{eqnarray}
where $\hat{a}$ and $\hat{b}$ are the annihilation operators of the cavity mode $a$ and $b$, respectively.
$\epsilon_{k}$ and $\zeta_k^{a}$ ($\zeta_k^{b}$) are respectively the deviations of the laser fields and the cavity mode $a~(b)$ due to the inhomogeneities of spatial distribution.

\begin{figure*}[htbp]
	\centering
	\includegraphics[scale=0.066]{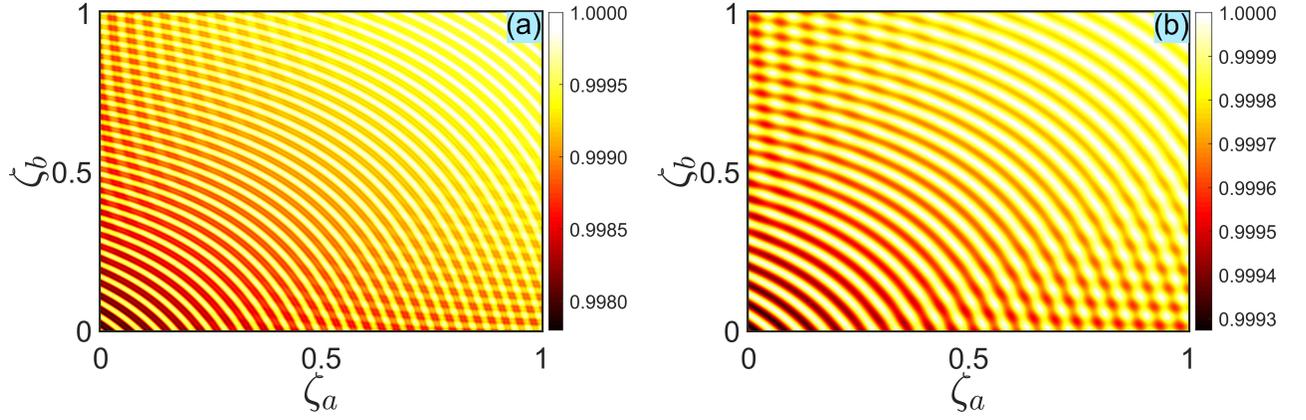}
	\caption{ Fidelity $F_{S}$ vs the deviations $\zeta_a$ and $\zeta_b$ in different schemes, where $\zeta^a_k=\zeta_a$ and $\zeta^b_k=\zeta_b$ ($k=1,2,3$). (a) The three-pulse sequence scheme. (b) The resonant pulses scheme shown in Ref.~\cite{Shao2010}. The fidelity is defined by $F_{S}=|\langle \Psi_S|\Psi\rangle|^2$, where $|\Psi\rangle$ is the system state. The parameters of the three-pulse sequence scheme are $\lambda_k=30\Omega_1$, $\Omega_2=(1\!-\!\sqrt{2}/2)\Omega_{1}$, $\epsilon_1=\epsilon_2=0$, $\{\alpha_n\}=\{0, 2.5382, 4.9009\}$, and $\{\beta_n\}=\{5.6372, 0.8226, 5.9370\}$. The results show that both schemes still maintain a high fidelity in presence of the deviations in the coupling constants.}  \label{fig:05c}
\end{figure*}

Apparently, the excited number is a conserved quantity in this system since $[H, \hat{N}_e]=0$, where the excited number operator is defined by $\hat{N}_e=\sum_{k=1}^3(|e\rangle_{kk}\langle e|+|1\rangle_{kk}\langle 1|)+ \hat{a}^\dag\hat{a} +\hat{b}^\dag\hat{b}$. When the condition $\Omega_k\ll \lambda_{k}^{a}$ ($\lambda_{k}^{b}$) is satisfied, we can restrict the system dynamics into the single-excited subspace and the Hamiltonian is approximated by \cite{Shao2010}
\begin{eqnarray}  \label{heff}
H'&=&\Big[\frac{\sqrt{3}}{3}(1+\epsilon_1)\Omega_1\exp(i\alpha_1)|\Psi_1\rangle\langle\Psi_3|  \nonumber\\[.3ex] &&+\frac{\sqrt{6}}{3}(1+\epsilon_2)\Omega_2\exp(i\beta_1)|\Psi_3\rangle\langle\Psi_2|+\mathrm{H.c.}\Big],~~
\end{eqnarray}
where we set $\Omega_3=\Omega_2$, $\alpha_2=\alpha_3=\beta_1$, $\epsilon_{3}=\epsilon_2$, and
\begin{eqnarray}
|\Psi_1\rangle&=&{1}/{\sqrt{2}}\big(|123\rangle-|132\rangle\big),   \nonumber\\[.1ex]
|\Psi_2\rangle&=&{1}/{2}\big(|231\rangle-|213\rangle+|312\rangle-|321\rangle\big),   \nonumber\\[.1ex]
|\Psi_3\rangle&=&{1}/{\sqrt{6}}\big(|e23\rangle\!-\!|2e3\rangle\!-\!|32e\rangle\!+\!|23e\rangle
\!+\!|3e2\rangle\!-\!|e32\rangle\big).  \nonumber
\end{eqnarray}
The state $|lmn\rangle$ represents the first atom in $|l\rangle$, the second atom in $|m\rangle$, and the third atom in $|n\rangle$. Since the cavity mode $a$ ($b$) is in vacuum state, we have ignored it in the Hamiltonian given by Eq.~(\ref{heff}).

It is easily found from Eq.~(\ref{heff}) that the atom-cavity system can be reduced to the three-level physical model studied in Sec.~\ref{II}. Hence, if the initial state of this system is $|\Psi_1\rangle$,
by fixing the ratio of two coupling coefficients and modulating the phases according to the above composite pulses theory, one can obtain the following superposition state
\begin{eqnarray}
|\Psi\rangle=\cos\theta|\Psi_1\rangle+e^{-i\gamma}\sin\theta|\Psi_2\rangle.
\end{eqnarray}
By setting $\gamma=0$ and $\theta=\arctan\sqrt{2}$, the state $|\Psi\rangle$ becomes
\begin{eqnarray}
|\Psi_{S}\rangle&=&1/\sqrt{3}|\Psi_1\rangle+\sqrt{6}/3|\Psi_2\rangle  \nonumber\\[.1ex]
&=&{1}/{\sqrt{6}}\big(|123\rangle\!-\!|132\rangle\!+\!|231\rangle\!-\!|213\rangle\!+\!|312\rangle\!-\!|321\rangle \big), \nonumber
\end{eqnarray}
which is actually the three-atom singlet state.

\begin{figure}[b]
	\centering
	\includegraphics[scale=0.056]{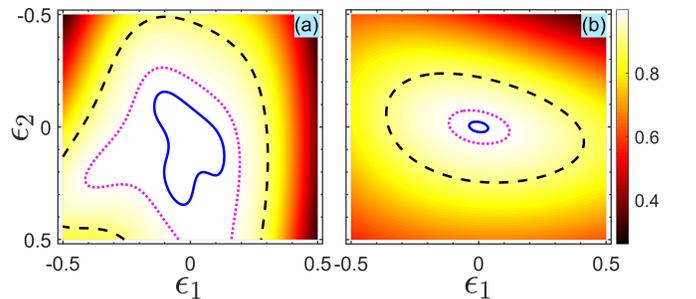}
	\caption{Fidelity $F_{S}$ vs the deviations $\epsilon_1$ and $\epsilon_2$ in different schemes. (a) The three-pulse sequence scheme. (b) The resonant pulses scheme shown in Ref.~\cite{Shao2010}. The blue-solid curve, the pink-dotted curve and the black-dashed curve correspond to $F_S$=0.999, 0.99, and 0.9, respectively. Here, $\eta^a_k=\eta^b_k=0$ and the other parameters are the same as in Fig.~\ref{fig:05c}. }  \label{fig:05a}
\end{figure}

\begin{figure*}[htbp]
	\centering
	\includegraphics[scale=0.6]{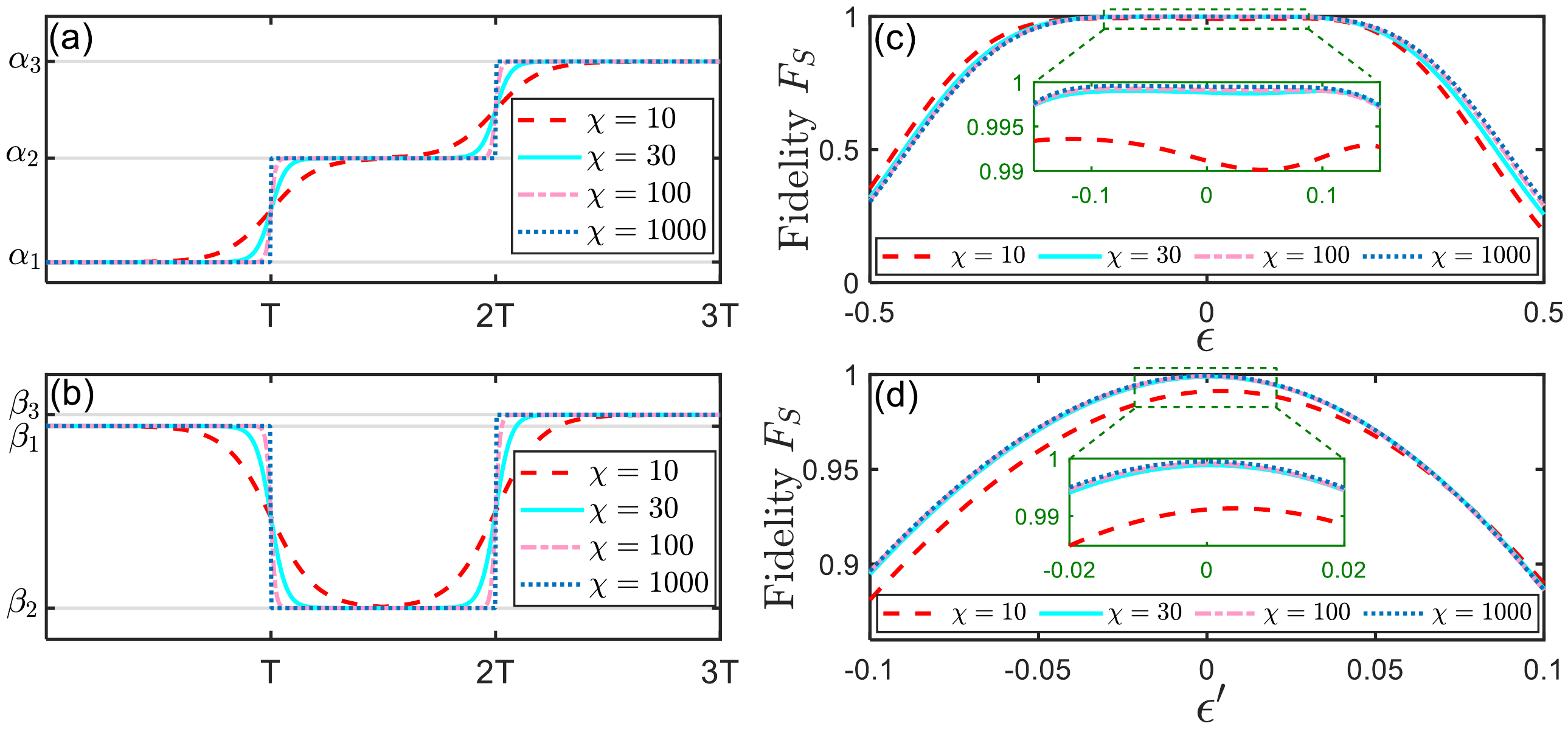}
	\caption{The waveform of the phases (a) $\alpha(t)$ and (b) $\beta(t)$ with different $\chi$ in the three-pulse sequence. (c) Fidelity $F_S$ vs the derivation $\epsilon$ in the coupling strengths, where we set $\epsilon_1=\epsilon_2=\epsilon$. (d) Fidelity $F_S$ vs the derivation $\epsilon'$ in the phases, where we set $\epsilon_\alpha=\epsilon_\beta=\epsilon'$. The other parameters are the same as in Fig.~\ref{fig:05c}. }  \label{fig:05b}
\end{figure*}

We first explore how the deviations of two coupling constants $\lambda_{k}^{a}$ and $\lambda_{k}^{b}$ affect the fidelity $F_S$ of the singlet state.
The numerical results simulated by the three-pulse sequence scheme are shown in Fig.~\ref{fig:05c}(a), where we choose $\zeta_a, \zeta_b\in[0,1]$ to guarantee the validity of the effective Hamiltonian given by Eq.~(\ref{heff}) (i.e., the condition $\Omega_k\ll\lambda_k^a, \lambda_k^b$ is well satisfied).
As a comparison, we also plot in Fig.~\ref{fig:05c}(b) the fidelity $F_S$ as a function of the deviations $\zeta_a$ and $\zeta_a$ by the resonant pulses scheme shown in Ref.~\cite{Shao2010}.
The results demonstrate that both schemes are robust against the deviations in two coupling constants, since the impact on the fidelity $F_S$ can be negligible no matter how large the deviations are, cf., $F_S\geq0.9978$ in Fig.~\ref{fig:05c}.
The reason for this phenomenon is that both $\zeta_a$ and $\zeta_b$ are absent in the Hamiltonian given by Eq.~(\ref{heff}).
Besides,
an interesting finding is that the fidelity $F_{S}$ tends to a higher value when the deviations $\zeta_a$ and $\zeta_b$ are much larger.
A reasonable interpretation is given as follows.
During the process of deriving the Hamiltonian given by Eq.~(\ref{heff}),
the strong coupling condition $\Omega_k\ll \lambda_{k}^{a}$ ($\lambda_{k}^{b}$) is exploited.
The coupling constant $\lambda_{k}^{a}$ ($\lambda_{k}^{b}$) increases with the raising of the deviation $\zeta_a~(\zeta_b)$.
As a result, the dynamics of this system is more and more restricted in the single-excited subspace so that the Hamiltonian given by Eq.~(\ref{heff}) ensures a better validity for suppressing the leakage to other subspaces.
Thus, a higher fidelity of the singlet state can be achieved when a larger deviation $\zeta_a~(\zeta_b)$ appears.

Next, we address the influence of the two deviations $\epsilon_1$ and $\epsilon_2$ on the fidelity $F_{S}$.
In Fig.~\ref{fig:05a}(a),
we display the fidelity $F_{S}$  versus the deviations $\epsilon_1$ and $\epsilon_2$ for the three-pulse sequence.
For comparison,
we also plot in Fig.~\ref{fig:05a}(b) the performance of the fidelity for the resonant pulses scheme shown in Ref.~\cite{Shao2010}.
As shown by the area surrounded by the blue-solid (or pink-dotted) curve in Fig.~\ref{fig:05a}(b),
it is clear that the scheme shown in Ref.~\cite{Shao2010} is highly susceptible to the deviations $\epsilon_1$ and $\epsilon_2$.
However,
the fidelity of the singlet state obtained by the three-pulse sequence remains in a high value ($F_S\geq0.99$) even when a very large deviation occurs.
This benefits from the fact that the second-order coefficients of systematic errors are restricted to an extremely low value by the specific set of phases.
Furthermore, as the pulse number increases,
the higher-order coefficients can be restricted to a low value that ensures a robust behavior against the deviations $\epsilon_1$ and $\epsilon_2$.
Thus, the CPs sequence can further improve an error-tolerant performance of the singlet state.

Another experimental issue, which inevitably impacts the performance of the final fidelity, is the waveform distortion.
In the ideal condition, our input pulse shape is associated with a perfect square waveform.
However, in practice, the input square wave always produces a tiny smooth rising and falling edge.
Let us study this issue by taking the three-pulse sequence as an example.
First of all, we design a group of functions in which the waveform distortion can be arbitrarily adjustable.
The functions are given by
\begin{equation}\label{aw}
\alpha(t)\!=\!\left\{
\begin{aligned}
    \displaystyle\!&\alpha_2-\frac{\alpha_2-\alpha_1}{1+\exp[\chi(t\!-\!T)]},  ~~~0\leq t\leq3T/2, \\[1.5ex]
    \displaystyle\!&\alpha_3-\frac{\alpha_3-\alpha_2}{1+\exp[\chi(t\!-\!2T)]},  ~3T/2< t\leq3T, \\
\end{aligned}
\right.
\end{equation}
\begin{equation}\label{bw}
\beta(t)\!=\!\left\{
\begin{aligned}
    \displaystyle\!&\beta_2-\frac{\beta_2-\beta_1}{1+\exp[\chi(t\!-\!T)]},  ~~~0\leq t\leq3T/2, \\[1.5ex]
    \displaystyle\!&\beta_3-\frac{\beta_3-\beta_2}{1+\exp[\chi(t\!-\!2T)]},  ~3T/2< t\leq3T, \\
\end{aligned}
\right.
\end{equation}
where $T=\pi/\sqrt{\Omega_1^2+\Omega_2^2}$.
Here, $\chi$ is a dimensionless parameter that defines the magnitude of the rising and falling edge of the square wave.

\begin{figure*}[htbp]
	\centering
	\includegraphics[scale=0.075]{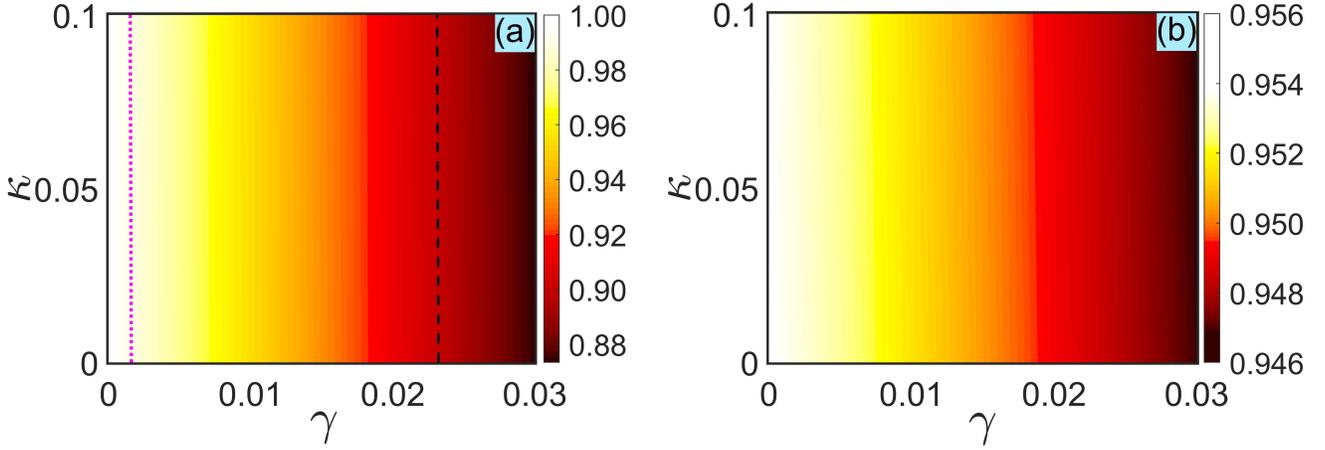}
	\caption{ Fidelity $F_{S}$ vs the dissipation rate $\gamma$ and the decay rate $\kappa$ in different schemes. (a) The three-pulse sequence scheme. (b) The resonant pulses scheme shown in Ref.~\cite{Shao2010}. The pink-dotted line and the black-dashed line correspond to $F_S$=0.99, and 0.9, respectively. $\epsilon_1=\epsilon_2=0.15$, $\eta^a_k=\eta^b_k=0$, and the other parameters are the same as in Fig.~\ref{fig:05c}. }  \label{fig:10}
\end{figure*}

As $\chi$ increases, the functions (\ref{aw}) and (\ref{bw})
gradually become a square wave. Specifically from Fig.~\ref{fig:05b}(a) and \ref{fig:05b}(b), we see
that the functions are almost square waveforms for $\chi$=1000 (the blue-dotted curves).
For $\chi\rightarrow\infty$,
they describe perfect square waves.
However,
when $\chi$ is small, e.g., $\chi$=10,
the duration of the rising and falling edge of the square wave becomes excessively long and the waveform has a serious shape change.
Then, functions (\ref{aw}) and (\ref{bw}) turn into a smooth time-dependent pulse, as shown by the red-dashed curves in Fig.~\ref{fig:05b}(a) and \ref{fig:05b}(b).
Obviously,
this kind of waveform does not satisfy the request of our CPs scheme.
As illustrated in Fig.~\ref{fig:05b}(c), even though the waveform suffers from serious distortion (the red-dashed curve),
the fidelity $F_{S}$ can still maintain a high value ($>$ 0.99).
Moreover,
as revealed by the cyan-solid and pink-dashed-dotted curves in Fig.~\ref{fig:05b}(c),
a slight distortion in the waveform hardly makes an impact on the final fidelity.
Hence,
the CPs sequence has the ability to acquire a precise and robust singlet state even with a distorted waveform.
Finally, we investigate the influence of the phase shift errors on the fidelity. In presence of the deviations in the phases, the actual phases can be rewritten as
\begin{eqnarray}
\alpha^a(t)=(1+\epsilon_\alpha)\alpha(t),~~~~~~\beta^a(t)=(1+\epsilon_\beta)\beta(t),
\end{eqnarray}
where $\epsilon_\alpha$ and $\epsilon_\beta$ represent the phase shift errors.
We plot in Fig.~\ref{fig:05b}(d) the relation between the fidelity and the phase shift errors, which shows that the current CPs sequence is sensitive to phase shift errors \cite{PhysRevA.103.033110}.
The reason is that we do not consider the phase shift errors in the Hamiltonian given by Eq.~(\ref{h3}).
To tackle this problem, one requires to regard the phase shift errors as variables in Eq.~(\ref{h3}), and then constructs a phase-error-corrected composite pulses sequence \cite{PhysRevA.100.023410}.

In a more realistic situation, one has to consider the influence of decoherence on the fidelity of the singlet state.
In this system, the decoherence mainly comes from two aspects:
($i$) the atomic spontaneous emission from the excited state to the ground states.
($ii$) the decay of two cavity modes.
Under the decoherence environment,
the dynamics of this system are governed by the Lindblad master equation and the form can be written as
\begin{eqnarray}
\dot{\rho}\!=\!-i[H,\rho]\!+\!\sum_{k,l=1}^3\!\!\gamma_k^{le}\mathcal{L}(\hat{\sigma}_k^{le})\rho \!+\!\kappa_a\mathcal{L}(\hat{a})\rho\!+\!\kappa_b\mathcal{L}(\hat{b})\rho,~~
\end{eqnarray}
where $\hat{\sigma}_k^{le}=|l\rangle_{kk}\langle e|$ and the Lindblad super-operator is
\begin{eqnarray}
\mathcal{L}(\hat{o})\rho=\hat{o}\rho \hat{o}^{\dag}-\frac{1}{2}(\hat{o}^{\dag}\hat{o}\rho+\rho\hat{o}^{\dag}\hat{o}),
\end{eqnarray}
where $\hat{o}$ is the standard Lindblad operator. $\gamma_k^{le}$ and $\kappa_a$ ($\kappa_b$) are the dissipation rate from the excited state $|e\rangle$ to the ground state $|l\rangle$ and the decay rate of the cavity mode $a$ ($b$), respectively.
For simplicity, we set $\gamma_k^{le}=\gamma/3$ and $\kappa_a=\kappa_b=\kappa$.

Figure~\ref{fig:10} displays the fidelity $F_{S}$ as a function of
the dissipation rate $\gamma$ and the decay rate $\kappa$ by both the three-pulse sequence scheme and the resonant pulses scheme shown in Ref.~\cite{Shao2010}.
We observe from Fig.~\ref{fig:10}(b) that the fidelity cannot maintain a relatively high value in the resonant pulses scheme even if the system does not suffer from the decoherence, cf., $F_S<0.956$ in Fig.~\ref{fig:10}(b). However, in the three-pulse sequence scheme, the fidelity can reach a high value ($\geq0.99$) when the dissipation rate is small, cf., the left region of the pink-dotted line in Fig.~\ref{fig:10}(a).
Another intriguing aspect in Fig.~\ref{fig:10} is that the fidelity does not sharply drop in the resonant pulses scheme
while the range of fidelity varies widely in the three-pulse sequence scheme.
This is because the evolution time in the resonant pulses scheme is much shorter than that in the three-pulse sequence scheme.
As a result, the influence of the decoherence on the system is relatively small by the resonant pulses scheme.
From this point, we find that the longer pulse sequence is not the optimal choice in a severe decoherence environment, even though the longer pulse sequence can help to improve the robustness of the system.
Furthermore, according to the pink-dotted line in Fig.~\ref{fig:10}(a), it is seen that
the influence of the cavity decay is almost negligible, since $F_S\geq0.99$ even when the decay rate $\kappa$ increases to 0.1.
This phenomenon can be explained by the fact that the strong coupling constant $\lambda_{k}^{a} (\lambda_{k}^{b})$ ensures the decoupling of the single-photon state to minimize the loss of the cavity decay.
From a physical point of view, the cavity is regarded as intermediary to realize indirect coupling between atoms, and it is almost in the vacuum state under the strong coupling condition.

On the other hand, for a given cavity decay rate, the fidelity $F_{S}$ decreases with the increase of the dissipation rate $\gamma$. Thus, the major influence on the preparation of single state is atomic spontaneous emission. The physical mechanism can be clarified as follows.
Under the strong coupling condition $\Omega_k\ll \lambda_{k}^{a}$ ($\lambda_{k}^{b}$),
the dynamics of the system can be approximately described by the Hamiltonian $H'$ in Eq.~(\ref{heff}).
To obtain the singlet state, we need to utilize the intermediate state $|\Psi_3\rangle$.
Nevertheless, we can see that $|\Psi_3\rangle$ contains the excited states $|e\rangle_{k}$ ($k=1,2,3$), which cannot be effectively eliminated during the evolution process.
Hence, the dynamics of the system is sensitive to the spontaneous emission.
An alternative approach to solve this problem is to adiabatically eliminate the excited state $|e\rangle_{k}$ \cite{Brion_2007,RevModPhys.89.015006},
i.e., adopting the Raman transition process.
To be specific, the one-photon detuning between the excited state and the ground states is large enough, while the system satisfies the two-photon resonance condition.
Then,
the population of the intermediate state is sharply suppressed, and the population can adiabatically transfer between ground states during the evolution.
As a result, the influence of the spontaneous emission is effectively eliminated.

\section{Conclusion}   \label{V}

In this work,
we achieved a high fidelity population transfer for the three-level system by the CPs sequence.
We derived the transition probability from the total propagator and used the Taylor expansion to disassemble it into a series of derivatives.
For a small number of pulses, since not enough variables (phases) can be employed to eliminate anticipated deviations,
we constructed a cost function to control the fidelity of the result.
Then, we showed that high fidelity of the target state,
or suppressing the population of the leakage state,
or both two objectives are robustly achievable by choosing different forms of cost functions.
Particularly, the method of the short pulse sequence can be further extended to a longer one if necessary.

As an example, we applied the CPs sequence to implement the three-atom singlet state with ultrahigh fidelity in the atom-cavity system,
where we only needed to individually control the phase of the inputting pulses.
The numerical results indicate that the final singlet state is robust against the deviations in two coupling coefficients.
Moreover, we found that significant distortion in the pulse sequence has a very small impact on the fidelity of the singlet state.
As the simulated results further reveal,
the fidelity of the final singlet state keeps a relatively high value in the decoherence environment.
Our CPs scheme provides a selectable way for robust controlling in the error-prone environment and shows an enormous potential application in various physical models, e.g., implementation of coherently convertible dual-type qubits with high fidelity \cite{yang2021realizing}.

\begin{acknowledgments}
	This work is supported by the National Natural Science Foundation of China under Grant Nos. 11874114, 12175033, 12147206, 11805036, the Natural Science Foundation of Fujian Province under Grant No.~2021J01575, the Natural Science Funds for Distinguished Young Scholar of Fujian Province under Grant 2020J06011, and the Project from Fuzhou University under Grant JG202001-2.
\end{acknowledgments}

\begin{widetext}
\begin{appendix}

\section{The detailed derivation of the total propagator without deviations in the composite $N$-pulse sequence}  \label{a3}

In this appendix, we first calculate the general expression of the propagator for each pulse, and then
give out the total propagator of the composite $N$-pulse sequence.

In absence of the deviations, the Hamiltonian of the $n$th pulse can be written as ($\hbar=1$)
\begin{equation}
H_n=\Delta_n|e\rangle\langle e|+\Omega_{1n}e^{i\alpha_n}|g\rangle\langle e|+\Omega_{2n}e^{i\beta_n}|r\rangle\langle e|+\mathrm{H.c.},   \nonumber
\end{equation}
By choosing the pulse duration $T_n=2\pi/\sqrt{\Delta_n^2+4(\Omega_{1n}^2+\Omega_{2n}^2)}$, the matrix form of the corresponding propagator $U(\theta_n,\gamma_n)=\exp(-iH_nT_n)$ in the basis $\{|g\rangle, |r\rangle, |e\rangle\}$ becomes (up to a global phase)
\begin{eqnarray}
U(\theta_n,\gamma_n)\!=\!\!\displaystyle \left(
                       \begin{array}{ccc}
                         \!\displaystyle\frac{i(\Omega_{1n}^2e^{-i\Theta_n}-\Omega_{2n}^2e^{i\Theta_n})}{\Omega_{1n}^2+\Omega_{2n}^2} & \!\displaystyle\frac{2i\Omega_{1n}\Omega_{2n}\cos\Theta_ne^{i(\alpha_n-\beta_n)}}{\Omega_{1n}^2+\Omega_{2n}^2} & 0 \\[2.5ex]
                         \displaystyle\frac{2i\Omega_{1n}\Omega_{2n}\cos\Theta_ne^{-i(\alpha_n-\beta_n)}}{\Omega_{1n}^2+\Omega_{2n}^2} & \!\displaystyle\frac{i(\Omega_{2n}^2e^{-i\Theta_n}-\Omega_{1n}^2e^{i\Theta_n})}{\Omega_{1n}^2+\Omega_{2n}^2} & 0 \\[2.5ex]
                         0 & 0 &  \!\!\!\!\!ie^{-i\Theta_n}\\
                       \end{array}
                     \right)\!\!=\!\!\left(
                               \begin{array}{ccc}
                                 \cos\theta_ne^{i\Phi_n} & \sin\theta_ne^{-i\gamma_n} & 0 \\[1ex]
                                 \!-\sin\theta_ne^{i\gamma_n} & \cos\theta_ne^{-i\Phi_n} & 0 \\[1ex]
                                 0 & 0 & \!\!ie^{-i\Theta_n} \\
                               \end{array}
                             \right),    \nonumber
\end{eqnarray}
where
\begin{eqnarray}
\Theta_n&=&\frac{\pi\Delta_n}{2\sqrt{\Delta_n^2+4(\Omega_{1n}^2+\Omega_{2n}^2)}},~~~~ \Phi_n=\arctan\frac{(\Omega_{1n}^2-\Omega_{2n}^2)\cos\Theta_n}{(\Omega_{1n}^2+\Omega_{2n}^2)\sin\Theta_n}, \nonumber\\[1ex]
\theta_n&=&\arcsin\frac{2\Omega_{1n}\Omega_{2n}\cos\Theta_n}{\Omega_{1n}^2+\Omega_{2n}^2},~~~~~~
\gamma_n=\beta_n-\alpha_n-\pi/2. \nonumber
\end{eqnarray}
Since the ground states are decoupled from the excited state $|e\rangle$ and the qubit subspace is of interest, we can only employ $2\times2$ matrix of the propagator $U(\theta_n,\gamma_n)$ in the following.

Next, we calculate the propagator of the first two pulses, which reads
\begin{eqnarray}
U(\theta_2,\gamma_2)U(\theta_1,\gamma_1)=\!\left(
                                           \begin{array}{cc}
                                             \cos\theta_2e^{i\Phi_2} & \sin\theta_2e^{-i\gamma_2}  \\
                                          -\sin\theta_2e^{i\gamma_2} & \cos\theta_2e^{-i\Phi_2} \\
                                           \end{array}
                                         \right)\!\!\left(
                                           \begin{array}{cc}
                                             \cos\theta_1e^{i\Phi_1} & \sin\theta_1e^{-i\gamma_1}  \\
                                          -\sin\theta_1e^{i\gamma_1} & \cos\theta_1e^{-i\Phi_1} \\
                                           \end{array}
                                         \right)\!=\!\left(
                                           \begin{array}{cc}
                                             \cos\vartheta_2e^{i\varphi_2} & \sin\vartheta_2e^{-i\phi_2}  \\
                                          -\sin\vartheta_2e^{i\phi_2} & \cos\vartheta_2e^{-i\varphi_2} \\
                                           \end{array}
                                         \right),~~~~ \nonumber
\end{eqnarray}
where
\begin{eqnarray}
\varphi_{2}&=&\arctan\frac{\cos\theta_2\cos\theta_{1}\sin(\Phi_2+\Phi_1)+ \sin\theta_2\sin\theta_{1}\sin(\gamma_2-\gamma_1)}{ \cos\theta_2\cos\theta_{1}\cos(\Phi_2+\Phi_1)- \sin\theta_2\sin\theta_{1}\cos(\gamma_2-\gamma_1)},  \nonumber\\[.5ex]
\phi_{2}&=&\arctan\frac{\cos\theta_2\sin\theta_{1}\sin(\gamma_1-\Phi_2)+ \sin\theta_{2}\cos\theta_{1}\sin(\gamma_2+\Phi_{1})}{\cos\theta_{2}\sin\theta_{1}\cos(\gamma_1-\Phi_2)+ \sin\theta_2\cos\theta_{1}\cos(\gamma_2+\Phi_1)},    \nonumber\\[.5ex]
\vartheta_{2}&=&\arccos\sqrt{\cos^2\theta_{2}\cos^2\theta_{1}+\sin^2\theta_{2}\sin^2\theta_{1}- \frac{1}{2}\sin2\theta_2\sin2\theta_{1}\cos(\gamma_2\!+\Phi_2+\Phi_1-\gamma_1)}.  \nonumber
\end{eqnarray}
One can easily find that the propagator $U(\theta_2,\gamma_2)U(\theta_1,\gamma_1)$ has the same form as that of the single pulse. Therefore, we can assume that the propagator of the first $(n-1)$ pulses has the following form ($n=2,3,\dots$)
\begin{eqnarray}
U(\theta_{n-1},\gamma_{n-1})\cdots U(\theta_2,\gamma_2)U(\theta_1,\gamma_1)=\left(
                                           \begin{array}{cc}
                                             \cos\vartheta_{n-1}e^{i\varphi_{n-1}} & \sin\vartheta_{n-1}e^{-i\phi_{n-1}}  \\
                                          -\sin\vartheta_{n-1}e^{i\phi_{n-1}} & \cos\vartheta_{n-1}e^{-i\varphi_{n-1}} \\
                                           \end{array}
                                         \right).~~~~ \nonumber
\end{eqnarray}
Then, the propagator of the first $n$ pulses becomes
\begin{eqnarray}  \label{aa6}
U(\theta_{n},\gamma_{n})U(\theta_{n-1},\gamma_{n-1})\cdots U(\theta_2,\gamma_2)U(\theta_1,\gamma_1)\!&=&\!\left(
                                           \begin{array}{cc}
                                             \cos\theta_ne^{i\Phi_n} & \sin\theta_ne^{-i\gamma_n}  \\
                                          \!-\sin\theta_ne^{i\gamma_n} & \cos\theta_ne^{-i\Phi_n} \\
                                           \end{array}
                                         \right)\!\left(
                                           \begin{array}{cc}
                                             \cos\vartheta_{n-1}e^{i\varphi_{n-1}} & \sin\vartheta_{n-1}e^{-i\phi_{n-1}}  \\
                                          \!-\sin\vartheta_{n-1}e^{i\phi_{n-1}} & \cos\vartheta_{n-1}e^{-i\varphi_{n-1}} \\
                                           \end{array}
                                         \right),  \nonumber\\[1ex]
                                         \!&=&\!\left(
                                           \begin{array}{cc}
                                             \cos\vartheta_{n}e^{i\varphi_{n}} & \sin\vartheta_{n}e^{-i\phi_{n}}  \\
                                          \!-\sin\vartheta_{n}e^{i\phi_{n}} & \cos\vartheta_{n}e^{-i\varphi_{n}} \\
                                           \end{array}
                                         \right).
\end{eqnarray}
From Eq.~(\ref{aa6}), we have the following recursive relations:
\begin{eqnarray}  \label{a21}
\varphi_{n}&=&\arctan\frac{\cos\theta_n\cos\vartheta_{n-1}\sin(\Phi_n+\varphi_{n-1})+ \sin\theta_n\sin\vartheta_{n-1}\sin(\gamma_n-\phi_{n-1})}{ \cos\theta_n\cos\vartheta_{n-1}\cos(\Phi_n+\varphi_{n-1})- \sin\theta_n\sin\vartheta_{n-1}\cos(\gamma_n-\phi_{n-1})},  \nonumber\\[.5ex]
\phi_{n}&=&\arctan\frac{\cos\theta_n\sin\vartheta_{n-1}\sin(\phi_{n-1}-\Phi_n)+ \sin\theta_{n}\cos\vartheta_{n-1}\sin(\gamma_n+\varphi_{n-1})}{\cos\theta_{n}\sin\vartheta_{n-1}\cos(\phi_{n-1}-\Phi_n)+ \sin\theta_n\cos\vartheta_{n-1}\cos(\gamma_n+\varphi_{n-1})},    \nonumber\\[.5ex]
\vartheta_{n}&=&\arccos\sqrt{\cos^2\theta_{n}\cos^2\vartheta_{n\!-\!1}+\sin^2\theta_{n}\sin^2\vartheta_{n\!-\!1}- \frac{1}{2}\sin2\theta_n\sin2\vartheta_{n\!-\!1}\cos(\gamma_n+\Phi_n+\varphi_{n\!-\!1}-\phi_{n\!-\!1})}.
\end{eqnarray}
Therefore, the total propagator of the composite $N$-pulse sequence becomes
\begin{eqnarray}        \label{r1}
U(\theta_{N},\gamma_{N})U(\theta_{N-1},\gamma_{N-1})\cdots U(\theta_2,\gamma_2)U(\theta_1,\gamma_1)=\left(
                                           \begin{array}{cc}
                                             \cos\vartheta_{N}e^{i\varphi_{N}} & \sin\vartheta_{N}e^{-i\phi_{N}}  \\
                                          \!-\sin\vartheta_{N}e^{i\phi_{N}} & \cos\vartheta_{N}e^{-i\varphi_{N}} \\
                                           \end{array}
                                         \right),
\end{eqnarray}
where $\varphi_{N}$, $\phi_{N}$, and $\vartheta_{N}$ satisfy the recursive relations given by Eq.~(\ref{a21}).
The population of the ground state $|r\rangle$ reads
\begin{eqnarray}
P^{\sst(0)}_r=\sin^2\vartheta_{N}=\sin^2\theta_{N}\cos^2\vartheta_{N\!-\!1}+\cos^2\theta_{N}\sin^2\vartheta_{N\!-\!1} +\frac{1}{2}\sin2\theta_N\sin2\vartheta_{N\!-\!1}\cos(\gamma_N+\Phi_N+\varphi_{N\!-\!1}-\phi_{N\!-\!1}).~~~
\end{eqnarray}

\section{The second-order coefficients in the two-pulse sequence}  \label{a1}

In this appendix, we give the expressions of the second-order coefficients $C^{\sst(2)}_{2,k}$ and $D^{\sst(2)}_{2,k}$ ($k=1,2,3$), in terms of $\alpha_{12}$, which read
\begin{eqnarray}
C^{\sst(2)}_{2,1}&=&\sin\theta\Big[\frac{\sqrt{2}}{8}\pi^2\sin(\alpha_{12}-\theta)-\frac{\pi^2-16}{4}\sin\theta\Big], \nonumber\\[0.1ex]
C^{\sst(2)}_{2,2}&=&\sin\theta\Big[\frac{3\sqrt{2}+4}{16}\pi^2\sin(\alpha_{12}-\theta)-\frac{(2\sqrt{2}+3)\pi^2+16}{8}\sin\theta\Big], \nonumber\\[0.1ex]
C^{\sst(2)}_{2,3}&=&\sin\theta\Big[\frac{3\sqrt{2}-4}{16}\pi^2\sin(\alpha_{12}-\theta)+\frac{(2\sqrt{2}-3)\pi^2-16}{8}\sin\theta\Big], \nonumber\\[0.1ex]
D^{\sst(2)}_{2,1}&=&\frac{\pi^2}{16}\Big[2(1+\sqrt{2})\cos\alpha_{12}+2\cos(\alpha_{12}-2\theta)+\sqrt{2}\cos2\theta+4+\sqrt{2}\Big], \nonumber\\[0.1ex]
D^{\sst(2)}_{2,2}&=&\frac{\pi^2}{32}\Big[2(5\sqrt{2}+7)\cos\alpha_{12}+ (6+4\sqrt{2})\cos(\alpha_{12}-2\theta)+(3\sqrt{2}+4)\cos2\theta+16+11\sqrt{2}\Big], \nonumber\\[0.1ex]
D^{\sst(2)}_{2,3}&=&\frac{\pi^2}{32}\Big[2(\sqrt{2}-1)\cos\alpha_{12}+(6-4\sqrt{2})\cos(\alpha_{12}-2\theta) +(3\sqrt{2}-4)\cos2\theta+8-5\sqrt{2}\Big].  \nonumber
\end{eqnarray}

With these coefficients, we can evaluate the cost functions.
First, the expression of the cost function $F=F_c^{\sst(2)}$ is
\begin{eqnarray}
F_c^{\sst(2)}&=&|C^{\sst(2)}_{2,1}|^2+|C^{\sst(2)}_{2,2}|^2+|C^{\sst(2)}_{2,3}|^2 \nonumber\\[0.1ex]
&=&\frac{\sin^2\theta}{64}\Big[19\pi^4\sin^2(\alpha_{12}-\theta)-2\sqrt{2}\pi^2(16+19\pi^2)\sin(\alpha_{12}-\theta)\sin\theta +2(768+32\pi^2+19\pi^4)\sin^2\theta\Big].    \nonumber
\end{eqnarray}
As a result, it is not hard to calculate that the cost function $F=F_c^{\sst(2)}$ is minimal when
\begin{equation}
\alpha_{12}\!=\!\left\{
\begin{aligned}
    \displaystyle\!&\theta+\arcsin\frac{\sqrt{2}(16+19\pi^2)\sin\theta}{19\pi^2},  ~~~~~0\leq\theta\leq\theta', \\[1.1ex]
    \displaystyle\!&\theta+\frac{\pi}{2},  ~~~~~~~~~~~~~~~~~~~~~~~~~~~~~~~~~~~~~\theta'<\theta\leq\frac{\pi}{2}, \\
\end{aligned}
\right.     \nonumber
\end{equation}
where $\theta'=\arcsin\big[\!{19\pi^2}/{\!\sqrt{2}(16\!+\!19\pi^2)}\big]$.
Second, the expression of the cost function $F=F_d^{\sst(2)}$ is
\begin{eqnarray}
F_d^{\sst(2)}&=&|D^{\sst(2)}_{2,1}|^2+|D^{\sst(2)}_{2,2}|^2+|D^{\sst(2)}_{2,3}|^2 \nonumber\\[0.1ex]
&=&\frac{19\pi^4(6+4\sqrt{2})}{256}\Big[\cos\alpha_{12}+(\sqrt{2}-1)\cos(\alpha_{12}-2\theta) +\frac{2+\sqrt{2}\cos^2\theta}{1+\sqrt{2}}\Big]^2.    \nonumber
\end{eqnarray}
Obviously, $F_d^{\sst(2)}$ is minimal when
\begin{eqnarray}
\alpha_{12}=\pi-\displaystyle\arctan\frac{(\sqrt{2}-1)\sin 2\theta}{(\sqrt{2}-1)\cos2\theta+1}.    \nonumber
\end{eqnarray}
Finally, the cost function $F=F_c^{\sst(2)}+F_d^{\sst(2)}$ is minimal when
\begin{eqnarray}
\alpha_{12}=2\arctan \Theta.    \nonumber
\end{eqnarray}
Here, $\Theta$ is the solution of the following quartic equation:
\begin{eqnarray}
a_1\Theta^4+a_2\Theta^3+a_3\Theta^2+a_4\Theta+a_5=0,   \nonumber
\end{eqnarray}
where the coefficients $a_k$ ($k=1,\dots,5$) are:
\begin{eqnarray}
a_1&=&\big[19\pi^2(\sqrt{2}-1)^2-16\sqrt{2}\big]\sin^2\theta\sin2\theta,     \nonumber\\[.1ex]
a_2&=&76\pi^2\big[\sqrt{2}+1+(\sqrt{2}-1)^2\cos2\theta\big]\sin^2\theta+\!64\sqrt{2}\sin^4\theta,    \nonumber\\[.1ex]
a_3&=&57\pi^2(2\sqrt{2}+1+3\cos2\theta)\sin2\theta,    \nonumber\\[.1ex]
a_4&=&64\sqrt{2}\sin^4\theta+19\pi^2\big[13+11\sqrt{2}+4(2+\sqrt{2})\cos2\theta+(3+\sqrt{2})\cos4\theta)\big],   \nonumber\\[.1ex]
a_5&=&\sin2\theta\big[16\sqrt{2}\sin^2\theta-19\pi^2\big(1+(\sqrt{2}+1)^2\cos^2\theta\big)\big]. \nonumber
\end{eqnarray}

\section{The second-order coefficients in the three-pulse sequence}  \label{a}

In this appendix, we give the expressions of the second-order coefficients $C^{\sst(2)}_k$ and $D^{\sst(2)}_k$ ($k=1,2,3$) in the three-pulse sequence, which read
\begin{eqnarray}
C^{\sst(2)}_{3,1}&=&-\frac{1}{4}\Big[ 2\cos(\alpha_{12}-\beta_{12})+2\cos(\alpha_{23}-\beta_{23}) -5\cos(\alpha_{12}+\alpha_{23}-\beta_{12}-\beta_{23})+3\cos(\alpha_{12}-\alpha_{23}-\beta_{12}+\beta_{23})\Big]  \cr
&&-\frac{\pi^2}{32}\Big[ 6+2(1+\sqrt{2})\cos\alpha_{12}+2\cos\alpha_{23}+2\cos(\alpha_{12}+\alpha_{23})+ 2\sqrt{2}\cos(\alpha_{12}-\beta_{12})+\sqrt{2}\cos(\alpha_{12}-\alpha_{23}-\beta_{12})   \cr
&&-(\sqrt{2}-2)\cos(\alpha_{12}+\alpha_{23}-\beta_{12})+2\cos\beta_{12}+2\cos(\alpha_{23}+\beta_{12}) -2\sqrt{2}\cos(\alpha_{23}-\beta_{23})-\sqrt{2}\cos(\alpha_{12}+\alpha_{23}-\beta_{23}) \cr
&&+\sqrt{2}\cos(\alpha_{12}-\beta_{12}-\beta_{23})+(2+\sqrt{2})\cos(\alpha_{23}-\beta_{12}-\beta_{23}) +3\cos(\alpha_{12}+\alpha_{23}-\beta_{12}-\beta_{23})-2(\sqrt{2}-1)\cos\beta_{23} \cr
&&-\sqrt{2}\cos(\alpha_{23}+\beta_{12}-\beta_{23})-2\cos(\alpha_{12}+\beta_{23})- (2+\sqrt{2})\cos(\alpha_{12}-\alpha_{23}+\beta_{23}) +(\sqrt{2}-2)\cos(\alpha_{12}-\beta_{12}+\beta_{23})   \cr
&&-3\cos(\alpha_{12}-\alpha_{23}-\beta_{12}+\beta_{23})+2\cos(\beta_{12}+\beta_{23}) \Big],
\cr
C^{\sst(2)}_{3,2}&=& \frac{1}{8}\Big[ 2(1+\sqrt{2})\cos(\alpha_{12}-\beta_{12})+2(1+\sqrt{2})\cos(\alpha_{23}-\beta_{23}) -(5+\sqrt{2})\cos(\alpha_{12}+\alpha_{23}-\beta_{12}-\beta_{23}) \cr
&&-(\sqrt{2}-3)\cos(\alpha_{12}-\alpha_{23}-\beta_{12}+\beta_{23})\Big] -\frac{\pi^2}{64}\Big[ 6(3+2\sqrt{2}) + 2(7+5\sqrt{2})\cos\alpha_{12}+(6+4\sqrt{2})\cos\alpha_{23}\cr
&&+(6+4\sqrt{2})\cos(\alpha_{12}+\alpha_{23}) +(8+6\sqrt{2})\cos(\alpha_{12}-\beta_{12})+(4+3\sqrt{2})\cos(\alpha_{12}-\alpha_{23}-\beta_{12})   \cr
&&+(\sqrt{2}+2)\cos(\alpha_{12}+\alpha_{23}-\beta_{12})+(6+4\sqrt{2})\cos\beta_{12}+2(3+2\sqrt{2})\cos(\alpha_{23}+\beta_{12}) -2(4+3\sqrt{2})\cos(\alpha_{23}-\beta_{23})\cr
&&-(4+3\sqrt{2})\cos(\alpha_{12}+\alpha_{23}-\beta_{23}) +(4+3\sqrt{2})\cos(\alpha_{12}-\beta_{12}-\beta_{23})+(10+7\sqrt{2})\cos(\alpha_{23}-\beta_{12}-\beta_{23}) \cr
&& +3(3+2\sqrt{2})\cos(\alpha_{12}+\alpha_{23}-\beta_{12}-\beta_{23})-2(1+\sqrt{2})\cos\beta_{23} -(4+3\sqrt{2})\cos(\alpha_{23}+\beta_{12}-\beta_{23}) \cr
&&-2(3+2\sqrt{2})\cos(\alpha_{12}+\beta_{23})- (10+7\sqrt{2})\cos(\alpha_{12}-\alpha_{23}+\beta_{23})-(\sqrt{2}+2)\cos(\alpha_{12}-\beta_{12}+\beta_{23}) \cr
&&-3(3+2\sqrt{2})\cos(\alpha_{12}-\alpha_{23}-\beta_{12}+\beta_{23})+2(3+2\sqrt{2})\cos(\beta_{12}+\beta_{23}) \Big],
\cr
C^{\sst(2)}_{3,3}&=& \frac{1}{8}\Big[ 2(1-\sqrt{2})\cos(\alpha_{12}-\beta_{12})+2(1-\sqrt{2})\cos(\alpha_{23}-\beta_{23}) -(5-\sqrt{2})\cos(\alpha_{12}+\alpha_{23}-\beta_{12}-\beta_{23}) \cr
&&+(\sqrt{2}+3)\cos(\alpha_{12}-\alpha_{23}-\beta_{12}+\beta_{23})\Big] +\frac{\pi^2}{64}\Big[ 6(2\sqrt{2}-3) + 2(1-\sqrt{2})\cos\alpha_{12}+(4\sqrt{2}-6)\cos\alpha_{23}\cr
&&+(4\sqrt{2}-6)\cos(\alpha_{12}+\alpha_{23}) +(8-6\sqrt{2})\cos(\alpha_{12}-\beta_{12})+(4-3\sqrt{2})\cos(\alpha_{12}-\alpha_{23}-\beta_{12})   \cr
&&+(7\sqrt{2}\!-\!10)\cos(\alpha_{12}+\alpha_{23}-\beta_{12})+(4\sqrt{2}\!-\!6)\cos\beta_{12} +2(2\sqrt{2}\!-\!3)\cos(\alpha_{23}+\beta_{12}) +2(3\sqrt{2}\!-\!4)\cos(\alpha_{23}-\beta_{23})\cr
&&-(4-3\sqrt{2})\cos(\alpha_{12}+\alpha_{23}-\beta_{23}) +(4-3\sqrt{2})\cos(\alpha_{12}-\beta_{12}-\beta_{23})+(\sqrt{2}-2)\cos(\alpha_{23}-\beta_{12}-\beta_{23}) \cr
&& +3(2\sqrt{2}-3)\cos(\alpha_{12}+\alpha_{23}-\beta_{12}-\beta_{23})+2(5\sqrt{2}-7)\cos\beta_{23} -(4-3\sqrt{2})\cos(\alpha_{23}+\beta_{12}-\beta_{23}) \cr
&&+2(3-2\sqrt{2})\cos(\alpha_{12}+\beta_{23})+ (2-\sqrt{2})\cos(\alpha_{12}-\alpha_{23}+\beta_{23})+(10-7\sqrt{2})\cos(\alpha_{12}-\beta_{12}+\beta_{23}) \cr
&&+3(3-2\sqrt{2})\cos(\alpha_{12}-\alpha_{23}-\beta_{12}+\beta_{23})-2(3-2\sqrt{2})\cos(\beta_{12}+\beta_{23}) \Big],
\cr
D^{\sst(2)}_{3,1}&=&\frac{\pi^2}{16}\Big[6+\sqrt{2}+2(1+\sqrt{2})\cos\alpha_{12}
+(2+\sqrt{2})\cos(\alpha_{12}+\alpha_{23})+2\sqrt{2}\cos(\alpha_{12}-\beta_{12})+\cos(\alpha_{12}+\alpha_{23}-\beta_{12})  \cr
&&+(1\!+\!\sqrt{2})\cos(\alpha_{12}\!-\!\alpha_{23}\!-\!\beta_{12})+2\cos\beta_{12}+(2\!+\!\sqrt{2})\cos(\alpha_{23}+\beta_{12}) -\cos(\alpha_{12}\!-\!\beta_{12}\!-\!\beta_{23})+\sqrt{2}\cos(\alpha_{12}+\beta_{23})  \cr
&&+(\sqrt{2}-1)\cos(\alpha_{12}-\beta_{12}+\beta_{23})-\sqrt{2}\cos(\beta_{12}+\beta_{23})+ (2-\sqrt{2})\cos\beta_{23}+\sqrt{2}\sin\alpha_{23}\cos\beta_{23}\sin(\alpha_{12}-\beta_{12}) \cr
&&+(2+\sqrt{2})\cos\alpha_{23}-\sqrt{2}\sin\beta_{23}\cos\alpha_{23}\sin(\alpha_{12}-\beta_{12})  \Big],
\cr
D^{\sst(2)}_{3,2}&=&\frac{\pi^2}{32}\Big[22+15\sqrt{2}+2(7+5\sqrt{2})\cos\alpha_{12}+(10+7\sqrt{2})\cos\alpha_{23}
+(10+7\sqrt{2})\cos(\alpha_{12}+\alpha_{23}) +(6+4\sqrt{2})\cos\beta_{12} \cr
&&+(8+6\sqrt{2})\cos(\alpha_{12}-\beta_{12})+(3+2\sqrt{2})\cos(\alpha_{12}+\alpha_{23}-\beta_{12})  +(7+5\sqrt{2})\cos(\alpha_{12}-\alpha_{23}-\beta_{12})  \cr
&&+(10+7\sqrt{2})\cos(\alpha_{23}+\beta_{12})-(3+2\sqrt{2})\cos(\alpha_{12}-\beta_{12}-\beta_{23})+ (2+\sqrt{2})\cos\beta_{23}+(4+3\sqrt{2})\cos(\alpha_{12}+\beta_{23})  \cr
&&+(\sqrt{2}+1)\cos(\alpha_{12}-\beta_{12}+\beta_{23})-(4+3\sqrt{2})\cos(\beta_{12}+\beta_{23}) +(4+3\sqrt{2})\sin(\alpha_{12}-\beta_{12})\sin(\alpha_{23}-\beta_{23})  \Big],
\cr
D^{\sst(2)}_{3,3}&=& \frac{\pi^2}{32}\Big[14-9\sqrt{2}-2(1-\sqrt{2})\cos\alpha_{12}+(2-\sqrt{2})\cos\alpha_{23}
+(2-\sqrt{2})\cos(\alpha_{12}+\alpha_{23}) +(6-4\sqrt{2})\cos\beta_{12} \cr
&&-(8-6\sqrt{2})\cos(\alpha_{12}-\beta_{12})+(3-2\sqrt{2})\cos(\alpha_{12}+\alpha_{23}-\beta_{12})  -(1-\sqrt{2})\cos(\alpha_{12}-\alpha_{23}-\beta_{12})  \cr
&&-(2-\sqrt{2})\cos(\alpha_{23}+\beta_{12})-(3-2\sqrt{2})\cos(\alpha_{12}-\beta_{12}-\beta_{23})+ (10-7\sqrt{2})\cos\beta_{23}-(4-3\sqrt{2})\cos(\alpha_{12}+\beta_{23})  \cr
&&+(5\sqrt{2}-7)\cos(\alpha_{12}-\beta_{12}+\beta_{23})+(4-3\sqrt{2})\cos(\beta_{12}+\beta_{23}) -(4-3\sqrt{2})\sin(\alpha_{12}-\beta_{12})\sin(\alpha_{23}-\beta_{23})  \Big], \nonumber
\end{eqnarray}
where $\alpha_{mn}=\alpha_{m}-\alpha_{n}$ and $\beta_{mn}=\beta_{m}-\beta_{n}$ ($m,n=1,2,3$).

\section{The first-order coefficients in the five-pulse sequence}  \label{c}

In this appendix, we provide the expressions of the first-order coefficients  in the five-pulse sequence, which are
\begin{eqnarray}
C^{\sst(1)}_{5,1}&=&-C^{\sst(1)}_{5,2}=\frac{1}{8 \sqrt{2}}\Big[\!-\!3 \cos ({\alpha _{15}}\!-\!{\beta _{15}})\!+\!\cos ({\alpha _{12}}\!-\!{\alpha _{23}}\!+\!{\alpha _{35}}-\!{\beta _{12}}\!+\!{\beta _{23}}\!-\!{\beta _{35}})\!+\!\cos ({\alpha _{13}}\!-\!{\alpha _{34}}\!+\!{\alpha _{45}}\!-\!{\beta _{13}}\!+\!{\beta _{34}}\!-\!{\beta _{45}})\nonumber\\[0.1ex]
&&\!+\!\cos ({\alpha _{12}}\!-\!{\alpha _{24}}\!+\!{\alpha _{45}}\!-\!{\beta _{12}}\!+\!{\beta _{24}}\!-\!{\beta _{45}})\!+\!\cos ({\alpha _{14}}\!-\!{\alpha _{45}}\!-\!{\beta _{14}}\!+\!{\beta _{45}})\!-\!3 \cos ({\alpha _{12}}\!-\!{\alpha _{23}}\!+\!{\alpha _{34}}\!-\!{\alpha _{45}}\!-\!{\beta _{12}}\!+\!{\beta _{23}}\!-\!{\beta _{34}}\!+\!{\beta _{45}}) \nonumber\\[0.1ex]
&&\!+\!2 [\cos ({\alpha _{23}}\!-\!{\beta _{23}})\!-\!1] \cos ({\alpha _{12}}\!-\!{\alpha _{35}}\!-\!{\beta _{12}}\!+\!{\beta _{35}})\!+\!2 \cos ({\alpha _{14}}\!-\!{\beta _{14}})\!+\!2 \cos ({\alpha _{12}}\!-\!{\alpha _{23}}\!+\!{\alpha _{34}}\!-\!{\beta _{12}}\!+\!{\beta _{23}}\!-\!{\beta _{34}})\nonumber\\[0.1ex]
&&\!-\!2 \cos ({\alpha _{13}}\!-\!{\alpha _{34}}\!-\!{\beta _{13}}\!+\!{\beta _{34}})\!-\!2 \cos ({\alpha _{12}}\!-\!{\alpha _{24}}\!-\!{\beta _{12}}\!+\!{\beta _{24}})\!-\!2 \cos ({\alpha _{13}}\!+\!{\alpha _{45}}\!-\!{\beta _{13}}\!-\!{\beta _{45}})\!+\!2 \cos ({\alpha _{25}}\!-\!{\beta _{25}})\nonumber\\[0.1ex]
&&\!-\!2 \cos ({\alpha _{12}}\!-\!{\alpha _{23}}\!+\!{\alpha _{45}}\!-\!{\beta _{12}}\!+\!{\beta _{23}}\!-\!{\beta _{45}})\!-\!2 \cos ({\alpha _{13}}\!-\!{\alpha _{45}}\!-\!{\beta _{13}}\!+\!{\beta _{45}})\!-\!2 \cos ({\alpha _{12}}\!-\!{\alpha _{23}}\!-\!{\alpha _{45}}\!-\!{\beta _{12}}\!+\!{\beta _{23}}\!+\!{\beta _{45}})\nonumber\\[0.1ex]
&&\!-\!2 \cos ({\alpha _{12}}\!+\!{\alpha _{35}}\!-\!{\beta _{12}}\!-\!{\beta _{35}})\!-\!2 \cos ({\alpha _{12}}\!-\!{\alpha _{34}}\!+\!{\alpha _{45}}\!-\!{\beta _{12}}\!+\!{\beta _{34}}\!-\!{\beta _{45}})\!-\!2 \cos ({\alpha _{12}}\!+\!{\alpha _{34}}\!-\!{\alpha _{45}}\!-\!{\beta _{12}}\!-\!{\beta _{34}}\!+\!{\beta _{45}})\nonumber\\[0.1ex]
&&\!+\!4 \cos ({\alpha _{12}}\!+\!{\alpha _{34}}\!-\!{\beta _{12}}\!-\!{\beta _{34}})\!+\!4 \cos ({\alpha _{12}}\!-\!{\alpha _{34}}\!-\!{\beta _{12}}\!+\!{\beta _{34}})\!+\!4 \cos ({\alpha _{12}}\!+\!{\alpha _{45}}\!-\!{\beta _{12}}\!-\!{\beta _{45}})\!+\!4 \cos ({\alpha _{12}}\!-\!{\alpha _{45}}\!-\!{\beta _{12}}\!+\!{\beta _{45}})\nonumber\\[0.1ex]
&&\!+\!2 \cos ({\alpha _{23}}\!-\!{\alpha _{34}}\!+\!{\alpha _{45}}\!-\!{\beta _{23}}\!+\!{\beta _{34}}\!-\!{\beta _{45}})\!-\!2 \cos ({\alpha _{24}}\!-\!{\alpha _{45}}\!-\!{\beta _{24}}\!+\!{\beta _{45}})\!-\!2 \cos ({\alpha _{23}}\!-\!{\alpha _{35}}\!-\!{\beta _{23}}\!+\!{\beta _{35}})\nonumber\\[0.1ex]
&&\!+\!4 \cos ({\alpha _{23}}\!+\!{\alpha _{45}}\!-\!{\beta _{23}}\!-\!{\beta _{45}})\!+\!4 \cos ({\alpha _{23}}\!-\!{\alpha _{45}}\!-\!{\beta _{23}}\!+\!{\beta _{45}})\Big],  \nonumber
\end{eqnarray}
where $\alpha_{mn}=\alpha_{m}-\alpha_{n}$ and $\beta_{mn}=\beta_{m}-\beta_{n}$ ($m,n=1,2,3,4,5$).

\section{Detailed discussion on the performance of the accuracy and leakage in the four-pulse sequence}  \label{b}

In this appendix, taking the four-pulse sequence as an example, we discuss in detail the influence of different cost functions on the performance of the accuracy and the leakage. As stated in Sec.~\ref{II}, the cost functions are mainly divided into three types: ($i$) $F=\sum_{l=2}^{\infty}A_lF_c^{\sst(l)}$ is meant to eliminate the deviations in $P_r^a$;
($ii$) $F=\sum_{l=2}^{\infty}B_lF_d^{\sst(l)}$ is aimed to eliminate the population leakage $P_e^a$;
($iii$) $F=\sum_{l=2}^{\infty}\left[A_lF_c^{\sst(l)}+B_lF_d^{\sst(l)}\right]$ is a combination of both, intended to eliminate both the deviations in $P_r^a$ and the population leakage $P_e^a$.

Figure~\ref{fig:b01}, Figure~\ref{fig:b02}, and Figure~\ref{fig:b03} are the plotted infidelity and population leakage as a function of deviations through the cost functions $F=F_c^{\sst(2)}+F_c^{\sst(3)}$, $F=F_d^{\sst(2)}+F_d^{\sst(3)}+F_d^{\sst(4)}$, and $F=F_c^{\sst(2)}+F_d^{\sst(2)}$, respectively. The results demonstrate that $P_r^a$ in Fig.~\ref{fig:b01} is much more accurate than those in Figs.~\ref{fig:b02} and Fig.~\ref{fig:b03}, while the population leakage $P_e^a$ in Fig.~\ref{fig:b02} is much smaller than those in Figs.~\ref{fig:b01} and Fig.~\ref{fig:b03}. In Fig.~\ref{fig:b03}, there is a tradeoff between the accuracy and the leakage.
Therefore, different cost functions have different effects, and one can choose a specific cost function according to the particular task in quantum information processing.

\begin{figure}[htbp]
	\centering
	\includegraphics[scale=0.066]{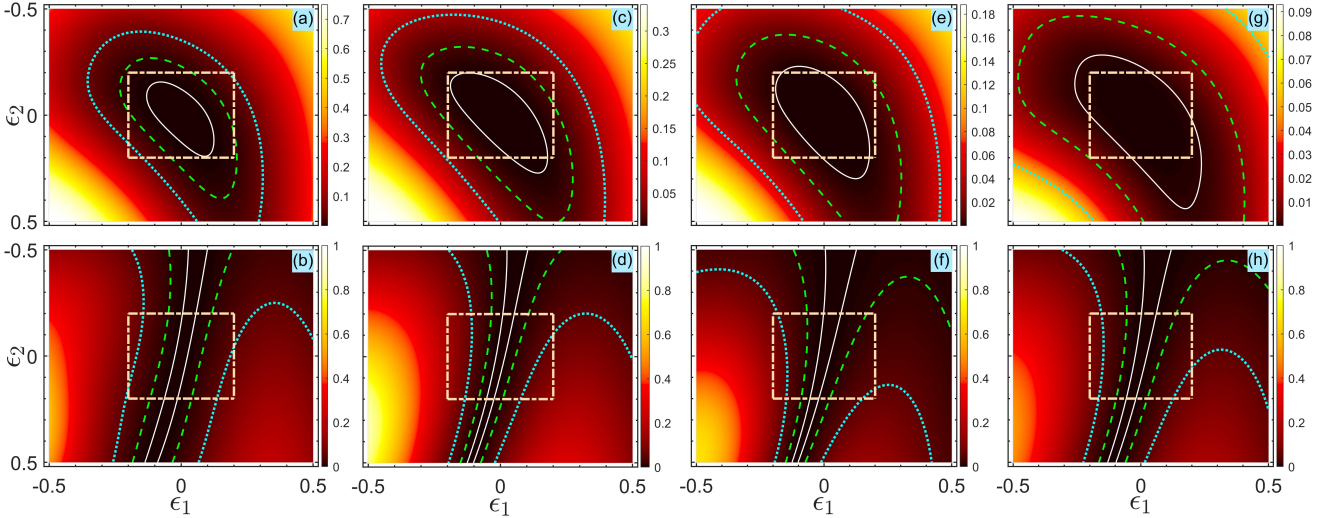}
	\caption{Infidelity $\mathcal{F}_r$ (top panels) and population leakage $P_e^a$ (bottom panels) vs the deviations $\epsilon_1$ and $\epsilon_2$ in the four-pulse sequence. The parameters are designed according to the cost function $F=F_c^{\sst(2)}+F_c^{\sst(3)}$, and the values are presented in Table \ref{bt1}. Henceforth, the white-solid curves, the green-dashed curves, and the cyan-dotted curves correspond to $\mathcal{F}_r(P_e^a)=0.001$, 0.01, and 0.05, respectively. The rectangular region (labeled by the yellow-dot-dashed lines) are: $|\epsilon_1|\leq0.2$ and $|\epsilon_2|\leq0.2$. The results demonstrate that the infidelity maintains a small value ($\leq0.001$, cf. the white-curves) in a very wide region. }  \label{fig:b01}
\end{figure}

\begin{figure}[htbp]
	\centering
	\includegraphics[scale=0.066]{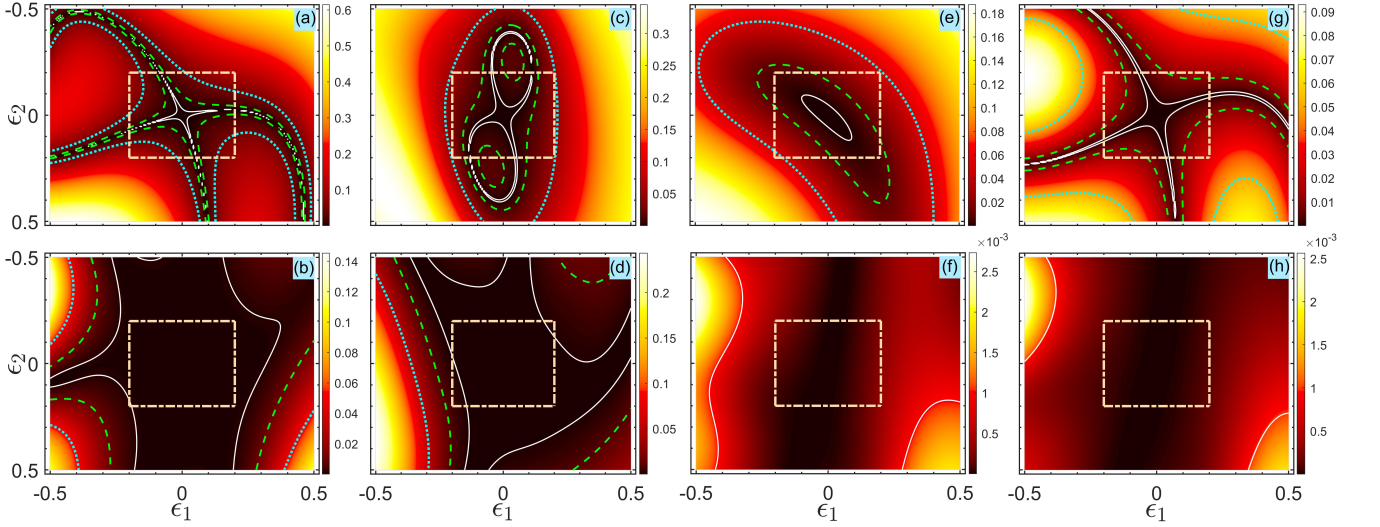}
	\caption{Infidelity $\mathcal{F}_r$ (top panels) and population leakage $P_e^a$ (bottom panels) vs the deviations $\epsilon_1$ and $\epsilon_2$ in the four-pulse sequence. The parameters are designed according to the cost function $F=F_d^{\sst(2)}+F_d^{\sst(3)}+F_d^{\sst(4)}$, and the values are presented in Table \ref{bt1}. The results demonstrate that the population leakage maintains a small value ($\leq0.001$, cf. the white-curves) in a very wide region. }  \label{fig:b02}
\end{figure}

\begin{figure}[htbp]
	\centering
	\includegraphics[scale=0.066]{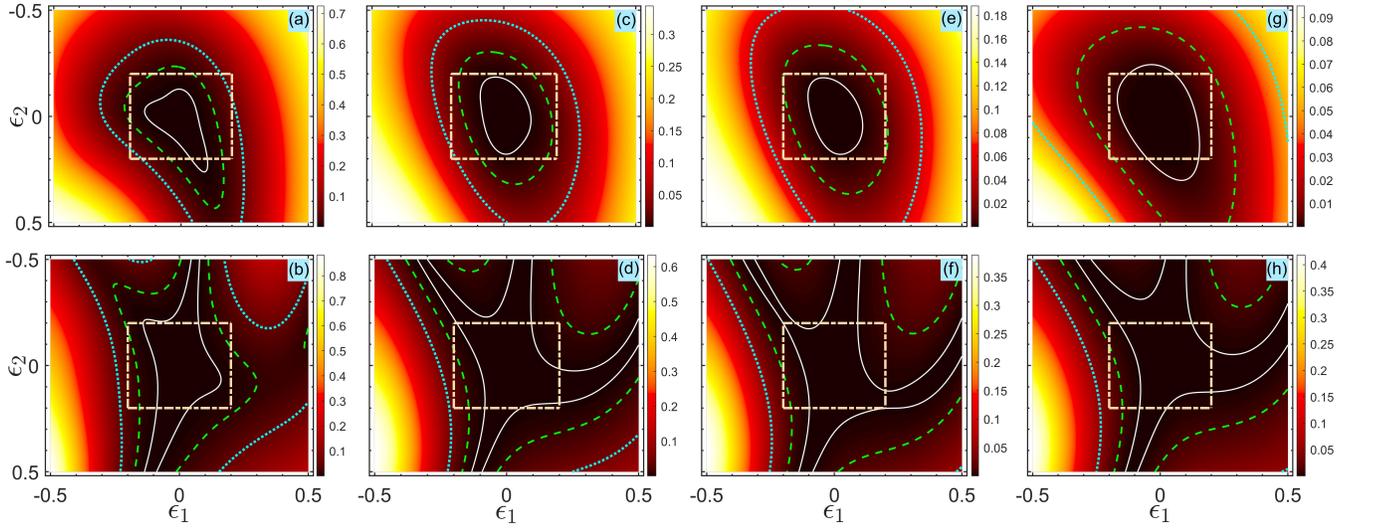}
	\caption{Infidelity $\mathcal{F}_r$ (top panels) and population leakage $P_e^a$ (bottom panels) vs the deviations $\epsilon_1$ and $\epsilon_2$ in the four-pulse sequence. The parameters are designed according to the cost function $F=F_c^{\sst(2)}+F_d^{\sst(2)}$, and the values are presented in Table \ref{bt1}. The results demonstrate that both the infidelity and the population leakage maintain a small value ($\leq0.001$, cf. the white-curves) in a very wide region, but the corresponding regions are not larger than those in Fig.~\ref{fig:b01} or Fig.~\ref{fig:b02}.}  \label{fig:b03}
\end{figure}

\renewcommand\arraystretch{1.2}
\begin{table}[htbp]
	\centering
	\caption{The parameters to achieve the predefined population $\mathcal{P}_{r}=\sin^2\theta$ in the four-pulse sequence, where $\alpha_1=\beta_1=0$.}
	\label{bt1}
	\begin{tabular}{ccccccccc}
		\hline
		\hline
          \multirow{2}{*}{} & \multirow{2}{*}{$\theta$} & \multirow{2}{*}{$F$} & \multirow{2}{*}{$\alpha_2$} & \multirow{2}{*}{$\alpha_3$} & \multirow{2}{*}{$\alpha_4$} & \multirow{2}{*}{$\beta_2$} & \multirow{2}{*}{$\beta_3$} & \multirow{2}{*}{$\beta_4$}  \\\\
		\hline
        \makecell[l]{Fig.~\!\ref{fig:04a}(b)} & $\pi$/4 &  0.0003  & 6.0928   & 4.6077   & 1.2835   & 2.1652   & 3.8217 & 1.2823 \\
        \makecell[l]{Fig.~\!\ref{fig:04a}(c)} & $\pi$/4 &  0.0004  & 3.6176   & 0.5585   & 3.1750   & 4.3222   & 4.4047 & 6.1550 \\
        \makecell[l]{Fig.~\!\ref{fig:04a}(d)} & $\pi$/4 &  0.0007  & 3.5541   & 2.3736   & 6.0291   & 2.7675   & 4.7286 & 6.0267 \\
        \makecell[l]{Figs.~\!\ref{fig:b01}(a)-\ref{fig:b01}(b)} & ~~~~$\pi$/3~~~~ & ~~~~0.0054~~~~  &  ~~~~3.0883~~~~  & ~~~~5.0265~~~~ & ~~~~1.7134~~~~ & ~~~~2.1955~~~~ & ~~~~0.9921~~~~ &~~~~0.9750~~~~\\
        \makecell[l]{Figs.~\!\ref{fig:b01}(c)-\ref{fig:b01}(d)} & $\pi$/5  &  0.0015  & 3.3013   & 0.0000   & 2.9063 &  2.5848 &  2.4251 &3.3583 \\
        \makecell[l]{Figs.~\!\ref{fig:b01}(e)-\ref{fig:b01}(f)} & $\pi$/7  &  0.0014  & 3.1948   & 5.1313   &  2.0471  & 4.7325 & 3.5274 & 1.0833\\
        \makecell[l]{Figs.~\!\ref{fig:b01}(g)-\ref{fig:b01}(h)} & $\pi$/10 &  0.0010  & 3.3013   & 0.6283   &  3.6638  & 4.4918   & 4.9604   & 2.2749 \\
        \makecell[l]{Figs.~\!\ref{fig:b02}(a)-\ref{fig:b02}(b)} & $\pi$/3  & 0.0021   &  3.9403  & 0.9425   &  3.0538  & 4.7064   & 4.8502   & 5.6332 \\
        \makecell[l]{Figs.~\!\ref{fig:b02}(c)-\ref{fig:b02}(d)} & $\pi$/5  & 0.0009   &3.5143    & 3.6652   &  1.2319  & 2.6600   & 5.9525   & 1.4083\\
        \makecell[l]{Figs.~\!\ref{fig:b02}(e)-\ref{fig:b02}(f)} & $\pi$/7  & 0.0010   &3.1948    & 0.1047   & 3.1443   & 3.6964   & 3.7479   & 0.1083\\
        \makecell[l]{Figs.~\!\ref{fig:b02}(g)-\ref{fig:b02}(h)} & $\pi$/10 & $2\!\times\!10^{-5}$ & 3.4078 & 0.2094 & 3.0884  & 3.6944  & 3.6376   &6.1749 \\
        \makecell[l]{Figs.~\!\ref{fig:b03}(a)-\ref{fig:b03}(b)} & $\pi$/3  & 0.0027  & 2.7689   &  5.5501   &  1.5253  &  5.7617  &  5.4013  &2.2749 \\
        \makecell[l]{Figs.~\!\ref{fig:b03}(c)-\ref{fig:b03}(d)} & $\pi$/5  & 0.0025   & 3.5143   & 1.8850   &  5.4201  &  2.7867  & 4.2990   &5.8499 \\
        \makecell[l]{Figs.~\!\ref{fig:b03}(e)-\ref{fig:b03}(f)} & $\pi$/7  & 0.0036   & 3.4603   & 1.9897   &  5.4838  &  2.8304  & 4.5014   &0.1848 \\
        \makecell[l]{Figs.~\!\ref{fig:b03}(g)-\ref{fig:b03}(h)} & $\pi$/10 & 0.0014   & 3.4078   & 0.7330   &  4.0386  &  2.8401  & 3.3069   &5.4165 \\
		\hline
		\hline
	\end{tabular}
\end{table}

\end{appendix}

\end{widetext}

\bibliographystyle{apsrev4-1}
\bibliography{references}

\begin{thebibliography}{82}%
\makeatletter
\providecommand \@ifxundefined [1]{%
 \@ifx{#1\undefined}
}%
\providecommand \@ifnum [1]{%
 \ifnum #1\expandafter \@firstoftwo
 \else \expandafter \@secondoftwo
 \fi
}%
\providecommand \@ifx [1]{%
 \ifx #1\expandafter \@firstoftwo
 \else \expandafter \@secondoftwo
 \fi
}%
\providecommand \natexlab [1]{#1}%
\providecommand \enquote  [1]{``#1''}%
\providecommand \bibnamefont  [1]{#1}%
\providecommand \bibfnamefont [1]{#1}%
\providecommand \citenamefont [1]{#1}%
\providecommand \href@noop [0]{\@secondoftwo}%
\providecommand \href [0]{\begingroup \@sanitize@url \@href}%
\providecommand \@href[1]{\@@startlink{#1}\@@href}%
\providecommand \@@href[1]{\endgroup#1\@@endlink}%
\providecommand \@sanitize@url [0]{\catcode `\\12\catcode `\$12\catcode
  `\&12\catcode `\#12\catcode `\^12\catcode `\_12\catcode `\%12\relax}%
\providecommand \@@startlink[1]{}%
\providecommand \@@endlink[0]{}%
\providecommand \url  [0]{\begingroup\@sanitize@url \@url }%
\providecommand \@url [1]{\endgroup\@href {#1}{\urlprefix }}%
\providecommand \urlprefix  [0]{URL }%
\providecommand \Eprint [0]{\href }%
\providecommand \doibase [0]{http://dx.doi.org/}%
\providecommand \selectlanguage [0]{\@gobble}%
\providecommand \bibinfo  [0]{\@secondoftwo}%
\providecommand \bibfield  [0]{\@secondoftwo}%
\providecommand \translation [1]{[#1]}%
\providecommand \BibitemOpen [0]{}%
\providecommand \bibitemStop [0]{}%
\providecommand \bibitemNoStop [0]{.\EOS\space}%
\providecommand \EOS [0]{\spacefactor3000\relax}%
\providecommand \BibitemShut  [1]{\csname bibitem#1\endcsname}%
\let\auto@bib@innerbib\@empty
\bibitem [{\citenamefont {Sali\`eres}\ \emph {et~al.}(1995)\citenamefont
  {Sali\`eres}, \citenamefont {L'Huillier},\ and\ \citenamefont
  {Lewenstein}}]{PhysRevLett.74.3776}%
  \BibitemOpen
  \bibfield  {author} {\bibinfo {author} {\bibfnamefont {P.}~\bibnamefont
  {Sali\`eres}}, \bibinfo {author} {\bibfnamefont {A.}~\bibnamefont
  {L'Huillier}}, \ and\ \bibinfo {author} {\bibfnamefont {M.}~\bibnamefont
  {Lewenstein}},\ }\bibinfo {title} {Coherence Control of High-Order
  Harmonics},\ \href {\doibase 10.1103/PhysRevLett.74.3776} {\bibfield
  {journal} {\bibinfo  {journal} {Phys. Rev. Lett.}\ }\textbf {\bibinfo
  {volume} {74}},\ \bibinfo {pages} {3776} (\bibinfo {year}
  {1995})}\BibitemShut {NoStop}%
\bibitem [{\citenamefont {Bergmann}\ \emph {et~al.}(1998)\citenamefont
  {Bergmann}, \citenamefont {Theuer},\ and\ \citenamefont
  {Shore}}]{RevModPhys.70.1003}%
  \BibitemOpen
  \bibfield  {author} {\bibinfo {author} {\bibfnamefont {K.}~\bibnamefont
  {Bergmann}}, \bibinfo {author} {\bibfnamefont {H.}~\bibnamefont {Theuer}}, \
  and\ \bibinfo {author} {\bibfnamefont {B.~W.}\ \bibnamefont {Shore}},\
  }\bibinfo {title} {Coherent population transfer among quantum states of atoms
  and molecules},\ \href {\doibase 10.1103/RevModPhys.70.1003} {\bibfield
  {journal} {\bibinfo  {journal} {Rev. Mod. Phys.}\ }\textbf {\bibinfo {volume}
  {70}},\ \bibinfo {pages} {1003} (\bibinfo {year} {1998})}\BibitemShut
  {NoStop}%
\bibitem [{\citenamefont {Makhlin}\ \emph {et~al.}(2001)\citenamefont
  {Makhlin}, \citenamefont {Sch\"on},\ and\ \citenamefont
  {Shnirman}}]{RevModPhys.73.357}%
  \BibitemOpen
  \bibfield  {author} {\bibinfo {author} {\bibfnamefont {Y.}~\bibnamefont
  {Makhlin}}, \bibinfo {author} {\bibfnamefont {G.}~\bibnamefont {Sch\"on}}, \
  and\ \bibinfo {author} {\bibfnamefont {A.}~\bibnamefont {Shnirman}},\
  }\bibinfo {title} {Quantum-state engineering with Josephson-junction
  devices},\ \href {\doibase 10.1103/RevModPhys.73.357} {\bibfield  {journal}
  {\bibinfo  {journal} {Rev. Mod. Phys.}\ }\textbf {\bibinfo {volume} {73}},\
  \bibinfo {pages} {357} (\bibinfo {year} {2001})}\BibitemShut {NoStop}%
\bibitem [{\citenamefont {Zhao}\ \emph {et~al.}(2006)\citenamefont {Zhao},
  \citenamefont {Loren}, \citenamefont {van Driel},\ and\ \citenamefont
  {Smirl}}]{PhysRevLett.96.246601}%
  \BibitemOpen
  \bibfield  {author} {\bibinfo {author} {\bibfnamefont {H.}~\bibnamefont
  {Zhao}}, \bibinfo {author} {\bibfnamefont {E.~J.}\ \bibnamefont {Loren}},
  \bibinfo {author} {\bibfnamefont {H.~M.}\ \bibnamefont {van Driel}}, \ and\
  \bibinfo {author} {\bibfnamefont {A.~L.}\ \bibnamefont {Smirl}},\ }\bibinfo
  {title} {Coherence Control of Hall Charge and Spin Currents},\ \href
  {\doibase 10.1103/PhysRevLett.96.246601} {\bibfield  {journal} {\bibinfo
  {journal} {Phys. Rev. Lett.}\ }\textbf {\bibinfo {volume} {96}},\ \bibinfo
  {pages} {246601} (\bibinfo {year} {2006})}\BibitemShut {NoStop}%
\bibitem [{\citenamefont {McCullough}\ \emph {et~al.}(2000)\citenamefont
  {McCullough}, \citenamefont {Shapiro},\ and\ \citenamefont
  {Brumer}}]{PhysRevA.61.041801}%
  \BibitemOpen
  \bibfield  {author} {\bibinfo {author} {\bibfnamefont {E.}~\bibnamefont
  {McCullough}}, \bibinfo {author} {\bibfnamefont {M.}~\bibnamefont {Shapiro}},
  \ and\ \bibinfo {author} {\bibfnamefont {P.}~\bibnamefont {Brumer}},\
  }\bibinfo {title} {Coherent control of refractive indices},\ \href {\doibase
  10.1103/PhysRevA.61.041801} {\bibfield  {journal} {\bibinfo  {journal} {Phys.
  Rev. A}\ }\textbf {\bibinfo {volume} {61}},\ \bibinfo {pages} {041801}
  (\bibinfo {year} {2000})}\BibitemShut {NoStop}%
\bibitem [{\citenamefont {Gordon}\ \emph {et~al.}(2002)\citenamefont {Gordon},
  \citenamefont {Heberle}, \citenamefont {Ramsay},\ and\ \citenamefont
  {Cleaver}}]{PhysRevA.65.051803}%
  \BibitemOpen
  \bibfield  {author} {\bibinfo {author} {\bibfnamefont {R.}~\bibnamefont
  {Gordon}}, \bibinfo {author} {\bibfnamefont {A.~P.}\ \bibnamefont {Heberle}},
  \bibinfo {author} {\bibfnamefont {A.~J.}\ \bibnamefont {Ramsay}}, \ and\
  \bibinfo {author} {\bibfnamefont {J.~R.~A.}\ \bibnamefont {Cleaver}},\
  }\bibinfo {title} {Experimental coherent control of lasers},\ \href {\doibase
  10.1103/PhysRevA.65.051803} {\bibfield  {journal} {\bibinfo  {journal} {Phys.
  Rev. A}\ }\textbf {\bibinfo {volume} {65}},\ \bibinfo {pages} {051803}
  (\bibinfo {year} {2002})}\BibitemShut {NoStop}%
\bibitem [{\citenamefont {Zeng}\ \emph {et~al.}(2020)\citenamefont {Zeng},
  \citenamefont {Li}, \citenamefont {Yang}, \citenamefont {Xiao},\ and\
  \citenamefont {Luo}}]{PhysRevA.102.012221}%
  \BibitemOpen
  \bibfield  {author} {\bibinfo {author} {\bibfnamefont {Z.-Y.}\ \bibnamefont
  {Zeng}}, \bibinfo {author} {\bibfnamefont {L.}~\bibnamefont {Li}}, \bibinfo
  {author} {\bibfnamefont {B.}~\bibnamefont {Yang}}, \bibinfo {author}
  {\bibfnamefont {J.}~\bibnamefont {Xiao}}, \ and\ \bibinfo {author}
  {\bibfnamefont {X.}~\bibnamefont {Luo}},\ }\bibinfo {title} {Coherent control
  of dissipative dynamics in a periodically driven lattice array},\ \href
  {\doibase 10.1103/PhysRevA.102.012221} {\bibfield  {journal} {\bibinfo
  {journal} {Phys. Rev. A}\ }\textbf {\bibinfo {volume} {102}},\ \bibinfo
  {pages} {012221} (\bibinfo {year} {2020})}\BibitemShut {NoStop}%
\bibitem [{\citenamefont {Devolder}\ \emph {et~al.}(2021)\citenamefont
  {Devolder}, \citenamefont {Brumer},\ and\ \citenamefont
  {Tscherbul}}]{PhysRevLett.126.153403}%
  \BibitemOpen
  \bibfield  {author} {\bibinfo {author} {\bibfnamefont {A.}~\bibnamefont
  {Devolder}}, \bibinfo {author} {\bibfnamefont {P.}~\bibnamefont {Brumer}}, \
  and\ \bibinfo {author} {\bibfnamefont {T.~V.}\ \bibnamefont {Tscherbul}},\
  }\bibinfo {title} {Complete Quantum Coherent Control of Ultracold Molecular
  Collisions},\ \href {\doibase 10.1103/PhysRevLett.126.153403} {\bibfield
  {journal} {\bibinfo  {journal} {Phys. Rev. Lett.}\ }\textbf {\bibinfo
  {volume} {126}},\ \bibinfo {pages} {153403} (\bibinfo {year}
  {2021})}\BibitemShut {NoStop}%
\bibitem [{\citenamefont {Levine}\ \emph {et~al.}(2018)\citenamefont {Levine},
  \citenamefont {Keesling}, \citenamefont {Omran}, \citenamefont {Bernien},
  \citenamefont {Schwartz}, \citenamefont {Zibrov}, \citenamefont {Endres},
  \citenamefont {Greiner}, \citenamefont {Vuleti\ifmmode~\acute{c}\else
  \'{c}\fi{}},\ and\ \citenamefont {Lukin}}]{PhysRevLett.121.123603}%
  \BibitemOpen
  \bibfield  {author} {\bibinfo {author} {\bibfnamefont {H.}~\bibnamefont
  {Levine}}, \bibinfo {author} {\bibfnamefont {A.}~\bibnamefont {Keesling}},
  \bibinfo {author} {\bibfnamefont {A.}~\bibnamefont {Omran}}, \bibinfo
  {author} {\bibfnamefont {H.}~\bibnamefont {Bernien}}, \bibinfo {author}
  {\bibfnamefont {S.}~\bibnamefont {Schwartz}}, \bibinfo {author}
  {\bibfnamefont {A.~S.}\ \bibnamefont {Zibrov}}, \bibinfo {author}
  {\bibfnamefont {M.}~\bibnamefont {Endres}}, \bibinfo {author} {\bibfnamefont
  {M.}~\bibnamefont {Greiner}}, \bibinfo {author} {\bibfnamefont
  {V.}~\bibnamefont {Vuleti\ifmmode~\acute{c}\else \'{c}\fi{}}}, \ and\
  \bibinfo {author} {\bibfnamefont {M.~D.}\ \bibnamefont {Lukin}},\ }\bibinfo
  {title} {High-Fidelity Control and Entanglement of Rydberg-Atom Qubits},\
  \href {\doibase 10.1103/PhysRevLett.121.123603} {\bibfield  {journal}
  {\bibinfo  {journal} {Phys. Rev. Lett.}\ }\textbf {\bibinfo {volume} {121}},\
  \bibinfo {pages} {123603} (\bibinfo {year} {2018})}\BibitemShut {NoStop}%
\bibitem [{\citenamefont {Daems}\ \emph {et~al.}(2013)\citenamefont {Daems},
  \citenamefont {Ruschhaupt}, \citenamefont {Sugny},\ and\ \citenamefont
  {Gu\'erin}}]{PhysRevLett.111.050404}%
  \BibitemOpen
  \bibfield  {author} {\bibinfo {author} {\bibfnamefont {D.}~\bibnamefont
  {Daems}}, \bibinfo {author} {\bibfnamefont {A.}~\bibnamefont {Ruschhaupt}},
  \bibinfo {author} {\bibfnamefont {D.}~\bibnamefont {Sugny}}, \ and\ \bibinfo
  {author} {\bibfnamefont {S.}~\bibnamefont {Gu\'erin}},\ }\bibinfo {title}
  {Robust Quantum Control by a Single-Shot Shaped Pulse},\ \href {\doibase
  10.1103/PhysRevLett.111.050404} {\bibfield  {journal} {\bibinfo  {journal}
  {Phys. Rev. Lett.}\ }\textbf {\bibinfo {volume} {111}},\ \bibinfo {pages}
  {050404} (\bibinfo {year} {2013})}\BibitemShut {NoStop}%
\bibitem [{\citenamefont {Allen}\ and\ \citenamefont
  {Eberly}(1975)}]{osti7365050}%
  \BibitemOpen
  \bibfield  {author} {\bibinfo {author} {\bibfnamefont {L.}~\bibnamefont
  {Allen}}\ and\ \bibinfo {author} {\bibfnamefont {J.~H.}\ \bibnamefont
  {Eberly}},\ }\href {https://www.osti.gov/biblio/7365050} {\emph {\bibinfo
  {title} {Optical resonance and two-level atoms}}}\ (\bibinfo  {publisher}
  {New York: Dover Publications},\ \bibinfo {year} {1975})\BibitemShut
  {NoStop}%
\bibitem [{\citenamefont {Vitanov}\ \emph {et~al.}(2001)\citenamefont
  {Vitanov}, \citenamefont {Halfmann}, \citenamefont {Shore},\ and\
  \citenamefont {Bergmann}}]{physchem.52.1.763}%
  \BibitemOpen
  \bibfield  {author} {\bibinfo {author} {\bibfnamefont {N.~V.}\ \bibnamefont
  {Vitanov}}, \bibinfo {author} {\bibfnamefont {T.}~\bibnamefont {Halfmann}},
  \bibinfo {author} {\bibfnamefont {B.~W.}\ \bibnamefont {Shore}}, \ and\
  \bibinfo {author} {\bibfnamefont {K.}~\bibnamefont {Bergmann}},\ }\bibinfo
  {title} {Laser-induced population transfer by adiabatic passage techniques},\
  \href {\doibase 10.1146/annurev.physchem.52.1.763} {\bibfield  {journal}
  {\bibinfo  {journal} {Annu. Rev. Phys. Chem.}\ }\textbf {\bibinfo {volume}
  {52}},\ \bibinfo {pages} {763} (\bibinfo {year} {2001})}\BibitemShut
  {NoStop}%
\bibitem [{\citenamefont {Gu\'ery-Odelin}\ \emph {et~al.}(2019)\citenamefont
  {Gu\'ery-Odelin}, \citenamefont {Ruschhaupt}, \citenamefont {Kiely},
  \citenamefont {Torrontegui}, \citenamefont {Mart\'{\i}nez-Garaot},\ and\
  \citenamefont {Muga}}]{RevModPhys.91.045001}%
  \BibitemOpen
  \bibfield  {author} {\bibinfo {author} {\bibfnamefont {D.}~\bibnamefont
  {Gu\'ery-Odelin}}, \bibinfo {author} {\bibfnamefont {A.}~\bibnamefont
  {Ruschhaupt}}, \bibinfo {author} {\bibfnamefont {A.}~\bibnamefont {Kiely}},
  \bibinfo {author} {\bibfnamefont {E.}~\bibnamefont {Torrontegui}}, \bibinfo
  {author} {\bibfnamefont {S.}~\bibnamefont {Mart\'{\i}nez-Garaot}}, \ and\
  \bibinfo {author} {\bibfnamefont {J.~G.}\ \bibnamefont {Muga}},\ }\bibinfo
  {title} {Shortcuts to adiabaticity: Concepts, methods, and applications},\
  \href {\doibase 10.1103/RevModPhys.91.045001} {\bibfield  {journal} {\bibinfo
   {journal} {Rev. Mod. Phys.}\ }\textbf {\bibinfo {volume} {91}},\ \bibinfo
  {pages} {045001} (\bibinfo {year} {2019})}\BibitemShut {NoStop}%
\bibitem [{\citenamefont {Scully}\ and\ \citenamefont
  {Zubairy}(1997)}]{scully97}%
  \BibitemOpen
  \bibfield  {author} {\bibinfo {author} {\bibfnamefont {M.~O.}\ \bibnamefont
  {Scully}}\ and\ \bibinfo {author} {\bibfnamefont {M.~S.}\ \bibnamefont
  {Zubairy}},\ }\href {\doibase 10.1017/cbo9780511813993} {\emph {\bibinfo
  {title} {Quantum Optics}}}\ (\bibinfo  {publisher} {Cambridge University
  Press},\ \bibinfo {year} {1997})\BibitemShut {NoStop}%
\bibitem [{\citenamefont {Gerry}\ and\ \citenamefont
  {Knight}(2004)}]{Gerry2004}%
  \BibitemOpen
  \bibfield  {author} {\bibinfo {author} {\bibfnamefont {C.}~\bibnamefont
  {Gerry}}\ and\ \bibinfo {author} {\bibfnamefont {P.}~\bibnamefont {Knight}},\
  }\href {\doibase 10.1017/CBO9780511791239} {\emph {\bibinfo {title}
  {Introductory Quantum Optics}}}\ (\bibinfo  {publisher} {Cambridge University
  Press},\ \bibinfo {year} {2004})\BibitemShut {NoStop}%
\bibitem [{\citenamefont {Shankar}(1994)}]{Shankar1994}%
  \BibitemOpen
  \bibfield  {author} {\bibinfo {author} {\bibfnamefont {R.}~\bibnamefont
  {Shankar}},\ }\href {\doibase 10.1007/978-1-4757-0576-8} {\emph {\bibinfo
  {title} {Principles of Quantum Mechanics}}}\ (\bibinfo  {publisher} {Springer
  {US}},\ \bibinfo {year} {1994})\BibitemShut {NoStop}%
\bibitem [{\citenamefont {Kr\'al}\ \emph {et~al.}(2007)\citenamefont {Kr\'al},
  \citenamefont {Thanopulos},\ and\ \citenamefont
  {Shapiro}}]{RevModPhys.79.53}%
  \BibitemOpen
  \bibfield  {author} {\bibinfo {author} {\bibfnamefont {P.}~\bibnamefont
  {Kr\'al}}, \bibinfo {author} {\bibfnamefont {I.}~\bibnamefont {Thanopulos}},
  \ and\ \bibinfo {author} {\bibfnamefont {M.}~\bibnamefont {Shapiro}},\
  }\bibinfo {title} {Colloquium: Coherently controlled adiabatic passage},\
  \href {\doibase 10.1103/RevModPhys.79.53} {\bibfield  {journal} {\bibinfo
  {journal} {Rev. Mod. Phys.}\ }\textbf {\bibinfo {volume} {79}},\ \bibinfo
  {pages} {53} (\bibinfo {year} {2007})}\BibitemShut {NoStop}%
\bibitem [{\citenamefont {Wimperis}(1991)}]{Wimperis1991}%
  \BibitemOpen
  \bibfield  {author} {\bibinfo {author} {\bibfnamefont {S.}~\bibnamefont
  {Wimperis}},\ }\bibinfo {title} {Iterative schemes for phase-distortionless
  composite 180{\textdegree} pulses},\ \href {\doibase
  10.1016/0022-2364(91)90043-s} {\bibfield  {journal} {\bibinfo  {journal} {J.
  Magn. Reson.}\ }\textbf {\bibinfo {volume} {93}},\ \bibinfo {pages} {199}
  (\bibinfo {year} {1991})}\BibitemShut {NoStop}%
\bibitem [{\citenamefont {Wimperis}(1994)}]{Wimperis1994}%
  \BibitemOpen
  \bibfield  {author} {\bibinfo {author} {\bibfnamefont {S.}~\bibnamefont
  {Wimperis}},\ }\bibinfo {title} {Broadband, Narrowband, and Passband
  Composite Pulses for Use in Advanced NMR Experiments},\ \href {\doibase
  https://doi.org/10.1006/jmra.1994.1159} {\bibfield  {journal} {\bibinfo
  {journal} {J. Magn. Reson.}\ }\textbf {\bibinfo {volume} {109}},\ \bibinfo
  {pages} {221} (\bibinfo {year} {1994})}\BibitemShut {NoStop}%
\bibitem [{\citenamefont {Levitt}\ and\ \citenamefont
  {Freeman}(1979)}]{Levitt1979}%
  \BibitemOpen
  \bibfield  {author} {\bibinfo {author} {\bibfnamefont {M.~H.}\ \bibnamefont
  {Levitt}}\ and\ \bibinfo {author} {\bibfnamefont {R.}~\bibnamefont
  {Freeman}},\ }\bibinfo {title} {{NMR} population inversion using a composite
  pulse},\ \href {\doibase 10.1016/0022-2364(79)90265-8} {\bibfield  {journal}
  {\bibinfo  {journal} {J. Magn. Reson.}\ }\textbf {\bibinfo {volume} {33}},\
  \bibinfo {pages} {473} (\bibinfo {year} {1979})}\BibitemShut {NoStop}%
\bibitem [{\citenamefont {Levitt}(1986)}]{Levitt1986}%
  \BibitemOpen
  \bibfield  {author} {\bibinfo {author} {\bibfnamefont {M.~H.}\ \bibnamefont
  {Levitt}},\ }\bibinfo {title} {Composite pulses},\ \href {\doibase
  10.1016/0079-6565(86)80005-x} {\bibfield  {journal} {\bibinfo  {journal}
  {Prog. NMR Spectrosc.}\ }\textbf {\bibinfo {volume} {18}},\ \bibinfo {pages}
  {61} (\bibinfo {year} {1986})}\BibitemShut {NoStop}%
\bibitem [{\citenamefont {Brown}\ \emph {et~al.}(2004)\citenamefont {Brown},
  \citenamefont {Harrow},\ and\ \citenamefont {Chuang}}]{PhysRevA.70.052318}%
  \BibitemOpen
  \bibfield  {author} {\bibinfo {author} {\bibfnamefont {K.~R.}\ \bibnamefont
  {Brown}}, \bibinfo {author} {\bibfnamefont {A.~W.}\ \bibnamefont {Harrow}}, \
  and\ \bibinfo {author} {\bibfnamefont {I.~L.}\ \bibnamefont {Chuang}},\
  }\bibinfo {title} {Arbitrarily accurate composite pulse sequences},\ \href
  {\doibase 10.1103/PhysRevA.70.052318} {\bibfield  {journal} {\bibinfo
  {journal} {Phys. Rev. A}\ }\textbf {\bibinfo {volume} {70}},\ \bibinfo
  {pages} {052318} (\bibinfo {year} {2004})}\BibitemShut {NoStop}%
\bibitem [{\citenamefont {Dridi}\ \emph {et~al.}(2020)\citenamefont {Dridi},
  \citenamefont {Mejatty}, \citenamefont {Glaser},\ and\ \citenamefont
  {Sugny}}]{PhysRevA.101.012321}%
  \BibitemOpen
  \bibfield  {author} {\bibinfo {author} {\bibfnamefont {G.}~\bibnamefont
  {Dridi}}, \bibinfo {author} {\bibfnamefont {M.}~\bibnamefont {Mejatty}},
  \bibinfo {author} {\bibfnamefont {S.~J.}\ \bibnamefont {Glaser}}, \ and\
  \bibinfo {author} {\bibfnamefont {D.}~\bibnamefont {Sugny}},\ }\bibinfo
  {title} {Robust control of a {NOT} gate by composite pulses},\ \href
  {\doibase 10.1103/PhysRevA.101.012321} {\bibfield  {journal} {\bibinfo
  {journal} {Phys. Rev. A}\ }\textbf {\bibinfo {volume} {101}},\ \bibinfo
  {pages} {012321} (\bibinfo {year} {2020})}\BibitemShut {NoStop}%
\bibitem [{\citenamefont {Torosov}\ and\ \citenamefont
  {Vitanov}(2011)}]{PhysRevA.83.053420}%
  \BibitemOpen
  \bibfield  {author} {\bibinfo {author} {\bibfnamefont {B.~T.}\ \bibnamefont
  {Torosov}}\ and\ \bibinfo {author} {\bibfnamefont {N.~V.}\ \bibnamefont
  {Vitanov}},\ }\bibinfo {title} {Smooth composite pulses for high-fidelity
  quantum information processing},\ \href {\doibase 10.1103/PhysRevA.83.053420}
  {\bibfield  {journal} {\bibinfo  {journal} {Phys. Rev. A}\ }\textbf {\bibinfo
  {volume} {83}},\ \bibinfo {pages} {053420} (\bibinfo {year}
  {2011})}\BibitemShut {NoStop}%
\bibitem [{\citenamefont {Genov}\ \emph {et~al.}(2014)\citenamefont {Genov},
  \citenamefont {Schraft}, \citenamefont {Halfmann},\ and\ \citenamefont
  {Vitanov}}]{PhysRevLett.113.043001}%
  \BibitemOpen
  \bibfield  {author} {\bibinfo {author} {\bibfnamefont {G.~T.}\ \bibnamefont
  {Genov}}, \bibinfo {author} {\bibfnamefont {D.}~\bibnamefont {Schraft}},
  \bibinfo {author} {\bibfnamefont {T.}~\bibnamefont {Halfmann}}, \ and\
  \bibinfo {author} {\bibfnamefont {N.~V.}\ \bibnamefont {Vitanov}},\ }\bibinfo
  {title} {Correction of Arbitrary Field Errors in Population Inversion of
  Quantum Systems by Universal Composite Pulses},\ \href {\doibase
  10.1103/PhysRevLett.113.043001} {\bibfield  {journal} {\bibinfo  {journal}
  {Phys. Rev. Lett.}\ }\textbf {\bibinfo {volume} {113}},\ \bibinfo {pages}
  {043001} (\bibinfo {year} {2014})}\BibitemShut {NoStop}%
\bibitem [{\citenamefont {Torosov}\ and\ \citenamefont
  {Vitanov}(2018)}]{PhysRevA.97.043408}%
  \BibitemOpen
  \bibfield  {author} {\bibinfo {author} {\bibfnamefont {B.~T.}\ \bibnamefont
  {Torosov}}\ and\ \bibinfo {author} {\bibfnamefont {N.~V.}\ \bibnamefont
  {Vitanov}},\ }\bibinfo {title} {Arbitrarily accurate twin composite
  $\ensuremath{\pi}$-pulse sequences},\ \href {\doibase
  10.1103/PhysRevA.97.043408} {\bibfield  {journal} {\bibinfo  {journal} {Phys.
  Rev. A}\ }\textbf {\bibinfo {volume} {97}},\ \bibinfo {pages} {043408}
  (\bibinfo {year} {2018})}\BibitemShut {NoStop}%
\bibitem [{\citenamefont {Jones}(2013)}]{PhysRevA.87.052317}%
  \BibitemOpen
  \bibfield  {author} {\bibinfo {author} {\bibfnamefont {J.~A.}\ \bibnamefont
  {Jones}},\ }\bibinfo {title} {Designing short robust not gates for quantum
  computation},\ \href {\doibase 10.1103/PhysRevA.87.052317} {\bibfield
  {journal} {\bibinfo  {journal} {Phys. Rev. A}\ }\textbf {\bibinfo {volume}
  {87}},\ \bibinfo {pages} {052317} (\bibinfo {year} {2013})}\BibitemShut
  {NoStop}%
\bibitem [{\citenamefont {Vitanov}(2011)}]{PhysRevA.84.065404}%
  \BibitemOpen
  \bibfield  {author} {\bibinfo {author} {\bibfnamefont {N.~V.}\ \bibnamefont
  {Vitanov}},\ }\bibinfo {title} {Arbitrarily accurate narrowband composite
  pulse sequences},\ \href {\doibase 10.1103/PhysRevA.84.065404} {\bibfield
  {journal} {\bibinfo  {journal} {Phys. Rev. A}\ }\textbf {\bibinfo {volume}
  {84}},\ \bibinfo {pages} {065404} (\bibinfo {year} {2011})}\BibitemShut
  {NoStop}%
\bibitem [{\citenamefont {Torosov}\ and\ \citenamefont
  {Vitanov}(2019{\natexlab{a}})}]{PhysRevA.99.013424}%
  \BibitemOpen
  \bibfield  {author} {\bibinfo {author} {\bibfnamefont {B.~T.}\ \bibnamefont
  {Torosov}}\ and\ \bibinfo {author} {\bibfnamefont {N.~V.}\ \bibnamefont
  {Vitanov}},\ }\bibinfo {title} {Robust high-fidelity coherent control of
  two-state systems by detuning pulses},\ \href {\doibase
  10.1103/PhysRevA.99.013424} {\bibfield  {journal} {\bibinfo  {journal} {Phys.
  Rev. A}\ }\textbf {\bibinfo {volume} {99}},\ \bibinfo {pages} {013424}
  (\bibinfo {year} {2019}{\natexlab{a}})}\BibitemShut {NoStop}%
\bibitem [{\citenamefont {Torosov}\ \emph
  {et~al.}(2020{\natexlab{a}})\citenamefont {Torosov}, \citenamefont {Ivanov},\
  and\ \citenamefont {Vitanov}}]{PhysRevA.102.013105}%
  \BibitemOpen
  \bibfield  {author} {\bibinfo {author} {\bibfnamefont {B.~T.}\ \bibnamefont
  {Torosov}}, \bibinfo {author} {\bibfnamefont {S.~S.}\ \bibnamefont {Ivanov}},
  \ and\ \bibinfo {author} {\bibfnamefont {N.~V.}\ \bibnamefont {Vitanov}},\
  }\bibinfo {title} {Narrowband and passband composite pulses for variable
  rotations},\ \href {\doibase 10.1103/PhysRevA.102.013105} {\bibfield
  {journal} {\bibinfo  {journal} {Phys. Rev. A}\ }\textbf {\bibinfo {volume}
  {102}},\ \bibinfo {pages} {013105} (\bibinfo {year}
  {2020}{\natexlab{a}})}\BibitemShut {NoStop}%
\bibitem [{\citenamefont {Genov}\ \emph {et~al.}(2020)\citenamefont {Genov},
  \citenamefont {Hain}, \citenamefont {Vitanov},\ and\ \citenamefont
  {Halfmann}}]{PhysRevA.101.013827}%
  \BibitemOpen
  \bibfield  {author} {\bibinfo {author} {\bibfnamefont {G.~T.}\ \bibnamefont
  {Genov}}, \bibinfo {author} {\bibfnamefont {M.}~\bibnamefont {Hain}},
  \bibinfo {author} {\bibfnamefont {N.~V.}\ \bibnamefont {Vitanov}}, \ and\
  \bibinfo {author} {\bibfnamefont {T.}~\bibnamefont {Halfmann}},\ }\bibinfo
  {title} {Universal composite pulses for efficient population inversion with
  an arbitrary excitation profile},\ \href {\doibase
  10.1103/PhysRevA.101.013827} {\bibfield  {journal} {\bibinfo  {journal}
  {Phys. Rev. A}\ }\textbf {\bibinfo {volume} {101}},\ \bibinfo {pages}
  {013827} (\bibinfo {year} {2020})}\BibitemShut {NoStop}%
\bibitem [{\citenamefont {Torosov}\ \emph {et~al.}(2015)\citenamefont
  {Torosov}, \citenamefont {Kyoseva},\ and\ \citenamefont
  {Vitanov}}]{PhysRevA.92.033406}%
  \BibitemOpen
  \bibfield  {author} {\bibinfo {author} {\bibfnamefont {B.~T.}\ \bibnamefont
  {Torosov}}, \bibinfo {author} {\bibfnamefont {E.~S.}\ \bibnamefont
  {Kyoseva}}, \ and\ \bibinfo {author} {\bibfnamefont {N.~V.}\ \bibnamefont
  {Vitanov}},\ }\bibinfo {title} {Composite pulses for ultrabroad-band and
  ultranarrow-band excitation},\ \href {\doibase 10.1103/PhysRevA.92.033406}
  {\bibfield  {journal} {\bibinfo  {journal} {Phys. Rev. A}\ }\textbf {\bibinfo
  {volume} {92}},\ \bibinfo {pages} {033406} (\bibinfo {year}
  {2015})}\BibitemShut {NoStop}%
\bibitem [{\citenamefont {Ichikawa}\ \emph {et~al.}(2011)\citenamefont
  {Ichikawa}, \citenamefont {Bando}, \citenamefont {Kondo},\ and\ \citenamefont
  {Nakahara}}]{PhysRevA.84.062311}%
  \BibitemOpen
  \bibfield  {author} {\bibinfo {author} {\bibfnamefont {T.}~\bibnamefont
  {Ichikawa}}, \bibinfo {author} {\bibfnamefont {M.}~\bibnamefont {Bando}},
  \bibinfo {author} {\bibfnamefont {Y.}~\bibnamefont {Kondo}}, \ and\ \bibinfo
  {author} {\bibfnamefont {M.}~\bibnamefont {Nakahara}},\ }\bibinfo {title}
  {Designing robust unitary gates: Application to concatenated composite
  pulses},\ \href {\doibase 10.1103/PhysRevA.84.062311} {\bibfield  {journal}
  {\bibinfo  {journal} {Phys. Rev. A}\ }\textbf {\bibinfo {volume} {84}},\
  \bibinfo {pages} {062311} (\bibinfo {year} {2011})}\BibitemShut {NoStop}%
\bibitem [{\citenamefont {Cohen}\ \emph {et~al.}(2016)\citenamefont {Cohen},
  \citenamefont {Rotem},\ and\ \citenamefont {Retzker}}]{PhysRevA.93.032340}%
  \BibitemOpen
  \bibfield  {author} {\bibinfo {author} {\bibfnamefont {I.}~\bibnamefont
  {Cohen}}, \bibinfo {author} {\bibfnamefont {A.}~\bibnamefont {Rotem}}, \ and\
  \bibinfo {author} {\bibfnamefont {A.}~\bibnamefont {Retzker}},\ }\bibinfo
  {title} {Refocusing two-qubit-gate noise for trapped ions by composite
  pulses},\ \href {\doibase 10.1103/PhysRevA.93.032340} {\bibfield  {journal}
  {\bibinfo  {journal} {Phys. Rev. A}\ }\textbf {\bibinfo {volume} {93}},\
  \bibinfo {pages} {032340} (\bibinfo {year} {2016})}\BibitemShut {NoStop}%
\bibitem [{\citenamefont {Kyoseva}\ \emph {et~al.}(2019)\citenamefont
  {Kyoseva}, \citenamefont {Greener},\ and\ \citenamefont
  {Suchowski}}]{PhysRevA.100.032333}%
  \BibitemOpen
  \bibfield  {author} {\bibinfo {author} {\bibfnamefont {E.}~\bibnamefont
  {Kyoseva}}, \bibinfo {author} {\bibfnamefont {H.}~\bibnamefont {Greener}}, \
  and\ \bibinfo {author} {\bibfnamefont {H.}~\bibnamefont {Suchowski}},\
  }\bibinfo {title} {Detuning-modulated composite pulses for high-fidelity
  robust quantum control},\ \href {\doibase 10.1103/PhysRevA.100.032333}
  {\bibfield  {journal} {\bibinfo  {journal} {Phys. Rev. A}\ }\textbf {\bibinfo
  {volume} {100}},\ \bibinfo {pages} {032333} (\bibinfo {year}
  {2019})}\BibitemShut {NoStop}%
\bibitem [{\citenamefont {Wang}\ \emph {et~al.}(2014)\citenamefont {Wang},
  \citenamefont {Bishop}, \citenamefont {Barnes}, \citenamefont {Kestner},\
  and\ \citenamefont {Sarma}}]{PhysRevA.89.022310}%
  \BibitemOpen
  \bibfield  {author} {\bibinfo {author} {\bibfnamefont {X.}~\bibnamefont
  {Wang}}, \bibinfo {author} {\bibfnamefont {L.~S.}\ \bibnamefont {Bishop}},
  \bibinfo {author} {\bibfnamefont {E.}~\bibnamefont {Barnes}}, \bibinfo
  {author} {\bibfnamefont {J.~P.}\ \bibnamefont {Kestner}}, \ and\ \bibinfo
  {author} {\bibfnamefont {S.~D.}\ \bibnamefont {Sarma}},\ }\bibinfo {title}
  {Robust quantum gates for singlet-triplet spin qubits using composite
  pulses},\ \href {\doibase 10.1103/PhysRevA.89.022310} {\bibfield  {journal}
  {\bibinfo  {journal} {Phys. Rev. A}\ }\textbf {\bibinfo {volume} {89}},\
  \bibinfo {pages} {022310} (\bibinfo {year} {2014})}\BibitemShut {NoStop}%
\bibitem [{\citenamefont {Mount}\ \emph {et~al.}(2015)\citenamefont {Mount},
  \citenamefont {Kabytayev}, \citenamefont {Crain}, \citenamefont {Harper},
  \citenamefont {Baek}, \citenamefont {Vrijsen}, \citenamefont {Flammia},
  \citenamefont {Brown}, \citenamefont {Maunz},\ and\ \citenamefont
  {Kim}}]{PhysRevA.92.060301}%
  \BibitemOpen
  \bibfield  {author} {\bibinfo {author} {\bibfnamefont {E.}~\bibnamefont
  {Mount}}, \bibinfo {author} {\bibfnamefont {C.}~\bibnamefont {Kabytayev}},
  \bibinfo {author} {\bibfnamefont {S.}~\bibnamefont {Crain}}, \bibinfo
  {author} {\bibfnamefont {R.}~\bibnamefont {Harper}}, \bibinfo {author}
  {\bibfnamefont {S.-Y.}\ \bibnamefont {Baek}}, \bibinfo {author}
  {\bibfnamefont {G.}~\bibnamefont {Vrijsen}}, \bibinfo {author} {\bibfnamefont
  {S.~T.}\ \bibnamefont {Flammia}}, \bibinfo {author} {\bibfnamefont {K.~R.}\
  \bibnamefont {Brown}}, \bibinfo {author} {\bibfnamefont {P.}~\bibnamefont
  {Maunz}}, \ and\ \bibinfo {author} {\bibfnamefont {J.}~\bibnamefont {Kim}},\
  }\bibinfo {title} {Error compensation of single-qubit gates in a
  surface-electrode ion trap using composite pulses},\ \href {\doibase
  10.1103/PhysRevA.92.060301} {\bibfield  {journal} {\bibinfo  {journal} {Phys.
  Rev. A}\ }\textbf {\bibinfo {volume} {92}},\ \bibinfo {pages} {060301}
  (\bibinfo {year} {2015})}\BibitemShut {NoStop}%
\bibitem [{\citenamefont {Demeter}(2016)}]{PhysRevA.93.023830}%
  \BibitemOpen
  \bibfield  {author} {\bibinfo {author} {\bibfnamefont {G.}~\bibnamefont
  {Demeter}},\ }\bibinfo {title} {Composite pulses for high-fidelity population
  inversion in optically dense, inhomogeneously broadened atomic ensembles},\
  \href {\doibase 10.1103/PhysRevA.93.023830} {\bibfield  {journal} {\bibinfo
  {journal} {Phys. Rev. A}\ }\textbf {\bibinfo {volume} {93}},\ \bibinfo
  {pages} {023830} (\bibinfo {year} {2016})}\BibitemShut {NoStop}%
\bibitem [{\citenamefont {Ivanov}\ and\ \citenamefont
  {Vitanov}(2011)}]{Ivanov2011}%
  \BibitemOpen
  \bibfield  {author} {\bibinfo {author} {\bibfnamefont {S.~S.}\ \bibnamefont
  {Ivanov}}\ and\ \bibinfo {author} {\bibfnamefont {N.~V.}\ \bibnamefont
  {Vitanov}},\ }\bibinfo {title} {High-fidelity local addressing of trapped
  ions and atoms by composite sequences of laser pulses},\ \href {\doibase
  10.1364/ol.36.001275} {\bibfield  {journal} {\bibinfo  {journal} {Opt.
  Lett.}\ }\textbf {\bibinfo {volume} {36}},\ \bibinfo {pages} {1275} (\bibinfo
  {year} {2011})}\BibitemShut {NoStop}%
\bibitem [{\citenamefont {Torosov}\ and\ \citenamefont
  {Vitanov}(2019{\natexlab{b}})}]{PhysRevA.99.013402}%
  \BibitemOpen
  \bibfield  {author} {\bibinfo {author} {\bibfnamefont {B.~T.}\ \bibnamefont
  {Torosov}}\ and\ \bibinfo {author} {\bibfnamefont {N.~V.}\ \bibnamefont
  {Vitanov}},\ }\bibinfo {title} {Arbitrarily accurate variable rotations on
  the Bloch sphere by composite pulse sequences},\ \href {\doibase
  10.1103/PhysRevA.99.013402} {\bibfield  {journal} {\bibinfo  {journal} {Phys.
  Rev. A}\ }\textbf {\bibinfo {volume} {99}},\ \bibinfo {pages} {013402}
  (\bibinfo {year} {2019}{\natexlab{b}})}\BibitemShut {NoStop}%
\bibitem [{\citenamefont {Vandersypen}\ and\ \citenamefont
  {Chuang}(2005)}]{RevModPhys.76.1037}%
  \BibitemOpen
  \bibfield  {author} {\bibinfo {author} {\bibfnamefont {L.~M.~K.}\
  \bibnamefont {Vandersypen}}\ and\ \bibinfo {author} {\bibfnamefont {I.~L.}\
  \bibnamefont {Chuang}},\ }\bibinfo {title} {NMR techniques for quantum
  control and computation},\ \href {\doibase 10.1103/RevModPhys.76.1037}
  {\bibfield  {journal} {\bibinfo  {journal} {Rev. Mod. Phys.}\ }\textbf
  {\bibinfo {volume} {76}},\ \bibinfo {pages} {1037} (\bibinfo {year}
  {2005})}\BibitemShut {NoStop}%
\bibitem [{\citenamefont {Fleischhauer}\ \emph {et~al.}(2005)\citenamefont
  {Fleischhauer}, \citenamefont {Imamoglu},\ and\ \citenamefont
  {Marangos}}]{RevModPhys.77.633}%
  \BibitemOpen
  \bibfield  {author} {\bibinfo {author} {\bibfnamefont {M.}~\bibnamefont
  {Fleischhauer}}, \bibinfo {author} {\bibfnamefont {A.}~\bibnamefont
  {Imamoglu}}, \ and\ \bibinfo {author} {\bibfnamefont {J.~P.}\ \bibnamefont
  {Marangos}},\ }\bibinfo {title} {Electromagnetically induced transparency:
  Optics in coherent media},\ \href {\doibase 10.1103/RevModPhys.77.633}
  {\bibfield  {journal} {\bibinfo  {journal} {Rev. Mod. Phys.}\ }\textbf
  {\bibinfo {volume} {77}},\ \bibinfo {pages} {633} (\bibinfo {year}
  {2005})}\BibitemShut {NoStop}%
\bibitem [{\citenamefont {Scully}\ \emph {et~al.}(1989)\citenamefont {Scully},
  \citenamefont {Zhu},\ and\ \citenamefont
  {Gavrielides}}]{PhysRevLett.62.2813}%
  \BibitemOpen
  \bibfield  {author} {\bibinfo {author} {\bibfnamefont {M.~O.}\ \bibnamefont
  {Scully}}, \bibinfo {author} {\bibfnamefont {S.-Y.}\ \bibnamefont {Zhu}}, \
  and\ \bibinfo {author} {\bibfnamefont {A.}~\bibnamefont {Gavrielides}},\
  }\bibinfo {title} {Degenerate quantum-beat laser: Lasing without inversion
  and inversion without lasing},\ \href {\doibase 10.1103/PhysRevLett.62.2813}
  {\bibfield  {journal} {\bibinfo  {journal} {Phys. Rev. Lett.}\ }\textbf
  {\bibinfo {volume} {62}},\ \bibinfo {pages} {2813} (\bibinfo {year}
  {1989})}\BibitemShut {NoStop}%
\bibitem [{\citenamefont {Levitt}\ \emph {et~al.}(1984)\citenamefont {Levitt},
  \citenamefont {Suter},\ and\ \citenamefont {Ernst}}]{Levitt1984}%
  \BibitemOpen
  \bibfield  {author} {\bibinfo {author} {\bibfnamefont {M.~H.}\ \bibnamefont
  {Levitt}}, \bibinfo {author} {\bibfnamefont {D.}~\bibnamefont {Suter}}, \
  and\ \bibinfo {author} {\bibfnamefont {R.~R.}\ \bibnamefont {Ernst}},\
  }\bibinfo {title} {Composite pulse excitation in three-level systems},\ \href
  {\doibase 10.1063/1.447142} {\bibfield  {journal} {\bibinfo  {journal} {J.
  Chem. Phys.}\ }\textbf {\bibinfo {volume} {80}},\ \bibinfo {pages} {3064}
  (\bibinfo {year} {1984})}\BibitemShut {NoStop}%
\bibitem [{\citenamefont {Ramamoorthy}\ and\ \citenamefont
  {Narasimhan}(1991)}]{Ramamoorthy1991}%
  \BibitemOpen
  \bibfield  {author} {\bibinfo {author} {\bibfnamefont {A.}~\bibnamefont
  {Ramamoorthy}}\ and\ \bibinfo {author} {\bibfnamefont {P.}~\bibnamefont
  {Narasimhan}},\ }\bibinfo {title} {Phase-alternated composite $\pi$/2 pulses
  for solid state quadrupole echo {NMR} spectroscopy},\ \href {\doibase
  10.1007/bf02847213} {\bibfield  {journal} {\bibinfo  {journal} {Pramana}\
  }\textbf {\bibinfo {volume} {36}},\ \bibinfo {pages} {399} (\bibinfo {year}
  {1991})}\BibitemShut {NoStop}%
\bibitem [{\citenamefont {Torosov}\ \emph {et~al.}(2011)\citenamefont
  {Torosov}, \citenamefont {Gu\'erin},\ and\ \citenamefont
  {Vitanov}}]{PhysRevLett.106.233001}%
  \BibitemOpen
  \bibfield  {author} {\bibinfo {author} {\bibfnamefont {B.~T.}\ \bibnamefont
  {Torosov}}, \bibinfo {author} {\bibfnamefont {S.}~\bibnamefont {Gu\'erin}}, \
  and\ \bibinfo {author} {\bibfnamefont {N.~V.}\ \bibnamefont {Vitanov}},\
  }\bibinfo {title} {High-Fidelity Adiabatic Passage by Composite Sequences of
  Chirped Pulses},\ \href {\doibase 10.1103/PhysRevLett.106.233001} {\bibfield
  {journal} {\bibinfo  {journal} {Phys. Rev. Lett.}\ }\textbf {\bibinfo
  {volume} {106}},\ \bibinfo {pages} {233001} (\bibinfo {year}
  {2011})}\BibitemShut {NoStop}%
\bibitem [{\citenamefont {Torosov}\ and\ \citenamefont
  {Vitanov}(2013)}]{PhysRevA.87.043418}%
  \BibitemOpen
  \bibfield  {author} {\bibinfo {author} {\bibfnamefont {B.~T.}\ \bibnamefont
  {Torosov}}\ and\ \bibinfo {author} {\bibfnamefont {N.~V.}\ \bibnamefont
  {Vitanov}},\ }\bibinfo {title} {Composite stimulated Raman adiabatic
  passage},\ \href {\doibase 10.1103/PhysRevA.87.043418} {\bibfield  {journal}
  {\bibinfo  {journal} {Phys. Rev. A}\ }\textbf {\bibinfo {volume} {87}},\
  \bibinfo {pages} {043418} (\bibinfo {year} {2013})}\BibitemShut {NoStop}%
\bibitem [{\citenamefont {Schraft}\ \emph {et~al.}(2013)\citenamefont
  {Schraft}, \citenamefont {Halfmann}, \citenamefont {Genov},\ and\
  \citenamefont {Vitanov}}]{PhysRevA.88.063406}%
  \BibitemOpen
  \bibfield  {author} {\bibinfo {author} {\bibfnamefont {D.}~\bibnamefont
  {Schraft}}, \bibinfo {author} {\bibfnamefont {T.}~\bibnamefont {Halfmann}},
  \bibinfo {author} {\bibfnamefont {G.~T.}\ \bibnamefont {Genov}}, \ and\
  \bibinfo {author} {\bibfnamefont {N.~V.}\ \bibnamefont {Vitanov}},\ }\bibinfo
  {title} {Experimental demonstration of composite adiabatic passage},\ \href
  {\doibase 10.1103/PhysRevA.88.063406} {\bibfield  {journal} {\bibinfo
  {journal} {Phys. Rev. A}\ }\textbf {\bibinfo {volume} {88}},\ \bibinfo
  {pages} {063406} (\bibinfo {year} {2013})}\BibitemShut {NoStop}%
\bibitem [{\citenamefont {Ishida}\ \emph {et~al.}(2018)\citenamefont {Ishida},
  \citenamefont {Nakamura}, \citenamefont {Tanaka}, \citenamefont {Mishima},
  \citenamefont {Kano}, \citenamefont {Kuroiwa}, \citenamefont {Sekiguchi},\
  and\ \citenamefont {Kosaka}}]{Ishida2018}%
  \BibitemOpen
  \bibfield  {author} {\bibinfo {author} {\bibfnamefont {N.}~\bibnamefont
  {Ishida}}, \bibinfo {author} {\bibfnamefont {T.}~\bibnamefont {Nakamura}},
  \bibinfo {author} {\bibfnamefont {T.}~\bibnamefont {Tanaka}}, \bibinfo
  {author} {\bibfnamefont {S.}~\bibnamefont {Mishima}}, \bibinfo {author}
  {\bibfnamefont {H.}~\bibnamefont {Kano}}, \bibinfo {author} {\bibfnamefont
  {R.}~\bibnamefont {Kuroiwa}}, \bibinfo {author} {\bibfnamefont
  {Y.}~\bibnamefont {Sekiguchi}}, \ and\ \bibinfo {author} {\bibfnamefont
  {H.}~\bibnamefont {Kosaka}},\ }\bibinfo {title} {Universal holonomic single
  quantum gates over a geometric spin with phase-modulated polarized light},\
  \href {\doibase 10.1364/ol.43.002380} {\bibfield  {journal} {\bibinfo
  {journal} {Opt. Lett.}\ }\textbf {\bibinfo {volume} {43}},\ \bibinfo {pages}
  {2380} (\bibinfo {year} {2018})}\BibitemShut {NoStop}%
\bibitem [{\citenamefont {Torosov}\ \emph
  {et~al.}(2020{\natexlab{b}})\citenamefont {Torosov}, \citenamefont
  {Drewsen},\ and\ \citenamefont {Vitanov}}]{PhysRevResearch.2.043235}%
  \BibitemOpen
  \bibfield  {author} {\bibinfo {author} {\bibfnamefont {B.~T.}\ \bibnamefont
  {Torosov}}, \bibinfo {author} {\bibfnamefont {M.}~\bibnamefont {Drewsen}}, \
  and\ \bibinfo {author} {\bibfnamefont {N.~V.}\ \bibnamefont {Vitanov}},\
  }\bibinfo {title} {Chiral resolution by composite {R}aman pulses},\ \href
  {\doibase 10.1103/PhysRevResearch.2.043235} {\bibfield  {journal} {\bibinfo
  {journal} {Phys. Rev. Research}\ }\textbf {\bibinfo {volume} {2}},\ \bibinfo
  {pages} {043235} (\bibinfo {year} {2020}{\natexlab{b}})}\BibitemShut
  {NoStop}%
\bibitem [{\citenamefont {Lehmann}(2018)}]{Lehmann2018}%
  \BibitemOpen
  \bibfield  {author} {\bibinfo {author} {\bibfnamefont {K.~K.}\ \bibnamefont
  {Lehmann}},\ }\bibinfo {title} {Influence of spatial degeneracy on rotational
  spectroscopy: Three-wave mixing and enantiomeric state separation of chiral
  molecules},\ \href {\doibase 10.1063/1.5045052} {\bibfield  {journal}
  {\bibinfo  {journal} {J. Chem. Phys.}\ }\textbf {\bibinfo {volume} {149}},\
  \bibinfo {pages} {094201} (\bibinfo {year} {2018})}\BibitemShut {NoStop}%
\bibitem [{\citenamefont {Leibscher}\ \emph {et~al.}(2019)\citenamefont
  {Leibscher}, \citenamefont {Giesen},\ and\ \citenamefont
  {Koch}}]{Leibscher2019}%
  \BibitemOpen
  \bibfield  {author} {\bibinfo {author} {\bibfnamefont {M.}~\bibnamefont
  {Leibscher}}, \bibinfo {author} {\bibfnamefont {T.~F.}\ \bibnamefont
  {Giesen}}, \ and\ \bibinfo {author} {\bibfnamefont {C.~P.}\ \bibnamefont
  {Koch}},\ }\bibinfo {title} {Principles of enantio-selective excitation in
  three-wave mixing spectroscopy of chiral molecules},\ \href {\doibase
  10.1063/1.5097406} {\bibfield  {journal} {\bibinfo  {journal} {J. Chem.
  Phys.}\ }\textbf {\bibinfo {volume} {151}},\ \bibinfo {pages} {014302}
  (\bibinfo {year} {2019})}\BibitemShut {NoStop}%
\bibitem [{\citenamefont {Genov}\ \emph {et~al.}(2011)\citenamefont {Genov},
  \citenamefont {Torosov},\ and\ \citenamefont {Vitanov}}]{PhysRevA.84.063413}%
  \BibitemOpen
  \bibfield  {author} {\bibinfo {author} {\bibfnamefont {G.~T.}\ \bibnamefont
  {Genov}}, \bibinfo {author} {\bibfnamefont {B.~T.}\ \bibnamefont {Torosov}},
  \ and\ \bibinfo {author} {\bibfnamefont {N.~V.}\ \bibnamefont {Vitanov}},\
  }\bibinfo {title} {Optimized control of multistate quantum systems by
  composite pulse sequences},\ \href {\doibase 10.1103/PhysRevA.84.063413}
  {\bibfield  {journal} {\bibinfo  {journal} {Phys. Rev. A}\ }\textbf {\bibinfo
  {volume} {84}},\ \bibinfo {pages} {063413} (\bibinfo {year}
  {2011})}\BibitemShut {NoStop}%
\bibitem [{\citenamefont {Torosov}\ and\ \citenamefont
  {Vitanov}(2020)}]{PhysRevResearch.2.043194}%
  \BibitemOpen
  \bibfield  {author} {\bibinfo {author} {\bibfnamefont {B.~T.}\ \bibnamefont
  {Torosov}}\ and\ \bibinfo {author} {\bibfnamefont {N.~V.}\ \bibnamefont
  {Vitanov}},\ }\bibinfo {title} {High-fidelity composite quantum gates for
  {R}aman qubits},\ \href {\doibase 10.1103/PhysRevResearch.2.043194}
  {\bibfield  {journal} {\bibinfo  {journal} {Phys. Rev. Research}\ }\textbf
  {\bibinfo {volume} {2}},\ \bibinfo {pages} {043194} (\bibinfo {year}
  {2020})}\BibitemShut {NoStop}%
\bibitem [{\citenamefont {Morris}\ and\ \citenamefont
  {Shore}(1983)}]{PhysRevA.27.906}%
  \BibitemOpen
  \bibfield  {author} {\bibinfo {author} {\bibfnamefont {J.~R.}\ \bibnamefont
  {Morris}}\ and\ \bibinfo {author} {\bibfnamefont {B.~W.}\ \bibnamefont
  {Shore}},\ }\bibinfo {title} {Reduction of degenerate two-level excitation to
  independent two-state systems},\ \href {\doibase 10.1103/PhysRevA.27.906}
  {\bibfield  {journal} {\bibinfo  {journal} {Phys. Rev. A}\ }\textbf {\bibinfo
  {volume} {27}},\ \bibinfo {pages} {906} (\bibinfo {year} {1983})}\BibitemShut
  {NoStop}%
\bibitem [{\citenamefont {Majorana}(1932)}]{Majorana1932}%
  \BibitemOpen
  \bibfield  {author} {\bibinfo {author} {\bibfnamefont {E.}~\bibnamefont
  {Majorana}},\ }\bibinfo {title} {Atomi orientati in campo magnetico
  variabile},\ \href {\doibase 10.1007/bf02960953} {\bibfield  {journal}
  {\bibinfo  {journal} {Nuovo Cimento}\ }\textbf {\bibinfo {volume} {9}},\
  \bibinfo {pages} {43} (\bibinfo {year} {1932})}\BibitemShut {NoStop}%
\bibitem [{\citenamefont {Ghosh}\ \emph {et~al.}(2017)\citenamefont {Ghosh},
  \citenamefont {Coppersmith},\ and\ \citenamefont
  {Friesen}}]{PhysRevB.95.241307}%
  \BibitemOpen
  \bibfield  {author} {\bibinfo {author} {\bibfnamefont {J.}~\bibnamefont
  {Ghosh}}, \bibinfo {author} {\bibfnamefont {S.~N.}\ \bibnamefont
  {Coppersmith}}, \ and\ \bibinfo {author} {\bibfnamefont {M.}~\bibnamefont
  {Friesen}},\ }\bibinfo {title} {Pulse sequences for suppressing leakage in
  single-qubit gate operations},\ \href {\doibase 10.1103/PhysRevB.95.241307}
  {\bibfield  {journal} {\bibinfo  {journal} {Phys. Rev. B}\ }\textbf {\bibinfo
  {volume} {95}},\ \bibinfo {pages} {241307} (\bibinfo {year}
  {2017})}\BibitemShut {NoStop}%
\bibitem [{\citenamefont {Shi}\ \emph {et~al.}(2021)\citenamefont {Shi},
  \citenamefont {Wu}, \citenamefont {Shen}, \citenamefont {Song}, \citenamefont
  {Xia}, \citenamefont {Yi},\ and\ \citenamefont
  {Zheng}}]{PhysRevA.103.052612}%
  \BibitemOpen
  \bibfield  {author} {\bibinfo {author} {\bibfnamefont {Z.-C.}\ \bibnamefont
  {Shi}}, \bibinfo {author} {\bibfnamefont {H.-N.}\ \bibnamefont {Wu}},
  \bibinfo {author} {\bibfnamefont {L.-T.}\ \bibnamefont {Shen}}, \bibinfo
  {author} {\bibfnamefont {J.}~\bibnamefont {Song}}, \bibinfo {author}
  {\bibfnamefont {Y.}~\bibnamefont {Xia}}, \bibinfo {author} {\bibfnamefont
  {X.~X.}\ \bibnamefont {Yi}}, \ and\ \bibinfo {author} {\bibfnamefont {S.-B.}\
  \bibnamefont {Zheng}},\ }\bibinfo {title} {Robust single-qubit gates by
  composite pulses in three-level systems},\ \href {\doibase
  10.1103/PhysRevA.103.052612} {\bibfield  {journal} {\bibinfo  {journal}
  {Phys. Rev. A}\ }\textbf {\bibinfo {volume} {103}},\ \bibinfo {pages}
  {052612} (\bibinfo {year} {2021})}\BibitemShut {NoStop}%
\bibitem [{\citenamefont {Leibfried}\ \emph {et~al.}(2003)\citenamefont
  {Leibfried}, \citenamefont {Blatt}, \citenamefont {Monroe},\ and\
  \citenamefont {Wineland}}]{RevModPhys.75.281}%
  \BibitemOpen
  \bibfield  {author} {\bibinfo {author} {\bibfnamefont {D.}~\bibnamefont
  {Leibfried}}, \bibinfo {author} {\bibfnamefont {R.}~\bibnamefont {Blatt}},
  \bibinfo {author} {\bibfnamefont {C.}~\bibnamefont {Monroe}}, \ and\ \bibinfo
  {author} {\bibfnamefont {D.}~\bibnamefont {Wineland}},\ }\bibinfo {title}
  {Quantum dynamics of single trapped ions},\ \href {\doibase
  10.1103/RevModPhys.75.281} {\bibfield  {journal} {\bibinfo  {journal} {Rev.
  Mod. Phys.}\ }\textbf {\bibinfo {volume} {75}},\ \bibinfo {pages} {281}
  (\bibinfo {year} {2003})}\BibitemShut {NoStop}%
\bibitem [{\citenamefont {Yang}\ \emph {et~al.}(2010)\citenamefont {Yang},
  \citenamefont {Xu}, \citenamefont {Feng},\ and\ \citenamefont
  {Du}}]{Yang2010}%
  \BibitemOpen
  \bibfield  {author} {\bibinfo {author} {\bibfnamefont {W.}~\bibnamefont
  {Yang}}, \bibinfo {author} {\bibfnamefont {Z.}~\bibnamefont {Xu}}, \bibinfo
  {author} {\bibfnamefont {M.}~\bibnamefont {Feng}}, \ and\ \bibinfo {author}
  {\bibfnamefont {J.}~\bibnamefont {Du}},\ }\bibinfo {title} {Entanglement of
  separate nitrogen-vacancy centers coupled to a whispering-gallery mode
  cavity},\ \href {\doibase 10.1088/1367-2630/12/11/113039} {\bibfield
  {journal} {\bibinfo  {journal} {New J. Phys.}\ }\textbf {\bibinfo {volume}
  {12}},\ \bibinfo {pages} {113039} (\bibinfo {year} {2010})}\BibitemShut
  {NoStop}%
\bibitem [{\citenamefont {Xiang}\ \emph {et~al.}(2013)\citenamefont {Xiang},
  \citenamefont {Ashhab}, \citenamefont {You},\ and\ \citenamefont
  {Nori}}]{RevModPhys.85.623}%
  \BibitemOpen
  \bibfield  {author} {\bibinfo {author} {\bibfnamefont {Z.-L.}\ \bibnamefont
  {Xiang}}, \bibinfo {author} {\bibfnamefont {S.}~\bibnamefont {Ashhab}},
  \bibinfo {author} {\bibfnamefont {J.~Q.}\ \bibnamefont {You}}, \ and\
  \bibinfo {author} {\bibfnamefont {F.}~\bibnamefont {Nori}},\ }\bibinfo
  {title} {Hybrid quantum circuits: Superconducting circuits interacting with
  other quantum systems},\ \href {\doibase 10.1103/RevModPhys.85.623}
  {\bibfield  {journal} {\bibinfo  {journal} {Rev. Mod. Phys.}\ }\textbf
  {\bibinfo {volume} {85}},\ \bibinfo {pages} {623} (\bibinfo {year}
  {2013})}\BibitemShut {NoStop}%
\bibitem [{\citenamefont {Xue}\ \emph {et~al.}(2017)\citenamefont {Xue},
  \citenamefont {Gu}, \citenamefont {Hong}, \citenamefont {Yang}, \citenamefont
  {Zhang}, \citenamefont {Hu},\ and\ \citenamefont
  {You}}]{PhysRevApplied.7.054022}%
  \BibitemOpen
  \bibfield  {author} {\bibinfo {author} {\bibfnamefont {Z.-Y.}\ \bibnamefont
  {Xue}}, \bibinfo {author} {\bibfnamefont {F.-L.}\ \bibnamefont {Gu}},
  \bibinfo {author} {\bibfnamefont {Z.-P.}\ \bibnamefont {Hong}}, \bibinfo
  {author} {\bibfnamefont {Z.-H.}\ \bibnamefont {Yang}}, \bibinfo {author}
  {\bibfnamefont {D.-W.}\ \bibnamefont {Zhang}}, \bibinfo {author}
  {\bibfnamefont {Y.}~\bibnamefont {Hu}}, \ and\ \bibinfo {author}
  {\bibfnamefont {J.~Q.}\ \bibnamefont {You}},\ }\bibinfo {title} {Nonadiabatic
  Holonomic Quantum Computation with Dressed-State Qubits},\ \href {\doibase
  10.1103/PhysRevApplied.7.054022} {\bibfield  {journal} {\bibinfo  {journal}
  {Phys. Rev. Applied}\ }\textbf {\bibinfo {volume} {7}},\ \bibinfo {pages}
  {054022} (\bibinfo {year} {2017})}\BibitemShut {NoStop}%
\bibitem [{\citenamefont {Kang}\ \emph {et~al.}(2016)\citenamefont {Kang},
  \citenamefont {Chen}, \citenamefont {Shi}, \citenamefont {Song},\ and\
  \citenamefont {Xia}}]{PhysRevA.94.052311}%
  \BibitemOpen
  \bibfield  {author} {\bibinfo {author} {\bibfnamefont {Y.-H.}\ \bibnamefont
  {Kang}}, \bibinfo {author} {\bibfnamefont {Y.-H.}\ \bibnamefont {Chen}},
  \bibinfo {author} {\bibfnamefont {Z.-C.}\ \bibnamefont {Shi}}, \bibinfo
  {author} {\bibfnamefont {J.}~\bibnamefont {Song}}, \ and\ \bibinfo {author}
  {\bibfnamefont {Y.}~\bibnamefont {Xia}},\ }\bibinfo {title} {Fast preparation
  of $W$ states with superconducting quantum interference devices by using
  dressed states},\ \href {\doibase 10.1103/PhysRevA.94.052311} {\bibfield
  {journal} {\bibinfo  {journal} {Phys. Rev. A}\ }\textbf {\bibinfo {volume}
  {94}},\ \bibinfo {pages} {052311} (\bibinfo {year} {2016})}\BibitemShut
  {NoStop}%
\bibitem [{\citenamefont {Kang}\ \emph {et~al.}(2017)\citenamefont {Kang},
  \citenamefont {Chen}, \citenamefont {Shi}, \citenamefont {Huang},
  \citenamefont {Song},\ and\ \citenamefont {Xia}}]{PhysRevA.96.022304}%
  \BibitemOpen
  \bibfield  {author} {\bibinfo {author} {\bibfnamefont {Y.-H.}\ \bibnamefont
  {Kang}}, \bibinfo {author} {\bibfnamefont {Y.-H.}\ \bibnamefont {Chen}},
  \bibinfo {author} {\bibfnamefont {Z.-C.}\ \bibnamefont {Shi}}, \bibinfo
  {author} {\bibfnamefont {B.-H.}\ \bibnamefont {Huang}}, \bibinfo {author}
  {\bibfnamefont {J.}~\bibnamefont {Song}}, \ and\ \bibinfo {author}
  {\bibfnamefont {Y.}~\bibnamefont {Xia}},\ }\bibinfo {title} {Complete
  Bell-state analysis for superconducting-quantum-interference-device qubits
  with a transitionless tracking algorithm},\ \href {\doibase
  10.1103/PhysRevA.96.022304} {\bibfield  {journal} {\bibinfo  {journal} {Phys.
  Rev. A}\ }\textbf {\bibinfo {volume} {96}},\ \bibinfo {pages} {022304}
  (\bibinfo {year} {2017})}\BibitemShut {NoStop}%
\bibitem [{\citenamefont {Hong}\ \emph {et~al.}(2018)\citenamefont {Hong},
  \citenamefont {Liu}, \citenamefont {Cai}, \citenamefont {Zhang},
  \citenamefont {Hu}, \citenamefont {Wang},\ and\ \citenamefont
  {Xue}}]{PhysRevA.97.022332}%
  \BibitemOpen
  \bibfield  {author} {\bibinfo {author} {\bibfnamefont {Z.-P.}\ \bibnamefont
  {Hong}}, \bibinfo {author} {\bibfnamefont {B.-J.}\ \bibnamefont {Liu}},
  \bibinfo {author} {\bibfnamefont {J.-Q.}\ \bibnamefont {Cai}}, \bibinfo
  {author} {\bibfnamefont {X.-D.}\ \bibnamefont {Zhang}}, \bibinfo {author}
  {\bibfnamefont {Y.}~\bibnamefont {Hu}}, \bibinfo {author} {\bibfnamefont
  {Z.~D.}\ \bibnamefont {Wang}}, \ and\ \bibinfo {author} {\bibfnamefont
  {Z.-Y.}\ \bibnamefont {Xue}},\ }\bibinfo {title} {Implementing universal
  nonadiabatic holonomic quantum gates with transmons},\ \href {\doibase
  10.1103/PhysRevA.97.022332} {\bibfield  {journal} {\bibinfo  {journal} {Phys.
  Rev. A}\ }\textbf {\bibinfo {volume} {97}},\ \bibinfo {pages} {022332}
  (\bibinfo {year} {2018})}\BibitemShut {NoStop}%
\bibitem [{\citenamefont {Chen}\ \emph {et~al.}(2020)\citenamefont {Chen},
  \citenamefont {Shen},\ and\ \citenamefont {Xue}}]{PhysRevApplied.14.034038}%
  \BibitemOpen
  \bibfield  {author} {\bibinfo {author} {\bibfnamefont {T.}~\bibnamefont
  {Chen}}, \bibinfo {author} {\bibfnamefont {P.}~\bibnamefont {Shen}}, \ and\
  \bibinfo {author} {\bibfnamefont {Z.-Y.}\ \bibnamefont {Xue}},\ }\bibinfo
  {title} {Robust and Fast Holonomic Quantum Gates with Encoding on
  Superconducting Circuits},\ \href {\doibase 10.1103/PhysRevApplied.14.034038}
  {\bibfield  {journal} {\bibinfo  {journal} {Phys. Rev. Applied}\ }\textbf
  {\bibinfo {volume} {14}},\ \bibinfo {pages} {034038} (\bibinfo {year}
  {2020})}\BibitemShut {NoStop}%
\bibitem [{\citenamefont {Greentree}\ \emph {et~al.}(2004)\citenamefont
  {Greentree}, \citenamefont {Cole}, \citenamefont {Hamilton},\ and\
  \citenamefont {Hollenberg}}]{PhysRevB.70.235317}%
  \BibitemOpen
  \bibfield  {author} {\bibinfo {author} {\bibfnamefont {A.~D.}\ \bibnamefont
  {Greentree}}, \bibinfo {author} {\bibfnamefont {J.~H.}\ \bibnamefont {Cole}},
  \bibinfo {author} {\bibfnamefont {A.~R.}\ \bibnamefont {Hamilton}}, \ and\
  \bibinfo {author} {\bibfnamefont {L.~C.~L.}\ \bibnamefont {Hollenberg}},\
  }\bibinfo {title} {Coherent electronic transfer in quantum dot systems using
  adiabatic passage},\ \href {\doibase 10.1103/PhysRevB.70.235317} {\bibfield
  {journal} {\bibinfo  {journal} {Phys. Rev. B}\ }\textbf {\bibinfo {volume}
  {70}},\ \bibinfo {pages} {235317} (\bibinfo {year} {2004})}\BibitemShut
  {NoStop}%
\bibitem [{\citenamefont {Bruderer}\ \emph {et~al.}(2012)\citenamefont
  {Bruderer}, \citenamefont {Franke}, \citenamefont {Ragg}, \citenamefont
  {Belzig},\ and\ \citenamefont {Obreschkow}}]{PhysRevA.85.022312}%
  \BibitemOpen
  \bibfield  {author} {\bibinfo {author} {\bibfnamefont {M.}~\bibnamefont
  {Bruderer}}, \bibinfo {author} {\bibfnamefont {K.}~\bibnamefont {Franke}},
  \bibinfo {author} {\bibfnamefont {S.}~\bibnamefont {Ragg}}, \bibinfo {author}
  {\bibfnamefont {W.}~\bibnamefont {Belzig}}, \ and\ \bibinfo {author}
  {\bibfnamefont {D.}~\bibnamefont {Obreschkow}},\ }\bibinfo {title}
  {Exploiting boundary states of imperfect spin chains for high-fidelity state
  transfer},\ \href {\doibase 10.1103/PhysRevA.85.022312} {\bibfield  {journal}
  {\bibinfo  {journal} {Phys. Rev. A}\ }\textbf {\bibinfo {volume} {85}},\
  \bibinfo {pages} {022312} (\bibinfo {year} {2012})}\BibitemShut {NoStop}%
\bibitem [{\citenamefont {Chen}\ and\ \citenamefont {Li}(2016)}]{Chen2016}%
  \BibitemOpen
  \bibfield  {author} {\bibinfo {author} {\bibfnamefont {B.}~\bibnamefont
  {Chen}}\ and\ \bibinfo {author} {\bibfnamefont {Y.}~\bibnamefont {Li}},\
  }\bibinfo {title} {Coherent state transfer through a multi-channel quantum
  network: Natural versus controlled evolution passage},\ \href {\doibase
  10.1007/s11433-016-5791-y} {\bibfield  {journal} {\bibinfo  {journal} {Sci.
  China Phys. Mech. Astron.}\ }\textbf {\bibinfo {volume} {59}},\ \bibinfo
  {pages} {640302} (\bibinfo {year} {2016})}\BibitemShut {NoStop}%
\bibitem [{\citenamefont {Li}\ and\ \citenamefont
  {Shao}(2018)}]{PhysRevA.98.062338}%
  \BibitemOpen
  \bibfield  {author} {\bibinfo {author} {\bibfnamefont {D.~X.}\ \bibnamefont
  {Li}}\ and\ \bibinfo {author} {\bibfnamefont {X.~Q.}\ \bibnamefont {Shao}},\
  }\bibinfo {title} {Unconventional {R}ydberg pumping and applications in
  quantum information processing},\ \href {\doibase 10.1103/PhysRevA.98.062338}
  {\bibfield  {journal} {\bibinfo  {journal} {Phys. Rev. A}\ }\textbf {\bibinfo
  {volume} {98}},\ \bibinfo {pages} {062338} (\bibinfo {year}
  {2018})}\BibitemShut {NoStop}%
\bibitem [{\citenamefont {Kang}\ \emph {et~al.}(2018)\citenamefont {Kang},
  \citenamefont {Chen}, \citenamefont {Shi}, \citenamefont {Huang},
  \citenamefont {Song},\ and\ \citenamefont {Xia}}]{PhysRevA.97.042336}%
  \BibitemOpen
  \bibfield  {author} {\bibinfo {author} {\bibfnamefont {Y.-H.}\ \bibnamefont
  {Kang}}, \bibinfo {author} {\bibfnamefont {Y.-H.}\ \bibnamefont {Chen}},
  \bibinfo {author} {\bibfnamefont {Z.-C.}\ \bibnamefont {Shi}}, \bibinfo
  {author} {\bibfnamefont {B.-H.}\ \bibnamefont {Huang}}, \bibinfo {author}
  {\bibfnamefont {J.}~\bibnamefont {Song}}, \ and\ \bibinfo {author}
  {\bibfnamefont {Y.}~\bibnamefont {Xia}},\ }\bibinfo {title} {Nonadiabatic
  holonomic quantum computation using Rydberg blockade},\ \href {\doibase
  10.1103/PhysRevA.97.042336} {\bibfield  {journal} {\bibinfo  {journal} {Phys.
  Rev. A}\ }\textbf {\bibinfo {volume} {97}},\ \bibinfo {pages} {042336}
  (\bibinfo {year} {2018})}\BibitemShut {NoStop}%
\bibitem [{\citenamefont {Shi}\ \emph {et~al.}(2018)\citenamefont {Shi},
  \citenamefont {Ran}, \citenamefont {Shen}, \citenamefont {Xia},\ and\
  \citenamefont {Yi}}]{Shi:18}%
  \BibitemOpen
  \bibfield  {author} {\bibinfo {author} {\bibfnamefont {Z.-C.}\ \bibnamefont
  {Shi}}, \bibinfo {author} {\bibfnamefont {D.}~\bibnamefont {Ran}}, \bibinfo
  {author} {\bibfnamefont {L.-T.}\ \bibnamefont {Shen}}, \bibinfo {author}
  {\bibfnamefont {Y.}~\bibnamefont {Xia}}, \ and\ \bibinfo {author}
  {\bibfnamefont {X.~X.}\ \bibnamefont {Yi}},\ }\bibinfo {title} {Quantum state
  engineering by periodical two-step modulation in an atomic system},\ \href
  {\doibase 10.1364/OE.26.034789} {\bibfield  {journal} {\bibinfo  {journal}
  {Opt. Express}\ }\textbf {\bibinfo {volume} {26}},\ \bibinfo {pages} {34789}
  (\bibinfo {year} {2018})}\BibitemShut {NoStop}%
\bibitem [{\citenamefont {Li}\ \emph {et~al.}(2020)\citenamefont {Li},
  \citenamefont {Yu}, \citenamefont {Su},\ and\ \citenamefont
  {Qian}}]{PhysRevA.101.042328}%
  \BibitemOpen
  \bibfield  {author} {\bibinfo {author} {\bibfnamefont {R.}~\bibnamefont
  {Li}}, \bibinfo {author} {\bibfnamefont {D.}~\bibnamefont {Yu}}, \bibinfo
  {author} {\bibfnamefont {S.-L.}\ \bibnamefont {Su}}, \ and\ \bibinfo {author}
  {\bibfnamefont {J.}~\bibnamefont {Qian}},\ }\bibinfo {title} {Periodically
  driven facilitated high-efficiency dissipative entanglement with Rydberg
  atoms},\ \href {\doibase 10.1103/PhysRevA.101.042328} {\bibfield  {journal}
  {\bibinfo  {journal} {Phys. Rev. A}\ }\textbf {\bibinfo {volume} {101}},\
  \bibinfo {pages} {042328} (\bibinfo {year} {2020})}\BibitemShut {NoStop}%
\bibitem [{\citenamefont {Shao}(2020)}]{PhysRevA.102.053118}%
  \BibitemOpen
  \bibfield  {author} {\bibinfo {author} {\bibfnamefont {X.-Q.}\ \bibnamefont
  {Shao}},\ }\bibinfo {title} {Selective {R}ydberg pumping via strong dipole
  blockade},\ \href {\doibase 10.1103/PhysRevA.102.053118} {\bibfield
  {journal} {\bibinfo  {journal} {Phys. Rev. A}\ }\textbf {\bibinfo {volume}
  {102}},\ \bibinfo {pages} {053118} (\bibinfo {year} {2020})}\BibitemShut
  {NoStop}%
\bibitem [{\citenamefont {Wu}\ \emph {et~al.}(2021)\citenamefont {Wu},
  \citenamefont {Wang}, \citenamefont {Han}, \citenamefont {Su}, \citenamefont
  {Xia}, \citenamefont {Jiang},\ and\ \citenamefont
  {Song}}]{PhysRevA.103.012601}%
  \BibitemOpen
  \bibfield  {author} {\bibinfo {author} {\bibfnamefont {J.-L.}\ \bibnamefont
  {Wu}}, \bibinfo {author} {\bibfnamefont {Y.}~\bibnamefont {Wang}}, \bibinfo
  {author} {\bibfnamefont {J.-X.}\ \bibnamefont {Han}}, \bibinfo {author}
  {\bibfnamefont {S.-L.}\ \bibnamefont {Su}}, \bibinfo {author} {\bibfnamefont
  {Y.}~\bibnamefont {Xia}}, \bibinfo {author} {\bibfnamefont {Y.}~\bibnamefont
  {Jiang}}, \ and\ \bibinfo {author} {\bibfnamefont {J.}~\bibnamefont {Song}},\
  }\bibinfo {title} {Resilient quantum gates on periodically driven Rydberg
  atoms},\ \href {\doibase 10.1103/PhysRevA.103.012601} {\bibfield  {journal}
  {\bibinfo  {journal} {Phys. Rev. A}\ }\textbf {\bibinfo {volume} {103}},\
  \bibinfo {pages} {012601} (\bibinfo {year} {2021})}\BibitemShut {NoStop}%
\bibitem [{\citenamefont {Taie}\ \emph {et~al.}(2020)\citenamefont {Taie},
  \citenamefont {Ichinose}, \citenamefont {Ozawa},\ and\ \citenamefont
  {Takahashi}}]{Taie2020}%
  \BibitemOpen
  \bibfield  {author} {\bibinfo {author} {\bibfnamefont {S.}~\bibnamefont
  {Taie}}, \bibinfo {author} {\bibfnamefont {T.}~\bibnamefont {Ichinose}},
  \bibinfo {author} {\bibfnamefont {H.}~\bibnamefont {Ozawa}}, \ and\ \bibinfo
  {author} {\bibfnamefont {Y.}~\bibnamefont {Takahashi}},\ }\bibinfo {title}
  {Spatial adiabatic passage of massive quantum particles in an optical Lieb
  lattice},\ \href {\doibase 10.1038/s41467-019-14165-3} {\bibfield  {journal}
  {\bibinfo  {journal} {Nat. Commun.}\ }\textbf {\bibinfo {volume} {11}},\
  \bibinfo {pages} {257} (\bibinfo {year} {2020})}\BibitemShut {NoStop}%
\bibitem [{\citenamefont {Yang}\ \emph {et~al.}(2021)\citenamefont {Yang},
  \citenamefont {Ma}, \citenamefont {Wu}, \citenamefont {Wang}, \citenamefont
  {Cao}, \citenamefont {Guo}, \citenamefont {Huang}, \citenamefont {Feng},
  \citenamefont {Zhou},\ and\ \citenamefont {Duan}}]{yang2021realizing}%
  \BibitemOpen
  \bibfield  {author} {\bibinfo {author} {\bibfnamefont {H.~X.}\ \bibnamefont
  {Yang}}, \bibinfo {author} {\bibfnamefont {J.~Y.}\ \bibnamefont {Ma}},
  \bibinfo {author} {\bibfnamefont {Y.~K.}\ \bibnamefont {Wu}}, \bibinfo
  {author} {\bibfnamefont {Y.}~\bibnamefont {Wang}}, \bibinfo {author}
  {\bibfnamefont {M.~M.}\ \bibnamefont {Cao}}, \bibinfo {author} {\bibfnamefont
  {W.~X.}\ \bibnamefont {Guo}}, \bibinfo {author} {\bibfnamefont {Y.~Y.}\
  \bibnamefont {Huang}}, \bibinfo {author} {\bibfnamefont {L.}~\bibnamefont
  {Feng}}, \bibinfo {author} {\bibfnamefont {Z.~C.}\ \bibnamefont {Zhou}}, \
  and\ \bibinfo {author} {\bibfnamefont {L.~M.}\ \bibnamefont {Duan}},\
  }\bibinfo {title} {Realizing coherently convertible dual-type qubits with the
  same ion species},\ \href {https://arxiv.org/abs/2106.14906v1} {\bibfield
  {journal} {\bibinfo  {journal} {arXiv: 2106.14906}\ } (\bibinfo {year}
  {2021})}\BibitemShut {NoStop}%
\bibitem [{\citenamefont {Shao}\ \emph {et~al.}(2010)\citenamefont {Shao},
  \citenamefont {Wang}, \citenamefont {Chen}, \citenamefont {Zhang},
  \citenamefont {Zhao},\ and\ \citenamefont {Yeon}}]{Shao2010}%
  \BibitemOpen
  \bibfield  {author} {\bibinfo {author} {\bibfnamefont {X.-Q.}\ \bibnamefont
  {Shao}}, \bibinfo {author} {\bibfnamefont {H.-F.}\ \bibnamefont {Wang}},
  \bibinfo {author} {\bibfnamefont {L.}~\bibnamefont {Chen}}, \bibinfo {author}
  {\bibfnamefont {S.}~\bibnamefont {Zhang}}, \bibinfo {author} {\bibfnamefont
  {Y.-F.}\ \bibnamefont {Zhao}}, \ and\ \bibinfo {author} {\bibfnamefont
  {K.-H.}\ \bibnamefont {Yeon}},\ }\bibinfo {title} {Converting two-atom
  singlet state into three-atom singlet state via quantum {Z}eno dynamics},\
  \href {\doibase 10.1088/1367-2630/12/2/023040} {\bibfield  {journal}
  {\bibinfo  {journal} {New J. Phys.}\ }\textbf {\bibinfo {volume} {12}},\
  \bibinfo {pages} {023040} (\bibinfo {year} {2010})}\BibitemShut {NoStop}%
\bibitem [{\citenamefont {Torosov}\ \emph {et~al.}(2021)\citenamefont
  {Torosov}, \citenamefont {Shore},\ and\ \citenamefont
  {Vitanov}}]{PhysRevA.103.033110}%
  \BibitemOpen
  \bibfield  {author} {\bibinfo {author} {\bibfnamefont {B.~T.}\ \bibnamefont
  {Torosov}}, \bibinfo {author} {\bibfnamefont {B.~W.}\ \bibnamefont {Shore}},
  \ and\ \bibinfo {author} {\bibfnamefont {N.~V.}\ \bibnamefont {Vitanov}},\
  }\bibinfo {title} {Coherent control techniques for two-state quantum systems:
  A comparative study},\ \href {\doibase 10.1103/PhysRevA.103.033110}
  {\bibfield  {journal} {\bibinfo  {journal} {Phys. Rev. A}\ }\textbf {\bibinfo
  {volume} {103}},\ \bibinfo {pages} {033110} (\bibinfo {year}
  {2021})}\BibitemShut {NoStop}%
\bibitem [{\citenamefont {Torosov}\ and\ \citenamefont
  {Vitanov}(2019{\natexlab{c}})}]{PhysRevA.100.023410}%
  \BibitemOpen
  \bibfield  {author} {\bibinfo {author} {\bibfnamefont {B.~T.}\ \bibnamefont
  {Torosov}}\ and\ \bibinfo {author} {\bibfnamefont {N.~V.}\ \bibnamefont
  {Vitanov}},\ }\bibinfo {title} {Composite pulses with errant phases},\ \href
  {\doibase 10.1103/PhysRevA.100.023410} {\bibfield  {journal} {\bibinfo
  {journal} {Phys. Rev. A}\ }\textbf {\bibinfo {volume} {100}},\ \bibinfo
  {pages} {023410} (\bibinfo {year} {2019}{\natexlab{c}})}\BibitemShut
  {NoStop}%
\bibitem [{\citenamefont {Brion}\ \emph {et~al.}(2007)\citenamefont {Brion},
  \citenamefont {Pedersen},\ and\ \citenamefont {M{\o}lmer}}]{Brion_2007}%
  \BibitemOpen
  \bibfield  {author} {\bibinfo {author} {\bibfnamefont {E.}~\bibnamefont
  {Brion}}, \bibinfo {author} {\bibfnamefont {L.~H.}\ \bibnamefont {Pedersen}},
  \ and\ \bibinfo {author} {\bibfnamefont {K.}~\bibnamefont {M{\o}lmer}},\
  }\bibinfo {title} {Adiabatic elimination in a lambda system},\ \href
  {\doibase 10.1088/1751-8113/40/5/011} {\bibfield  {journal} {\bibinfo
  {journal} {J. Phys. A: Math. Theor.}\ }\textbf {\bibinfo {volume} {40}},\
  \bibinfo {pages} {1033} (\bibinfo {year} {2007})}\BibitemShut {NoStop}%
\bibitem [{\citenamefont {Vitanov}\ \emph {et~al.}(2017)\citenamefont
  {Vitanov}, \citenamefont {Rangelov}, \citenamefont {Shore},\ and\
  \citenamefont {Bergmann}}]{RevModPhys.89.015006}%
  \BibitemOpen
  \bibfield  {author} {\bibinfo {author} {\bibfnamefont {N.~V.}\ \bibnamefont
  {Vitanov}}, \bibinfo {author} {\bibfnamefont {A.~A.}\ \bibnamefont
  {Rangelov}}, \bibinfo {author} {\bibfnamefont {B.~W.}\ \bibnamefont {Shore}},
  \ and\ \bibinfo {author} {\bibfnamefont {K.}~\bibnamefont {Bergmann}},\
  }\bibinfo {title} {Stimulated Raman adiabatic passage in physics, chemistry,
  and beyond},\ \href {\doibase 10.1103/RevModPhys.89.015006} {\bibfield
  {journal} {\bibinfo  {journal} {Rev. Mod. Phys.}\ }\textbf {\bibinfo {volume}
  {89}},\ \bibinfo {pages} {015006} (\bibinfo {year} {2017})}\BibitemShut
  {NoStop}%
\end{thebibliography}%

\end{document}